\documentclass[11pt]{article}

\usepackage{epsfig,graphicx,color}
\usepackage{amsmath,amsfonts,amssymb}

\newcommand{\unit}{\leavevmode\hbox{\small1\kern-3.6pt\normalsize1}}

\def\lsim{\raise0.3ex\hbox{$\;<$\kern-0.75em\raise-1.1ex\hbox{$\sim\;$}}}
\def\gsim{\raise0.3ex\hbox{$\;>$\kern-0.75em\raise-1.1ex\hbox{$\sim\;$}}}

\def\locald{\rho_0}

\def\mnucl{m_{N}}
\def\redN{\mu_N}
\def\redn{\mu_n}
\def\sigmaN{\sigma_{WN}}

\newcommand{\sigsi}{\sigma^{SI}}
\newcommand{\sigsd}{\sigma^{SD}}

\newcommand{\sigsdp}{\sigma^{SD,\,p}}
\newcommand{\sigsdn}{\sigma^{SD,\,n}}

\newcommand{\sigsinu}{\sigma^{SI,\,N}_0}
\newcommand{\sigsdnu}{\sigma^{SD,\,N}_0}

\newcommand{\mwimp}{m_\chi}
\def\msi{(\sigsi,\,\mwimp)}
\def\msd{(\sigsd,\,\mwimp)}
\def\sisd{(\sigsd,\,\sigsi)}

\allowdisplaybreaks

\begin{document}

\thispagestyle{empty}
\begin{flushright}
  IFT-UAM/CSIC-13-032\\
  FTUAM-13-6\\

  \vspace*{2.mm}{April 8, 2013}
\end{flushright}

\begin{center}
  {\Large \textbf{Complementarity of dark matter direct detection:\\ the role of bolometric targets } }  
  
  \vspace{0.5cm}
  D.G.~Cerde\~no$^{1,2}$,
  C.~Cuesta$^{3,a}$,
  M.~Fornasa$^{4}$,
  E.~Garc\'ia$^{3}$,
  C.~Ginestra$^3$,
  Ji-Haeng Huh$^{1,2,5}$, 
  M.~Mart\'inez$^{3,6}$,
  Y.~Ortigoza$^{3,b}$,
  M.~Peir\'o$^{1,2,a}$,
  J.~Puimed\'on$^3$,
  L.~Robledo$^{2}$,
  M.L.~Sarsa$^{3}$
  \\[0.2cm] 
    
  {\footnotesize{ 
      ${}^1$ 
      Instituto de F\'{\i}sica Te\'{o}rica
      UAM/CSIC, Universidad Aut\'{o}noma de Madrid,  28049
      Madrid, Spain\\[0pt] 
      ${}^2$ Departamento de F\'{\i}sica Te\'{o}rica,
      Universidad Aut\'{o}noma de Madrid, 28049
      Madrid, Spain\\[0pt] 
      ${}^3$
      Grupo de F\'\i sica Nuclear y Astropart\'\i culas,
      Universidad de Zaragoza, 
      50009 Zaragoza, Spain\\[0pt] 
      ${}^4$ School of Physics and Astronomy, University of Nottingham, 
      University Park, NG7 2RD,\\ Nottingham, United Kingdom\\[0pt] 
      ${}^5$ Department of Physics and Astronomy, UCLA, 475 Portola Plaza, Los Angeles, CA 90095, USA\\
      ${}^6$ Fundaci\'on ARAID, Mar\'ia de Luna 11, Edificio CEEI Arag\'on, 50018 Zaragoza, Spain\\
        }}
  
\vspace*{0.7cm}
\begin{abstract}
We study how the combined observation of dark matter in various direct detection experiments can be used to determine the phenomenological properties of WIMP dark matter: mass, spin-dependent (SD) and spin-independent (SI) scattering cross section off nucleons. A convenient choice of target materials, including nuclei that couple to dark matter particles through a significantly different ratio of SD vs SI interactions, could break the degeneracies in the determination of those parameters that a single experiment cannot discriminate. In this work we investigate different targets that can be used as scintillating bolometers and could provide complementary information to germanium and xenon detectors. We observe that Al$_2$O$_3$ and LiF bolometers could allow a good reconstruction of the DM properties over regions of the parameter space with a SD scattering cross section as small as $10^{-5}$ pb and a SI cross section as small as $5\times10^{-10}$~pb for a 50 GeV WIMP. In the case of a CaWO$_4$ bolometer the area in which full complementarity is obtained is smaller but we show that it can be used to determine the WIMP mass and its SI cross section.
For each target we study the required exposure and background.
\end{abstract}
\end{center}

\let\oldthefootnote\thefootnote
\renewcommand{\thefootnote}{\alph{footnote}}
\footnotetext[1]{MultiDark Scholar}
\footnotetext[2]{MultiDark Fellow}
\let\thefootnote\oldthefootnote
\setcounter{footnote}{0}

\newpage
	

\section{Introduction} 
\label{sec:intro}
The detection and identification of the dark matter (DM) is a challenging goal that is currently being pursued by a large number of experiments around the world using different techniques. DM can be searched for directly (attempting to observe its scattering off nuclei in a detector), indirectly (looking for the products of its annihilation or decay), and in particle colliders (which explore the nature of physics at the TeV scale, the typical scale for many models of particle DM). Such variety of strategies is sensitive to DM candidates with very different properties, allowing the exploration of a wide range of particle models. Among the various possibilities, a generic weakly-interacting massive particle (WIMP) is well motivated since it predicts quite naturally a value for the DM thermal relic density that is of the same order of magnitude as the observed cold DM relic abundance. 

Direct DM detection is currently undergoing an exciting situation, with several experimental collaborations (using different targets and techniques) reporting potential signals of WIMP DM that have, nevertheless, not showed up in other detectors. In particular, an annual modulation in the detection rate was observed by the DAMA collaboration using NaI(Tl) as target \cite{Bernabei:2003za} and was later confirmed by the extended experiment DAMA/LIBRA \cite{Bernabei:2008yi} reaching a statistical significance of 8.9\,$\sigma$. A new experiment, ANAIS, projected to be carried out at the Canfranc Underground Laboratory with up to 250 kg of NaI(Tl) would test this observation in a model independent way \cite{cebrian2012,Amare:2012ex}. Also, the CoGeNT collaboration (using a germanium target) observed an irreducible excess in their data \cite{Aalseth:2010vx} that, if interpreted in terms of WIMPs, would correspond to a very light particle, with mass in the range $7-12$~GeV, and a large elastic scattering cross section, of order $10^{-4}$~pb. Furthermore, hints of an annual modulation in the CoGeNT experiment were also claimed \cite{Aalseth:2011wp} but with a limited statistical significance of 2.8\,$\sigma$ in the first year data. Moreover, data presented by the CRESST collaboration (which uses CaWO$_4$ as target and beta/gamma background rejection) also displayed an excess that could be compatible with light WIMPs \cite{Angloher:2011uu}. Several analyses have investigated the compatibility of these three signals \cite{Schwetz:2011xm,Hooper:2011hd,Farina:2011pw,McCabe:2011sr,Belli:2011kw}, although this only seems possible if extreme assumptions are made for the different uncertainties, such as the inclusion of large quenching factors or channeling effects. 

These observations are challenged by the negative results obtained by other experimental collaborations. Most notably, CDMS II \cite{Ahmed:2009zw,Akerib:2010pv}, XENON10 \cite{Angle:2011th}, XENON100 \cite{Aprile:2011hi,Aprile:2012nq}, SIMPLE \cite{Felizardo:2011uw}, KIMS \cite{Kim:2012rz} and a combination of CDMS and EDELWEISS data \cite{Ahmed:2011gh} have set upper bounds on the spin-independent (SI) part of the WIMP-nucleon cross section that are in strong tension with the regions of the parameter space compatible with WIMP signals in the DAMA/LIBRA or CoGeNT experiments. Several alternatives to ordinary WIMP DM have been proposed in the literature, trying to account for this discrepancy. Moreover, CDMS~II has recently searched for annual modulation in a  reanalysis of their data and did not observe any indication for it \cite{Ahmed:2012vq}. This further strengthens its incompatibility with CoGeNT since both use the same target, leaving no room for model dependence affecting the comparison. Similarly, the spin-dependent (SD) contribution to the WIMP-nucleon cross section is also constrained from negative results in direct detection experiments. The leading constraints are due to XENON100 \cite{Aprile:2012nq,xenon100sd,Garny:2012it} (SD cross section with neutrons, $\sigsdn$) and COUPP \cite{Behnke:2010xt}, PICASSO \cite{Archambault:2012pm}, SIMPLE \cite{Felizardo:2011uw} and KIMS \cite{Kim:2012rz}  (SD cross section with protons, $\sigsdp$).

In the near future more sensitive experiments are going to continue probing the DM parameter space. This will allow us to clarify the current situation regarding light WIMPs and also explore DM candidates with smaller interaction cross sections. In particular, some of the existing experiments are involved in the upgrading of their detectors, moving towards several hundreds kilograms of target material or even reaching the 1 ton scale. This is the case of the XENON1T \cite{Aprile:2012zx} and LUX \cite{Akerib:2012ys} collaborations, with a xenon target, SuperCDMS \cite{Brink:2012zza}, using germanium, COUPP, with a C$_3$FI target, and ArDM, using liquid argon \cite{Marchionni:2010fi}.
On a similar timescale, the EURECA \cite{Kraus:2011zz} consortium has plans for a 1 ton scale experiment, able to operate different types of cryogenic detectors consisting of Ge bolometers measuring heat and ionization, and CaWO$_4$ scintillating bolometers measuring heat and light, although other scintillating targets could also be accommodated. The CRESST \cite{bavykina:2009} and ROSEBUD \cite{calleja:2008} collaborations have worked on R\&D and tested other potential scintillating targets, focusing especially on targets that contain nuclei with enhanced sensitivity to SD interactions and low mass WIMPs. According to the characterization and performances of the different scintillating bolometers and their complementarity with Ge and Xe for the determination of the DM properties, some of these materials could be incorporated as additional targets for EURECA, probably in the second phase. 

If DM is detected, the use of different targets is crucial, as it can serve to unambiguously determine some of the WIMP properties (e.g., its mass and interaction cross section off protons and neutrons), thus helping to discriminate among the various WIMP candidates. This idea was applied to the case of the COUPP experiment in Ref.\,\cite{Bertone:2007xj}, emphasizing the relevant role of targets which are sensitive to SD WIMP-nucleus interactions and showing how detection in two complementary targets (in that case C$_4$F$_{10}$ and C$_3$FI) could allow a better measurement of the WIMP couplings.
The idea of target complementarity has later been applied to the determination of the WIMP mass and cross section from different DM experiments \cite{Drees:2008bv,Pato:2010zk} and the relevance of targets sensitive to the SD cross section has been analytically studied in Ref.\,\cite{Cannoni:2010ae}.

In this paper we focus on the determination of the phenomenological parameters of a generic WIMP, namely its mass, $\mwimp$, the SI contribution to the WIMP-nucleon cross section, $\sigsi$, and the SD component, $\sigsd$. In doing so we take into account all known sources of uncertainties, including those in the astrophysical parameters describing the DM halo, but also considering uncertainties in the nuclear form factors along the lines of our previous analysis in Ref.\,\cite{Cerdeno:2012ix}. We follow an approach based on Bayesian statistics that allows us to incorporate these uncertainties in a consistent way, thereby quantifying their effect.

With these tools, we carry out a systematic study of the performance of various targets used in direct DM searches. We first show that 1 ton experiments based on germanium and xenon might be unable to determine some of the WIMP parameters (in particular failing to measure the SD component of the cross section). We also show that although C$_3$FI is a good alternative, the complementary capability of COUPP is limited by its detection technique since it does not provide information about the recoil spectrum. We then turn our attention to other possible experiments and study how their use in combination with germanium and xenon can serve to unambiguously determine $\mwimp$, $\sigsi$, and $\sigsd$ in certain scenarios, a situation that we define as {\em complementarity}. We concentrate on the scintillating bolometers currently used by CRESST (CaWO$_4$) and on those characterised by the ROSEBUD collaboration (Al$_2$O$_3$ and LiF), which contain nuclei which are sensitive to the WIMP SD coupling and are also optimal for searches for low mass WIMPs. We show how for a certain range in the phenomenological parameter space, Al$_2$O$_3$ and LiF can provide a good complementary measurement which allows the degeneracy to be disentangled in the SI and SD contributions to the total cross section. This is generally the case when the detection rate in germanium and xenon is dominated by the SI component. On the other hand, CaWO$_4$ is a more convenient target when the rate in germanium or xenon is mostly SD (since tungsten is heavier and more sensitive to the SI component) and can be used to provide a good measurement of $\sigsi$. 

The paper is organised as follows. In Section\,\ref{sec:dd} we summarise the physics of direct DM detection and the determination of WIMP parameters from the observed number of events and energy spectrum. We describe the reconstruction method used in this work, based on Bayesian statistics and show explicitly that conventional targets which are more sensitive to the SI WIMP-nucleus cross section (such as Xe and Ge) do not, in general, allow for a complete reconstruction of all the WIMP parameters. We also show how the situation changes when a target that is sensitive to SD interactions (using a hypothetical 1 ton version of the COUPP experiment).
In Section\,\ref{sec:rosebud} we study various possible targets used in the CRESST and ROSEBUD experiments and analyse the conditions under which complementarity is achieved. 
The conclusions are left for Section\,\ref{sec:conclusions}.

\section{Determination of WIMP properties from direct detection}
\label{sec:dd}
We start by summarising the basic expressions that describe the WIMP rate in a direct DM detection experiment \cite{Smith:1988kw} (for a recent review see Ref.\,\cite{Cerdeno:2010jj}).
The differential event rate for the elastic scattering of a WIMP with mass $\mwimp$ off a nucleus with mass $\mnucl$ and a detector mass $M_{det}$ is given by
\begin{equation}
  \frac{dR}{dE_R}=\frac{M_{det}\,\locald}{\mnucl\,\mwimp}\int_{v_{min}}^{v_{esc}}  v
  f(v)\, \frac{d\sigmaN}{dE_R}(v,E_R)\, d v\,,
  \label{drate}
\end{equation}
where $\locald$ is the local WIMP density, $f(v)$ is the WIMP speed distribution normalized to unity and velocities are expressed in the detector reference frame. The integration over the WIMP velocity is performed from the minimum WIMP speed needed to induce a recoil of energy $E_{R}$, $v_{min}=\sqrt{(\mnucl E_R)/(2\redN^2)}$ where $\redN$ is the WIMP-nucleus reduced mass, to a escape velocity $v_{ esc}$ also in the detector reference frame, above which WIMPs are not gravitationally bound to the Milky Way.
The WIMP-nucleus differential cross section $d\sigmaN/dE_R$ is computed from the Lagrangian that describes the interaction of a given WIMP with ordinary matter and encodes the particle physics input. 
The total event rate is then calculated by integrating the differential event rate over all the possible recoil energies,
 \begin{equation}
  R=\int_{E_T}^{E_{max}}
  dE_R\frac{M_{det}\,\locald}{\mnucl\,\mwimp}\int_{v_{min}}^{v_{esc}}
  v f(v)\, \frac{d\sigmaN}{dE_R}(v,E_R)\, d v \,.
  \label{rate}
\end{equation}
The window for WIMP searches is selected from a threshold energy, $E_{T}$, to a maximal recoil energy, $E_{max}$, and depends on the specific experiment. The total number of DM recoils is $N=R\,t$, where $t$ is the live time of the experiment. In the following, we will also refer to the exposure, defined as $\epsilon=M_{det}t$.

In general, the WIMP-nucleon scattering cross section can be separated into a SI and a SD contribution. The total WIMP-nucleus cross section is calculated by adding these coherently, using nuclear wave functions. The differential cross section thus reads
\begin{equation}
	\frac{d\sigmaN}{dE_R} = \frac{m_N}{2 \mu_N^2 v^2}
	\left(\sigsinu F^2_{SI}(E_R) + 
	\sigsdnu F^2_{SD}(E_R) \right),
	\label{eqn:SI_and_SD}
\end{equation}
where $\sigsinu$ and $\sigsdnu$ are the SI and SD components of the WIMP-nucleus cross sections at zero momentum transfer, and the form factors $F^2_{SI,\,SD}(E_R)$ account for the coherence loss which leads to a suppression in the event rate for heavy nuclei in the SI and SD contributions. See Ref.\,\cite{Lewin:1995rx} for a complete description of these prescriptions.

The WIMP-nucleus cross section at zero momentum transfer can be written as \cite{Lewin:1995rx,Jungman:1995df}
\begin{eqnarray}
\sigsinu &=& \frac{4\redN^2}{\pi}\,\left[Z f_p+(A-Z) f_n\right]^2\,,\nonumber\\
\sigsdnu &=& \frac{32\redN^2G_F^2}{\pi}\,\left[a_p S_p+ a_n S_n\right]^2\,\left(\frac{J+1}{J}\right)\,,
\end{eqnarray}
where $S_p$ and $S_n$ are the expectation values of the total proton and neutron spin operators; $f_p$, $f_n$ and $a_p$, $a_n$ are the effective WIMP couplings to protons and neutrons in the SI and SD case, respectively; $G_F$ is the Fermi coupling constant, 
and $J$ is the total nuclear spin.\footnote{Notice that for simplicity we have not included here a possible vector coupling (corresponding to non-Majorana DM particles), which would lead to an extra contribution in the expression for $\sigsinu$. }

In the following we will assume that $f_p=f_n$, so that
\begin{equation}
\sigsinu=\left(\frac{\redN}{\redn}\right)^2A^2 \sigsi\,,
\end{equation}
where $\sigsi$ is the WIMP-nucleon SI cross section and $\redn$ is the WIMP-nucleon reduced mass. In the SD dependent case we can also express $\sigsdnu$ in terms of the WIMP-proton and WIMP-neutron cross sections as \cite{Tovey:2000mm}
\begin{equation}
\sigsdnu=\frac{4}{3}\left(\frac{J+1}{J}\right)\left(\frac{\redN}{\redn}\right)^2 \left(S_p\,\sqrt{\sigsdp}+S_n\,\sqrt{\sigsdn}\right)^2\,,
\end{equation}
We will assume a specific relation between the couplings to protons and neutrons, namely $a_p/a_n=-1$ so that only one parameter $\sigsdp=\sigsdn\equiv\sigsd$ will be needed to describe SD interactions\footnote{The analysis can be extended to consider a larger parameter space, which would require further study. This simplification is motivated by the fact that particle models for DM matter generally predict $|\sigsdn|\approx|\sigsdp|$.}
 (see Section\,\ref{sec:complementarity}).

In this paper we follow a phenomenological approach, where no specific particle physics model for the DM is assumed. Instead, the WIMP is characterised simply by its mass $m_\chi$, spin-independent and spin-dependent WIMP-nucleon interaction cross sections, $\sigsi$ and $\sigsd$. 

If a DM signal is obtained in a direct detection experiment, the observed number of events and (if the experiment provides it) the energy dependence of the differential rate, i.e., the energy spectrum, can be used to reconstruct the properties of the DM particle \cite{Green:2007rb,Green:2008rd,Drees:2007hr,Drees:2008bv}.  In doing this, it is important to remember that expression (\ref{drate}) is subject to uncertainties in the nuclear form factors and in the parameters describing the DM halo. Determining the impact of these is crucial to understand the capability of a DM experiment to reconstruct the WIMP properties. In particular, astrophysical uncertainties are known to significantly affect the reconstruction of the mass and scattering cross section of the DM, see e.g., Refs.\,\cite{Green:2010gw,Green:2011bv}. Similarly, uncertainties in the SD structure functions can lead to a mis-reconstruction of the WIMP mass and SD scattering cross section \cite{Cerdeno:2012ix}. 

For a given experimental setup (we use the label $a$ to denote the target) we define an energy window for WIMP searches, from a threshold, $E_T^a$ to a maximum energy $E_{max}^a$. We divide that energy range into energy bins $\{E_i^a,E_i^a+\Delta E^a\}$ with a width $\Delta E^a$. We then compute, for a choice of DM parameters, the expected number of events $\{ \lambda_i^a\}$ in each energy bin, by integrating Eq.\,(\ref{drate}) in the corresponding interval for a given live time and adding a certain level of background events $B_i^a$. The specific energy windows, bin size and background assumed can be found in Appendix\,\ref{sec:targets} for the experimental set-ups considered.

We consider the quantities $\{ \lambda_i^a\}$ as the experimental information from which we attempt to reconstruct the DM parameters. Our analysis is based on the Bayes theorem, which determines the posterior probability distribution function (pdf) $p(\mathbf{\Theta}|\mathbf{D})$ of a set of parameters $\mathbf{\Theta}$ (for which a prior probability is assumed $p(\mathbf{\Theta})$) from a set of experimental data $\mathbf{D}$, encoded in the likelihood function $p(\mathbf{D}|\mathbf{\Theta})$ (or $\mathcal{L}(\mathbf{\Theta}$)),
\begin{equation}
	p(\mathbf{\Theta}|\mathbf{D}) = 
	\frac{p(\mathbf{D}|\mathbf{\Theta}) p(\mathbf{\Theta})}{p({\mathbf{D})}}\,.
	\label{eqn:Bayes}
\end{equation}
The evidence $p(\mathbf{D})$ in the denominator of Eq.\,(\ref{eqn:Bayes}) is a function of only the experimental data. For our purposes it works as a normalization factor and can therefore be ignored. The pdf in Eq.\,(\ref{eqn:Bayes}), in principle, depends on the priors $p(\mathbf{\Theta})$ and different choices of priors can affect the shape of the final pdf. 
However, should this happen, it would mean that the experimental data are not constraining enough, not being able to dominate the final probability distribution. Residual prior dependence can be seen, e.g., in Refs.\,\cite{Bertone:2011nj,Feroz:2011bj,deAustri:2006pe}.
The scans over the parameter space that allow us to retrieve the pdf are performed with MultiNest 2.9 \cite{Feroz:2008xx,Feroz:2007kg} interfaced with our own code for the computation of the number of recoil events and the likelihood. Scans are performed with 20000 live points and a tolerance of 0.0001 to reach a good sampling of the profile likelihood (see below) as found in Ref.\,\cite{Feroz:2011bj}.

In our case the experimental data consists of the predicted sets of binned WIMP induced nuclear recoils for each target, $\mathbf{D}=(\{\lambda_i^a\})$. The parameter space is $\mathbf{\Theta}=(\mwimp,\,\sigsi,\,\sigsd)$ and our scans span the following range: $\mwimp=1-10^5$~GeV, $\sigsi=10^{-12}-10^{-6}$~pb, and $\sigsd=10^{-8}-1$~pb. Logarithmic flat priors are assumed for the three variables.  
The parameter space is further extended with inputs describing the astrophysical and nuclear uncertainties, which are considered  nuisance parameters. 
See Appendix\,\ref{app:sdsf} for a detailed description of the parameterization of nuclear uncertainties for the different targets. Astrophysical uncertainties have been included as in Ref.\,\cite{Cerdeno:2012ix}. Unless otherwise stated, all the scans in this work include both astrophysical and nuclear uncertainties. 

The likelihood $\mathcal{L}(\mathbf{\Theta})$ is computed for each point in the scan, computing the number of recoil events $N_i^a$ in the $i$-th bin for the experiment $a$, and comparing it with the prediction of the benchmark model in the same bin, $\lambda_i^a$ for the same target, assuming that data from each experiment follow independent Poissonian distributions,
\begin{equation}
	\mathcal{L}(\Theta) = 
        \prod_a \mathcal{L}^a(\Theta) = \prod_a\prod_i \frac{N_i^a(\mathbf{\Theta})^{\lambda_i^a} 
          e^{-N_i^a(\mathbf{\Theta})}}{\lambda_i^a!}\,.
	\label{eqn:likelihood}
\end{equation}
Notice that this is equivalent to the product of the likelihoods for each experiment $\mathcal{L}^a$.
The number of recoil events $N_i^a$ in the $i$-th bin for each experiment is obtained by integrating Eq.\,(\ref{drate}) between $E_i^a$ and $E_i^a+\Delta E^a$, for a given live time and including a certain number of background events $b^a/\Delta E^a$. $b^a$ indicates the number of background events in the whole energy for target $a$, so that, assuming no energy dependence, $b^a/\Delta E^a$ is the predicted number of background events in each energy bin. We scan over $b^a$, included in the scan as a nuisance parameter. We assume the number of background events $b^a$ follows a Poissonian distribution function with a mean value that coincides with the nominal background level expected for the different experiments (see Appendix\,\ref{sec:targets}).

In the next sections, the results of our scans will be plotted by means of two-dimensional plots. When the probability for a subset of the original $\mathbf{\Theta}$ is considered, one can account for the presence of the hidden parameters in two different ways:
\begin{itemize}
\item by marginalizing over them, obtaining the pdf for the $j$-th parameter integrating over all the others
\begin{equation}
	p(\Theta_j|\mathbf{D}) = \int p(\mathbf{\Theta}|\mathbf{D}) \,
	d\Theta_1 ... \, d\Theta_{j-1} \, d\Theta_{j+1} \, d\Theta_n;
	\label{eqn:marginalization}
\end{equation}
\item by maximizing over them, obtaining the so-called profile likelihood (PL)
\begin{equation}
	\mathcal{L}(\Theta_j)= \max_{\Theta_1, ..., \Theta_{j-1},\Theta_{j+1},\Theta_n}
	\mathcal{L}(\mathbf{\Theta}).
\end{equation}
\end{itemize}

The PL is usually more sensitive to small fine-tuned regions with large likelihood, while the integration implemented for the pdf allows to account for volume effects. Thus, a parameter space characterized by a complicated likelihood function $\mathcal{L}(\mathbf{\Theta})$ may result in different pdf and PL for the same parameters. 
In our work, to avoid the inclusion of too many figures, we will only represent the results for the PL. 
In the following sections, we demand closed contours in the PL 68\% and 99\% confidence regions (i.e. a good reconstruction of the DM parameters) as a requirement for complementarity.
In general, when closed contours are obtained in the distribution of the PL for the DM parameters we have also observed closed contours in the pdf.

\subsection{Complementarity of direct dark matter experiments}
\label{sec:complementarity}
The detection of WIMP DM in more than one target could provide more information about the nature of this particle \cite{Bertone:2007xj,Drees:2008bv,Cannoni:2010ae,Pato:2010zk}. First, the consistency of the energy spectra measured by experiments using different target nuclei would confirm that the events were due to WIMP scattering (rather than, for instance, neutron background)~\cite{Lewin:1995rx}.
Furthermore, part of the astrophysical uncertainty can potentially be removed, and in fact the comparison of the differential rates could be used to determine the moments of the halo WIMP velocity distribution \cite{Drees:2007hr,Kavanagh:2013wba}. This is one of the motivations for the multitarget project EURECA.

The use of targets which are sensitive to both the SI and SD components of the WIMP-nucleus cross section might allow us to determine the WIMP couplings to matter in an unambiguous way, potentially discriminating between different DM models. In Ref.\,\cite{Bertone:2007xj} this idea was put to the test by studying two of the possible targets of the COUPP experiment (C$_3$FI and C$_4$F$_{10}$). These two materials are sensitive to different combinations of SI and SD WIMP-nucleon couplings and it was shown that this information could be used, in the case of a hypothetical detection, to discriminate the lightest supersymmetric particle (LSP) in the minimal supersymmetric extension of standard model (MSSM) from the lightest Kaluza-Klein particle (LKP) in the universal extra dimension scenario. 
This is a direction that we explore in more detail in this work. In particular, {\em we investigate how well we can reconstruct the DM properties assuming an observation in more than one direct detection experiment and define complementarity as the capability of a combination of targets to determine these properties with a certain precision}.

Let us start by assuming a future detection of WIMP DM in a single direct detection experiment. As explained in the former subsection, the phenomenological parameters defining the WIMP DM particle can be extracted from the study of the total number of events and the differential recoil rate. However, one can easily understand that this procedure does not lead to an unique solution and the parameter space is indeed affected by degeneracies, since in general the experimental information is not enough to constrain the three unknown quantities.

\begin{figure}[t!]
	\epsfig{file=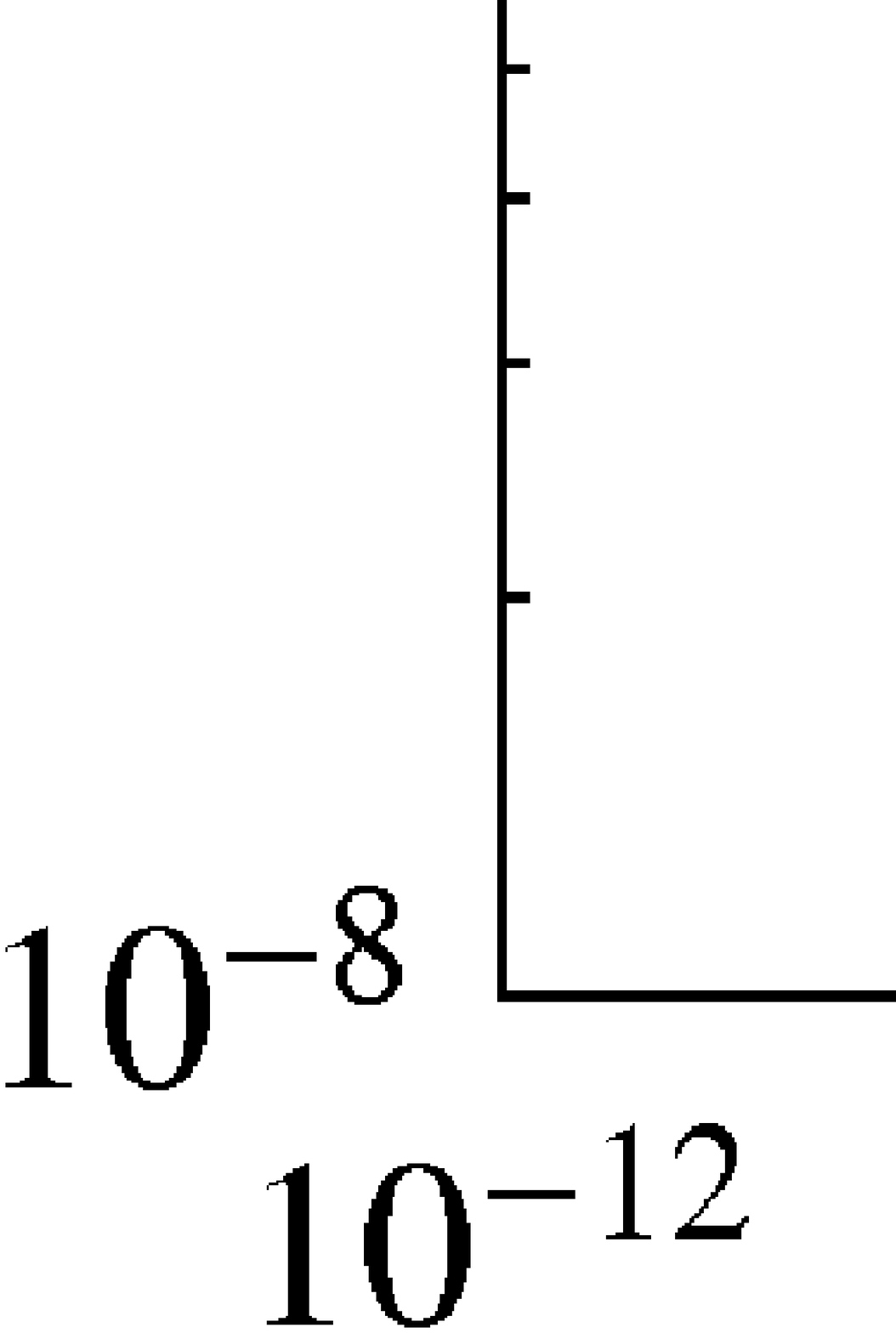,width=7cm}\hspace*{1cm}
	\epsfig{file=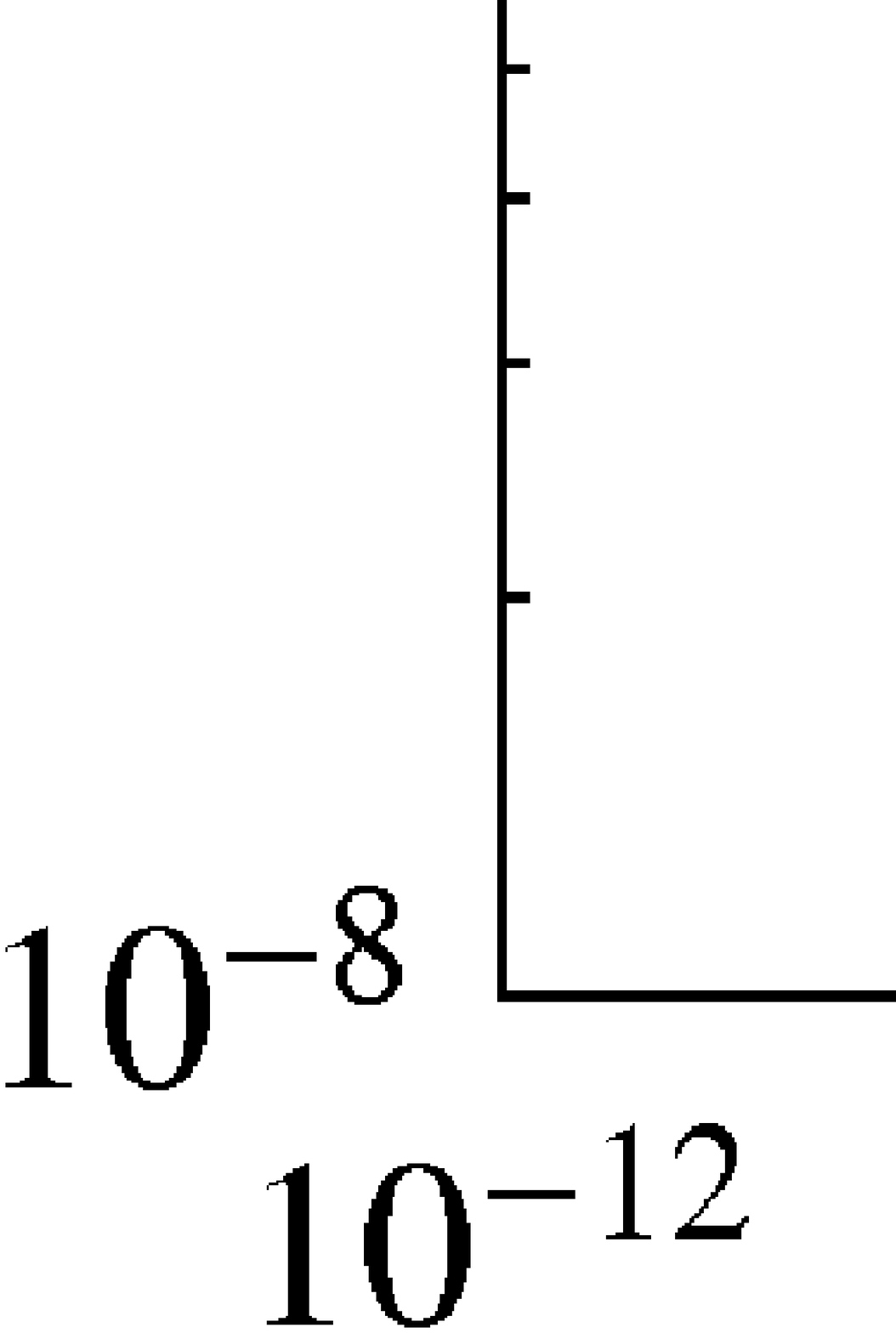,width=7cm}
  \caption{\small Schematic view of the reconstruction of $\sigsi$ and $\sigsd$ from the observed number of WIMP recoil events in two different DM experiments (orange and green shaded areas, respectively). Even assuming a precise measurement of the WIMP mass, a single experiment cannot unambiguously determine the SD and SI cross sections. The detection on a second experiment, however, provides extra information that may allow to further constrain these two parameters. On the left hand-side we display an example in which experiments are complementary, i.e. they intersect leaving only a closed region compatible with both data sets. On the right hand-side we illustrate the case in which the two experiments are not totally complementary, since the region where they overlap is not bounded from below. The nominal WIMP parameters for each case are shown with a dot.  }
  \label{fig:complementarity}
\end{figure}

To illustrate this more clearly, we note that the total number of WIMP recoil events can be expressed as
\begin{equation}
	N={\cal C}_{ SI}~{\sigsi}+{\cal C}_{ SD}~\left(2 S_p \sqrt{\sigsdp}+2 S_n\sqrt{ \sigsdn}\right)^2,
	\label{sigmadeg}
\end{equation}
where the coefficients ${\cal C}_{SI/SD}$ contain the integration in velocities and energies (and a dependence on the WIMP mass) for a given exposure, 
\begin{eqnarray}
	{\cal C}_{SI}&\equiv& \int dE_R\int \left(\frac{\epsilon\,\locald f(v)}{2\redn^2 \mwimp v} \right) A^2\,F^2_{SI}\,dv\,,\\
	{\cal C}_{SD}&\equiv& \int dE_R\int \left(\frac{\epsilon\,\locald f(v)}{2\redn^2 \mwimp v} \right)\left(\frac{J+1}{3J}\right) \,F^2_{SD}\,dv\,.
\end{eqnarray}
Notice that all the dependence on the astrophysical halo parameters and most of the dependence on experimental setup, such as target material, energy threshold, energy resolution, are contained in them (there is also a dependence on the target material in $S_p$ and $S_n$).
As previously stated, we particularise our analysis to the specific case $\sigsdp=\sigsdn$ and therefore Eq.\,(\ref{sigmadeg}) contains three unknown quantities: the SI and SD components of the WIMP scattering cross section and its mass. 
Even if we assume that the WIMP mass can be determined independently with a reasonable accuracy, we are still left with two parameters to reconstruct. Thus, given only one experimental result, the same number of events can be accounted for by different combinations of SI and SD couplings. This is illustrated in Fig.\,\ref{fig:complementarity}, where each of the color shaded areas corresponds schematically to the region in the $(\sigsi,\sigsd)$ plane that is compatible with the detection of a certain number of recoils in one particular detector. 
The detection of a WIMP in a second experiment with a different target can provide complementary information with which this degeneracy can be partially resolved, since changing target implies that also the $\mathcal{C}_{SI/SD}$ coefficients in Eq.\,(\ref{sigmadeg}) are different. 

Fig.\,\ref{fig:complementarity} depicts two possible situations. On the left hand-side we consider an example in which two targets are complementary and allow a good reconstruction of both the SD and SI couplings. This is the case, for example, if one target is mostly sensitive to SI interactions and the other one to SD ones. It can be seen that the region compatible with both experiments is closed, a situation that we call complementary.
On the right hand-side we show another case in which complementarity is not present since the overlapping region is unbounded from below. This is generically the case when the two targets are mostly sensitive to the SI coupling, and therefore, is a very common situation. In particular, as we will argue in Section\,\ref{sec:gexe}, this could happen when combining signals from germanium and xenon detectors. Notice that this example suggests that in such a case the SI coupling can be relatively well reconstructed but only an upper bound can be derived for the SD one. 

The extension of the overlapping region is a result of the various uncertainties. First of all, there is an obvious statistical component: if the detected number of events is small, it is subject to large statistical fluctuations and therefore the reconstruction of parameters is poor. Thus, the larger the exposure of the experiment (or alternatively, the larger the scattering cross section), the narrower the reconstructed band is.
On top of this, astrophysical and nuclear uncertainties further contribute to worsening the reconstruction. This is clearly a limiting factor that needs a careful implementation in order to describe a realistic experimental situation. 
Finally, in the examples above we have assumed that the mass of the WIMP is known, but in the following the WIMP mass is part of the parameter space we scan over and it will be determined along with the rest of the parameters. 

\subsection{Combination of signals from germanium and xenon experiments}
\label{sec:gexe}

Two technologies which have been very successfully applied to direct DM searches are semiconductor cryogenic detectors (e.g., the germanium detectors of CDMS and EDELWEISS) and noble liquid detectors (a xenon target in the case of XENON10 and XENON100). Both techniques show great potential and there are plans to extend target masses up to several hundred kilograms or even one ton within the next years (e.g., SuperCDMS, EURECA and XENON1T).
It is therefore conceivable that a future WIMP observation could take place in any of these targets. Let us therefore start by contemplating that possibility and assessing how well the DM properties can be reconstructed in germanium and xenon targets for different WIMP scenarios. For concreteness we consider the energy window for WIMP searches corresponding to CDMS-II in the case of germanium, XENON100 in the case of a xenon target (see Appendix\,\ref{sec:targets}).

We consider the set of WIMP benchmark points from Table\,\ref{tab:bm}, which take into account various mass choices and include cases in which either the SI or the SD contribution dominates the detection rate. 
We simulate the expected differential spectrum by considering a 1 ton experiment with a 30\% live time operating for a year, i.e. a total exposure of $\epsilon=300$~kg\,yr. 
The expected DM signal is computed as described in the previous section, and implemented in the scan as experimental information, in the attempt to reconstruct the DM phenomenological parameters.

\def\bmvlsd{VL-SD}
\def\bmlsd{L-SD}
\def\bmmsd{M-SD}
\def\bmhsd{H-SD}
\def\bmvlsi{VL-SI}
\def\bmlsi{L-SI}
\def\bmmsi{M-SI}
\def\bmhsi{H-SI}

\begin{table}
\begin{center}
\begin{tabular}{|c||c|c|c||c|
c|c|}
\hline
Benchmark Point &    $\mwimp$ (GeV) & $\sigsi$ (pb) & $\sigsd$ (pb) & $N_{\rm Ge}$ & $N_{\rm Xe}$ & $N_{{\rm C}_3{\rm FI}}$\\
\hline
\hline
\bmmsi &     $100$ &     $10^{-9}$ &    $10^{-5}$ &37 (37)    &54 (52)   &42 (32)\\
\bmlsi &     $50$ &     $10^{-9}$ &    $10^{-5}$ &40 (40)    &64 (62)   &49 (35)\\
\bmlsd    &     $50$ &    $4\times10^{-10}$    &$6\times10^{-4}$&44 (16)    &93 (14) &560 (16)\\
\bmvlsi &     $15$ &     $10^{-8}$ &    $10^{-5}$ &29 (29)    &28 (28)    &15 (9)\\
\hline
\end{tabular}
\caption{\small Set of benchmark points used in this work.
We consider three regimes for the WIMP mass, very light (VL), light (L) and medium (M). The label SI or SD in each benchmark indicates which component of the scattering cross section dominates the detection rate in germanium and xenon targets. For reference we also include the expected number of WIMP recoil events in Ge, Xe and C$_3$FI for an exposure $\epsilon=300$~kg\,yr, where the number in parenthesis corresponds to the contribution from only SI interactions. See Appendix\,\ref{sec:targets} for details on the experimental set-ups.}
\label{tab:bm}
\end{center}
\end{table}

\begin{figure}[t!]
	\hspace*{-1cm}
	\epsfig{file=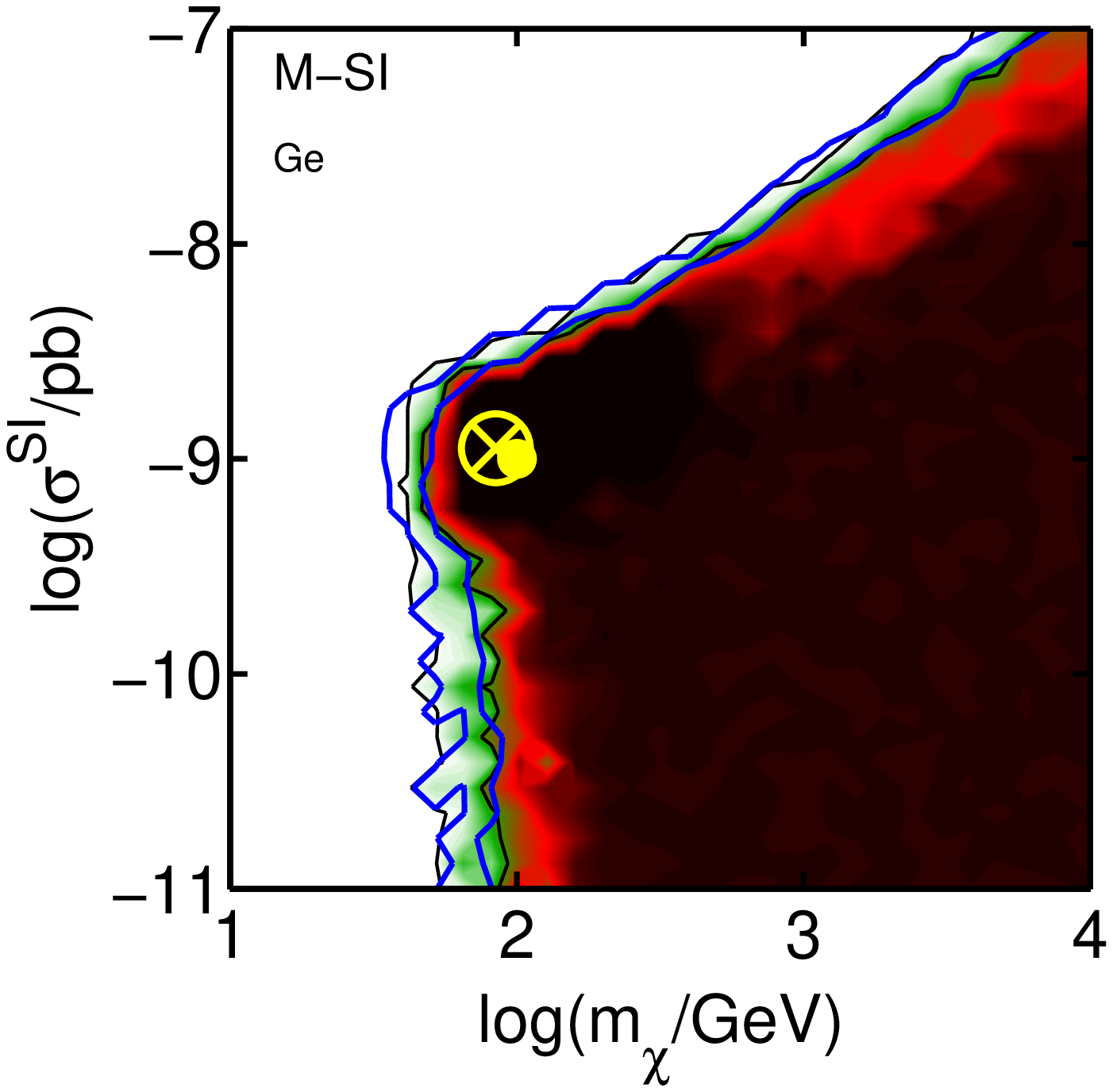,width=5.9cm}\hspace*{-0.6cm}
	\epsfig{file=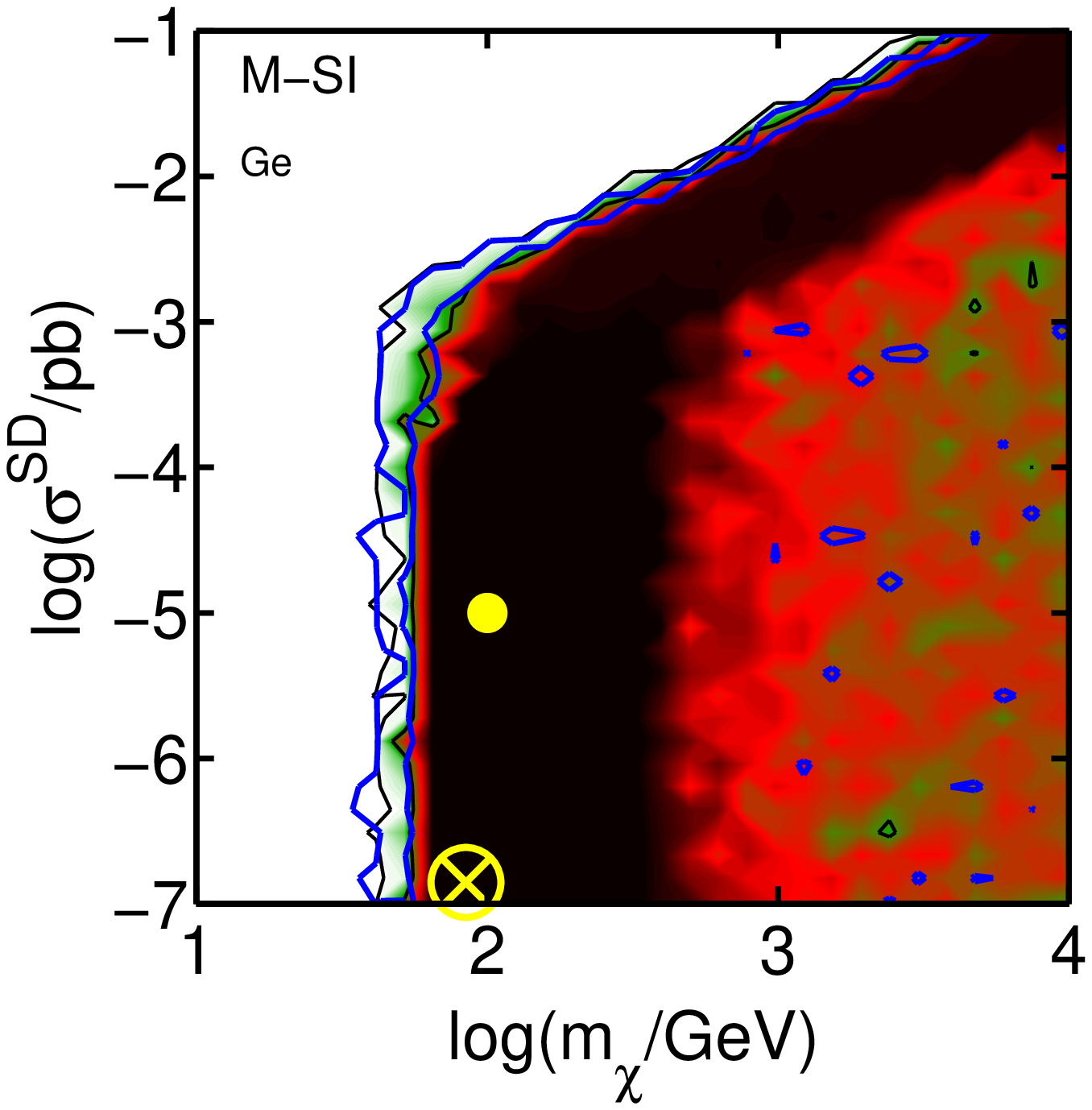,width=5.9cm}\hspace*{-0.6cm}
	\epsfig{file=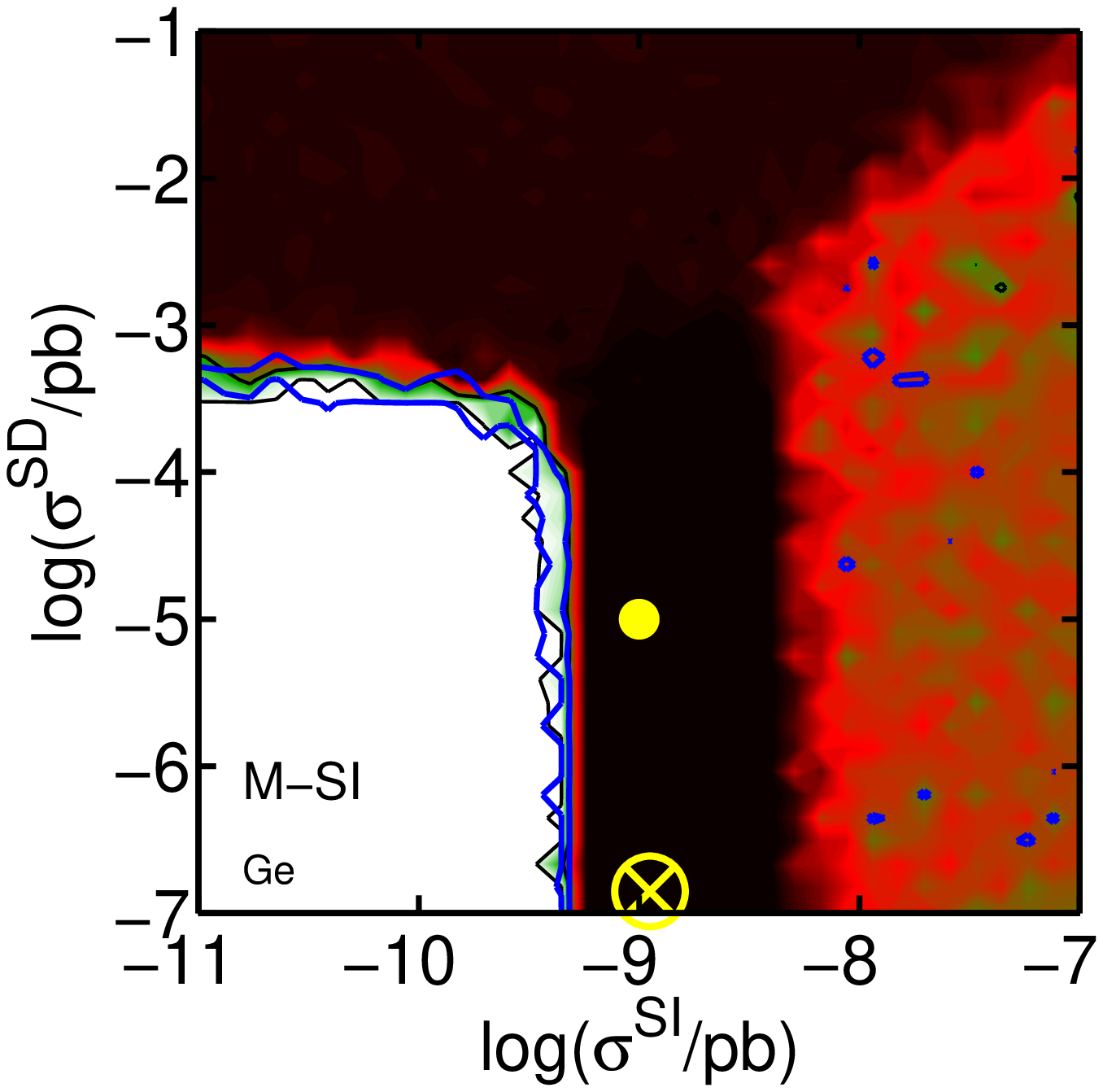,width=5.9cm}\\[-0.5cm]
	\hspace*{-1cm}
	\epsfig{file=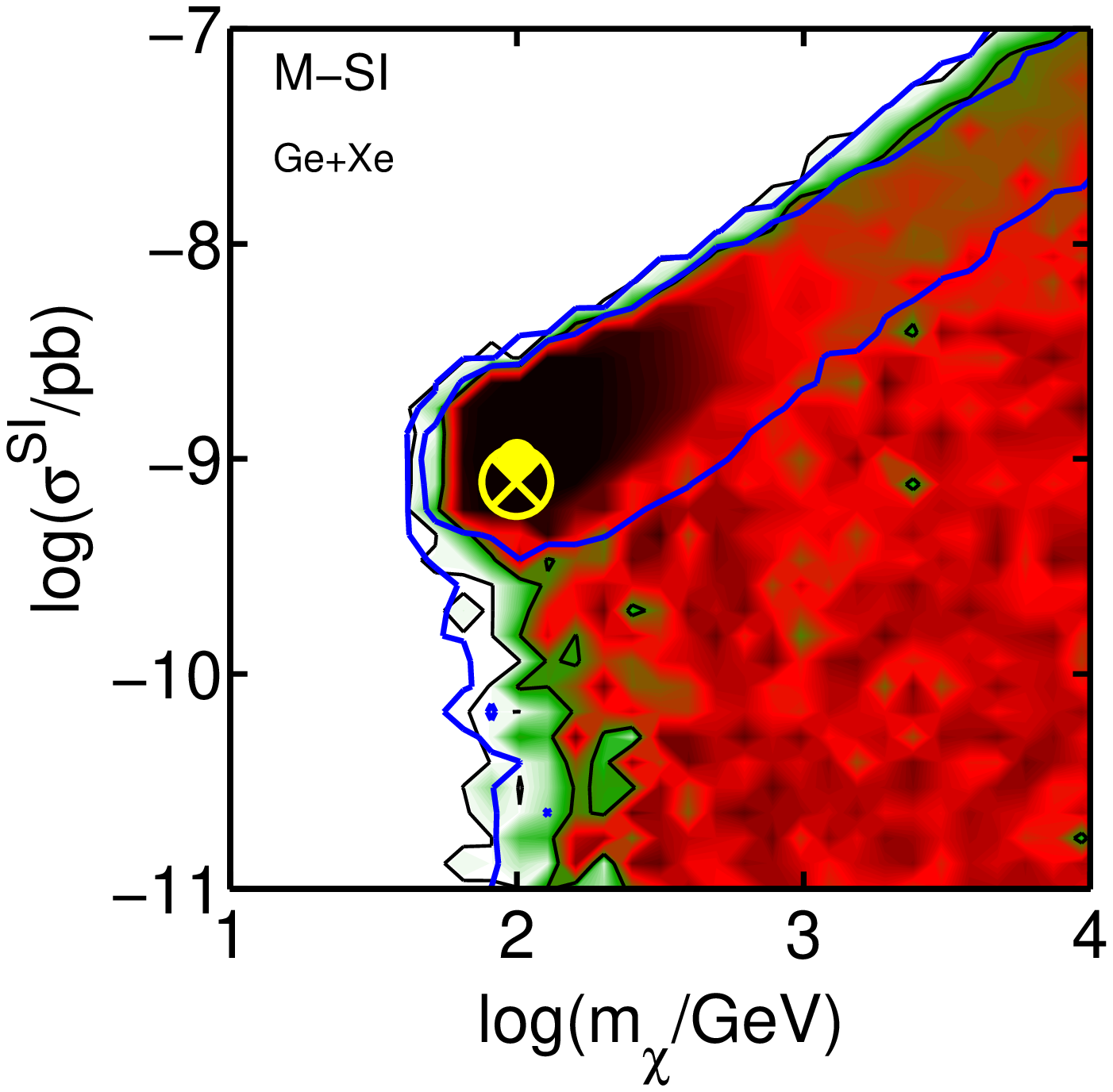,width=5.9cm}\hspace*{-0.6cm}
	\epsfig{file=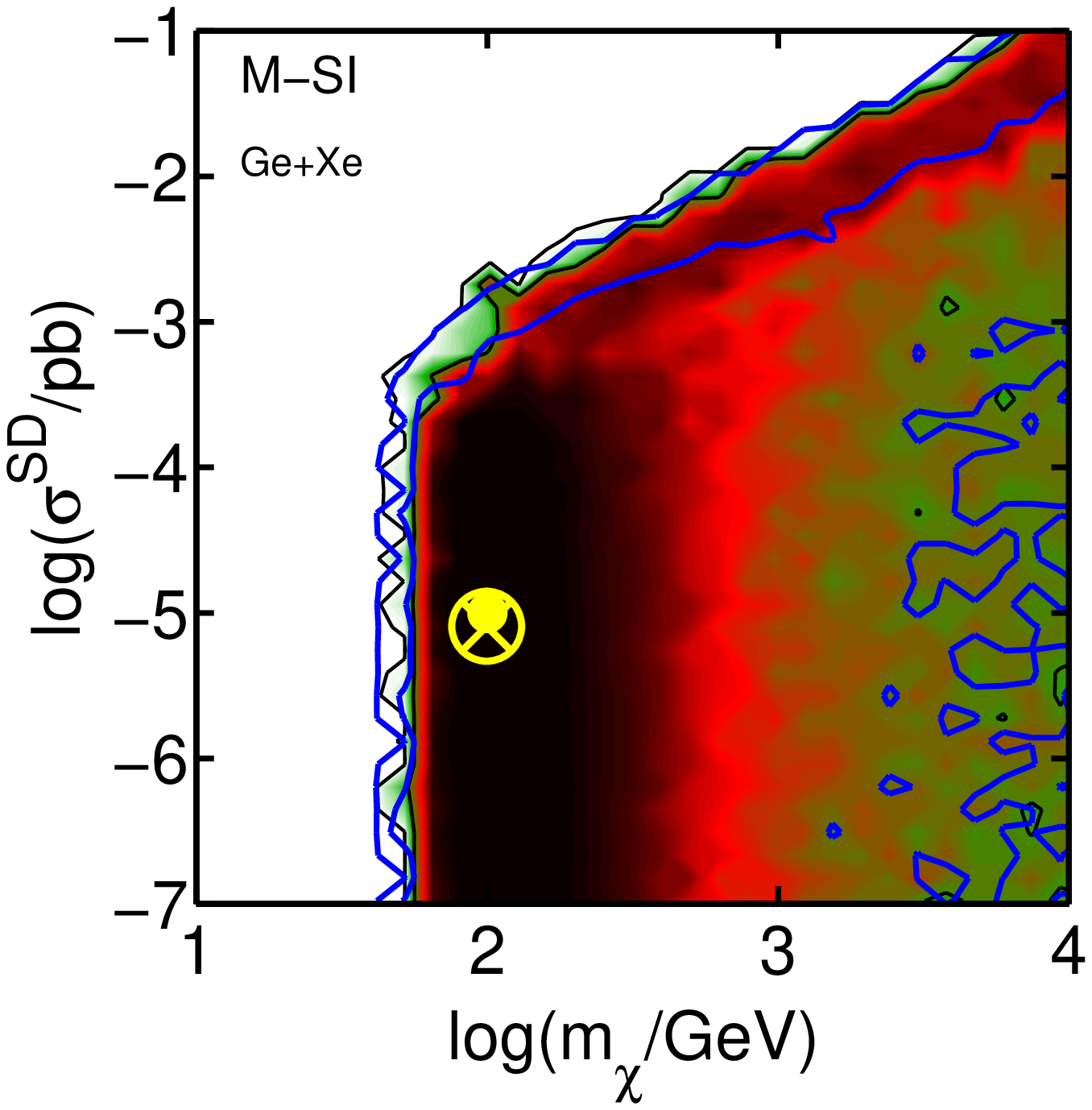,width=5.9cm}\hspace*{-0.6cm}
	\epsfig{file=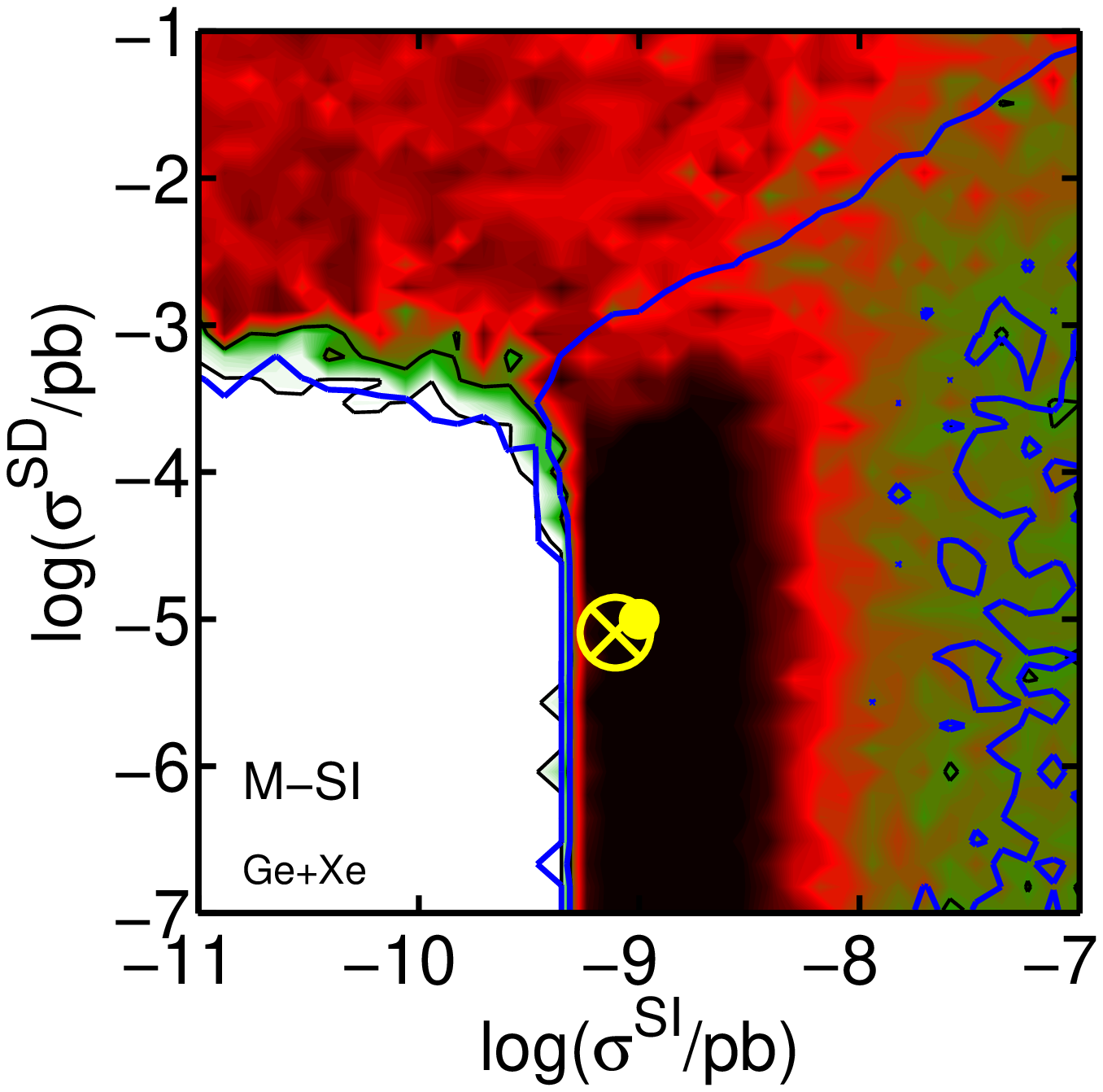,width=5.9cm}\\[-0.5cm]
	\hspace*{-1cm}
	\epsfig{file=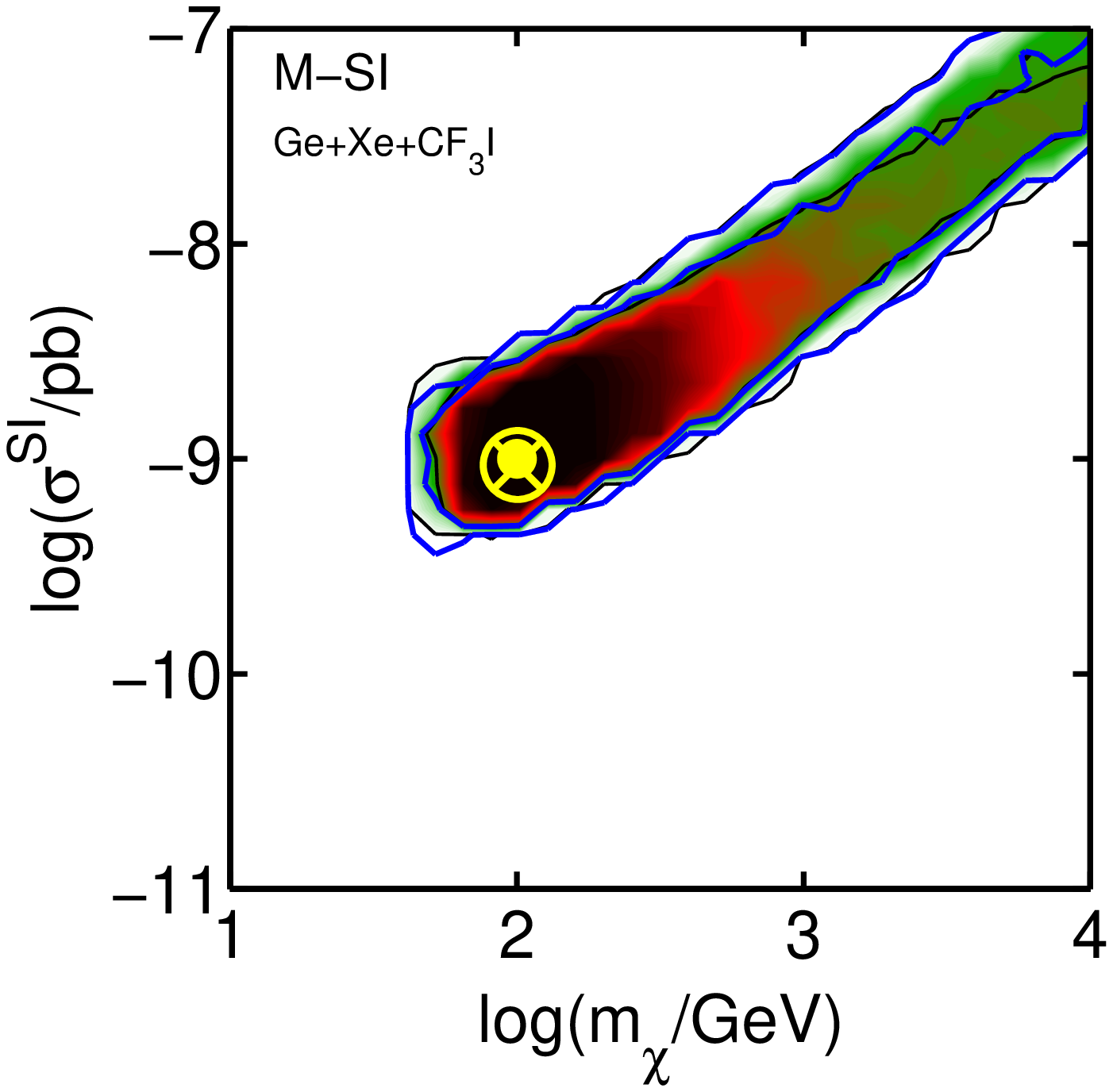,width=5.9cm}\hspace*{-0.6cm}
	\epsfig{file=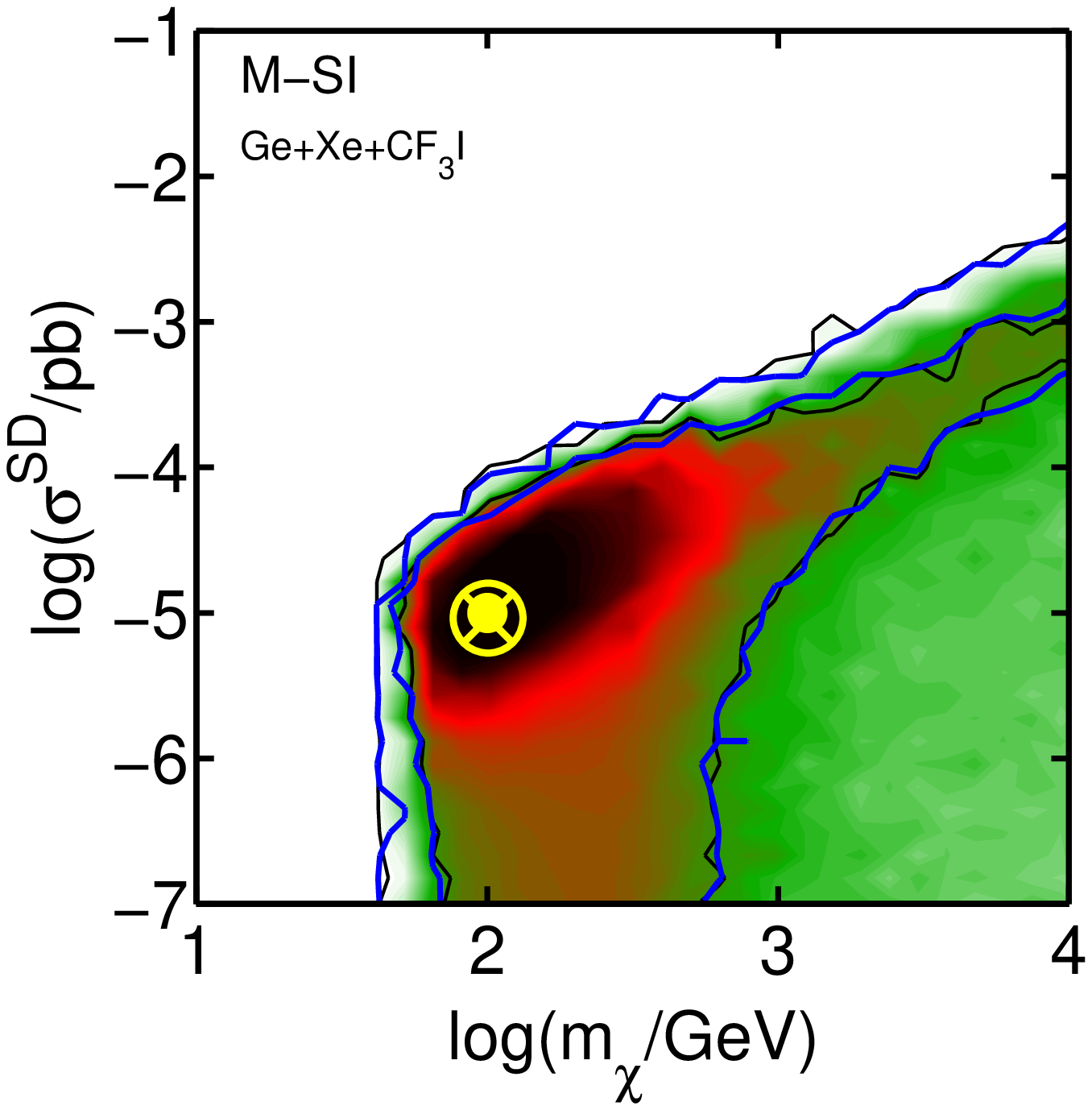,width=5.9cm}\hspace*{-0.6cm}
	\epsfig{file=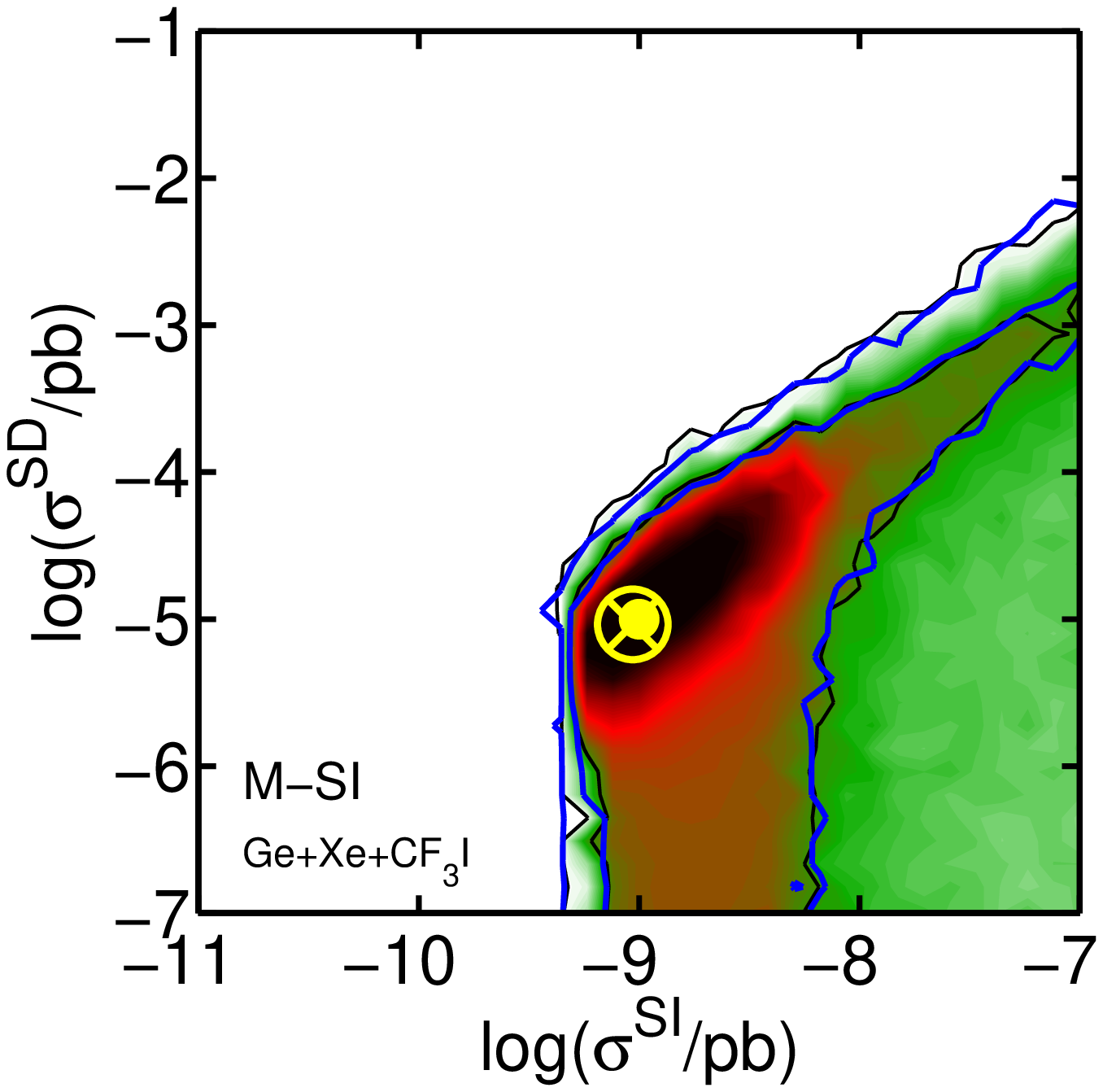,width=5.9cm}
  \caption{\small Profile likelihood for the DM parameters in the $\msi$, $\msd$, and $\sisd$ planes for the benchmark point \bmmsi. The plots in the first row are obtained assuming a detection only in a Ge experiment, the second row shows the combination between Ge and Xe targets, and the third row corresponds to Ge+Xe+C$_3$FI. From the inside out, contours are 68\% and 99\% confidence intervals. The yellow dot represents the nominal values for the benchmark point and the yellow circled cross the best-fit point. The coloured regions bounded by black contours correspond to the case where nuclear uncertainties are included, whereas the blue empty contours are the results of a scan with fixed nuclear parameters.}
  \label{fig:cdmsxenon-bmmsi-pl}
\end{figure}

Figure\,\ref{fig:cdmsxenon-bmmsi-pl} shows the PL of the three phenomenological parameters $\mwimp$, $\sigsi$, and $\sigsd$ projected onto the three two-dimensional planes $\msi$, $\msd$, and $\sisd$, for a DM particle corresponding to benchmark \bmmsi. The first row shows the determination of parameters after a detection in a germanium detector alone. The yellow dot indicates the nominal value of the benchmark point and the circled cross the best-fit point. Consistently with the previous discussion, we observe a large degeneracy in the three parameters, with a significant uncertainty in both components of the cross section and the mass. As already pointed out in a previous work \cite{Cerdeno:2012ix}, the reconstructed values of the SD or SI cross section show no upper bound for large values of the WIMP mass. Consequently, the 68\% confidence level contours are not closed, but extend beyond the limits of the plots. Moreover, the scan does not manage to reconstruct the correct value of the $\sigsd$.

The second row in Fig.\,\ref{fig:cdmsxenon-bmmsi-pl} corresponds to the combination of data from germanium and xenon detectors. An improvement is visible since large regions of the parameter space are associated with a smaller value of the PL and the best-fit point is now closer to the nominal value.
However, both the shape and the area of the outer contours remain very similar to the case with only germanium.

Finally, the third row of Fig.\,\ref{fig:cdmsxenon-bmmsi-pl} considers the combination with data from a hypothetical 1 ton version of COUPP. Since the target material, C$_3$FI, incorporates fluorine (which has a large nuclear spin) it has a larger sensitivity to the SD component. Thus, the allowed region is drastically reduced. In particular, the solutions with large SD cross section are not compatible anymore with the data, a more stringent upper limit on $\sigsd$ is produced (at least for the 99\% confidence region), which translates into a lower limit on $\sigsi$. 
Note that the experimental setup in COUPP does not allow the recoil energy to be measured and therefore does not provide more information on the WIMP mass.

Blue contours in Fig.\,\ref{fig:cdmsxenon-bmmsi-pl} indicate the 68\% and 99\% confidence regions for the PL of scans performed without including the nuclear uncertainties associated with the SD structure factors and fixing their parameters to the central value of the range considered in our scans (see Appendix\,\ref{app:sdsf}). We observe that the effect of these uncertainties can greatly affect the reconstruction of parameters, significantly enlarging the areas compatible with the observation in two experiments (see the second row of Fig.\,\ref{fig:cdmsxenon-bmmsi-pl}). 

\begin{figure}[t!]
	\hspace*{-1cm}
	\epsfig{file=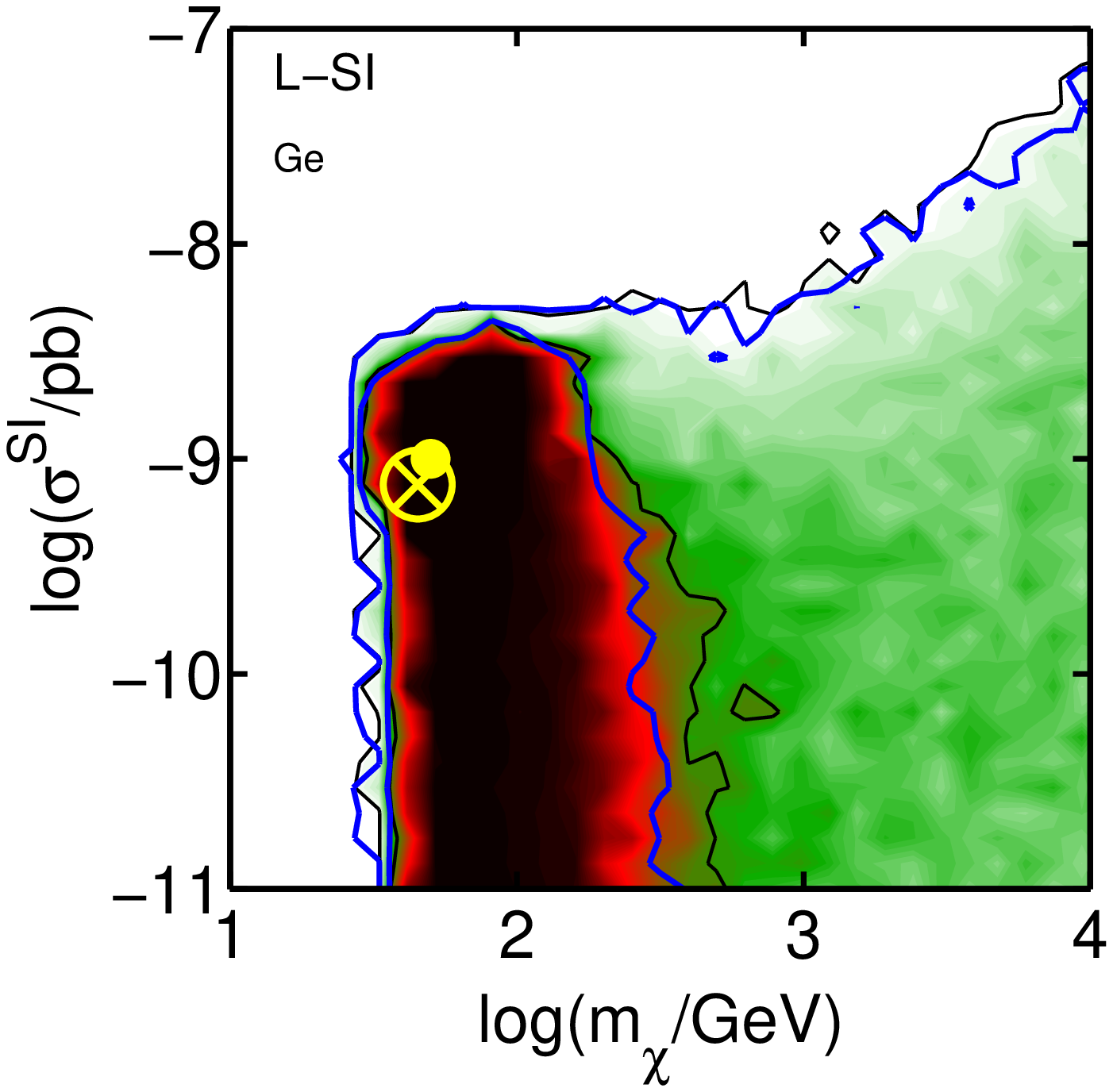,width=5.9cm}\hspace*{-0.6cm}
	\epsfig{file=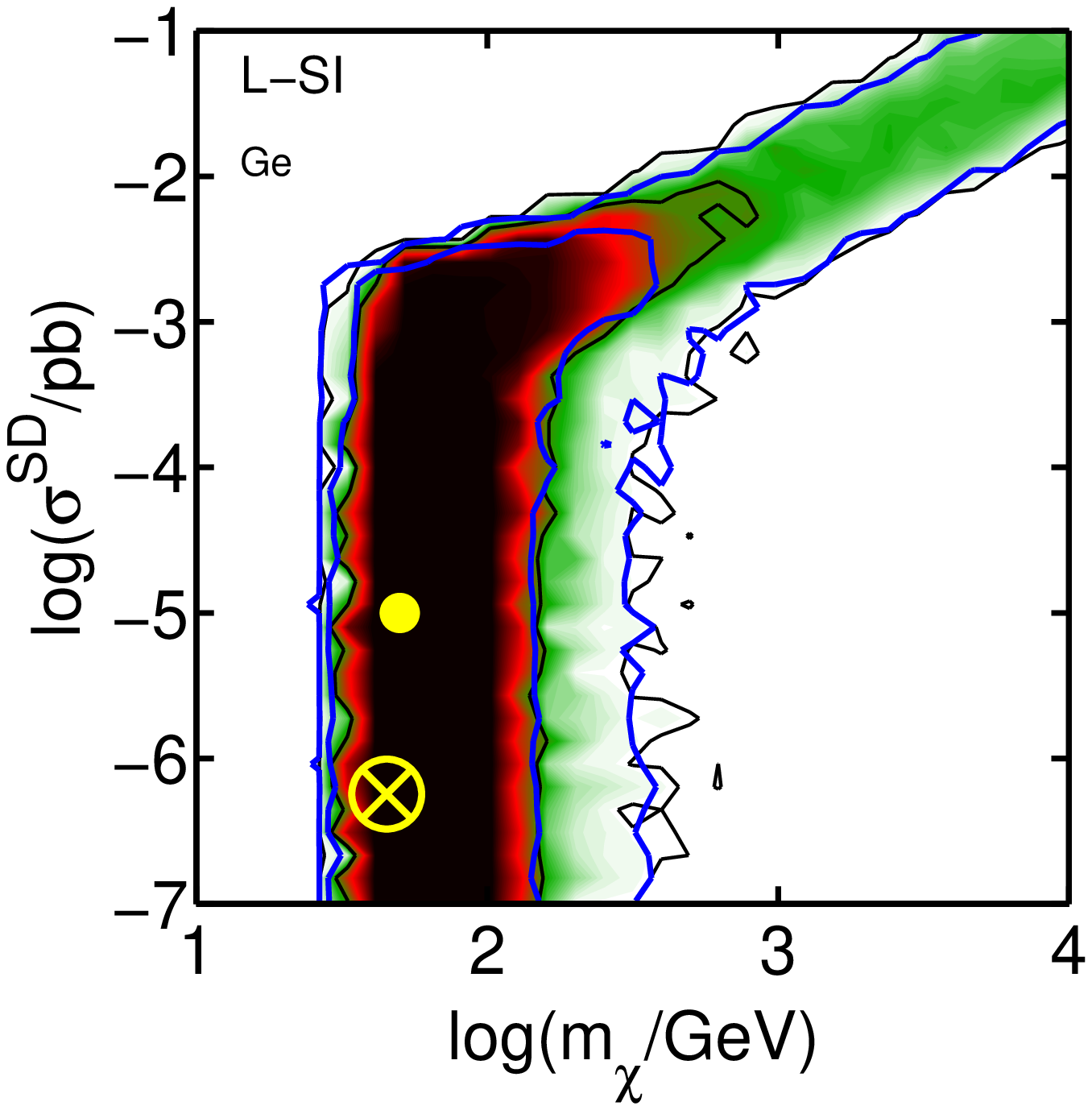,width=5.9cm}\hspace*{-0.6cm}
	\epsfig{file=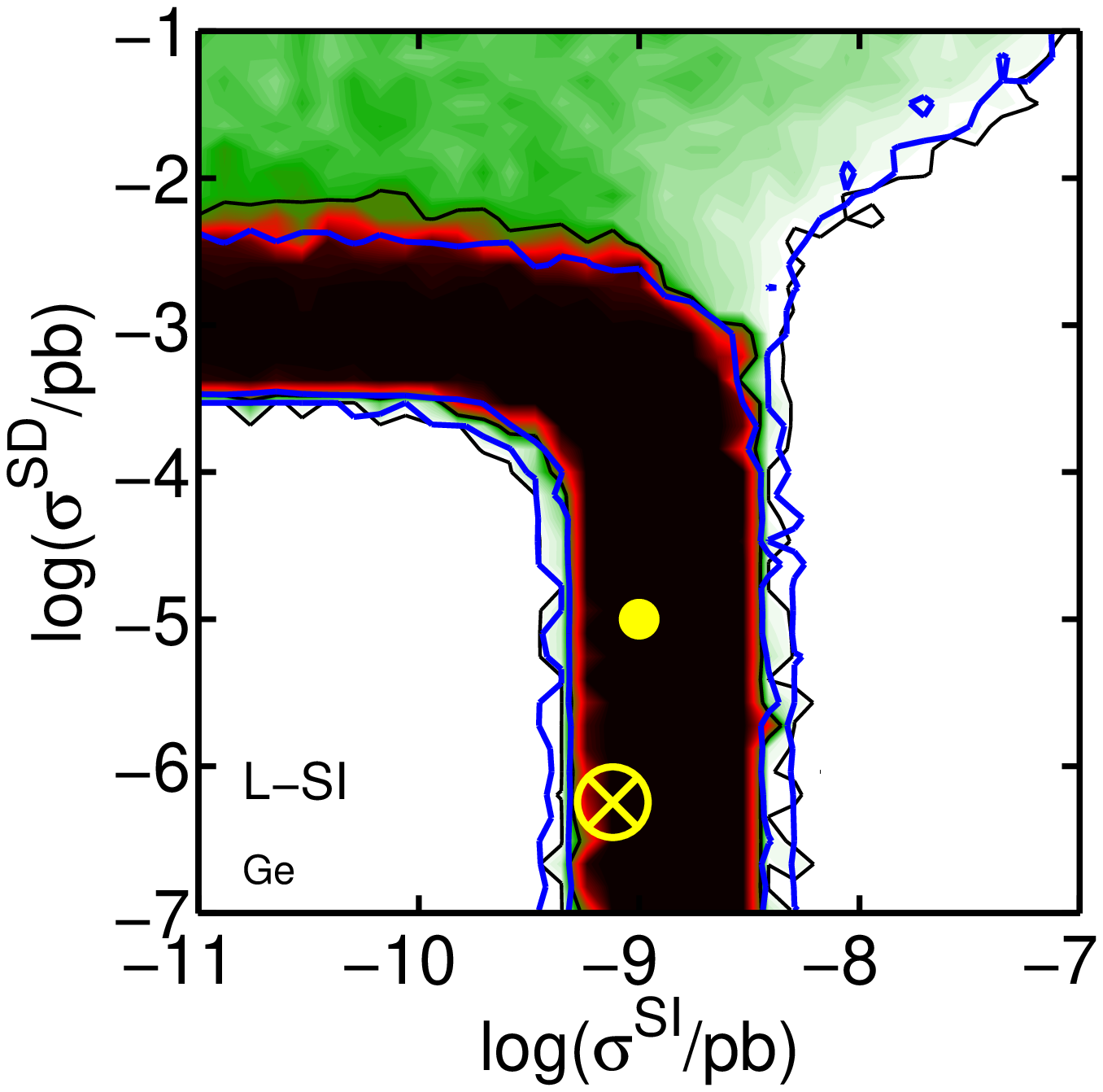,width=5.9cm}\\[-0.5cm]
	\hspace*{-1cm}
	\epsfig{file=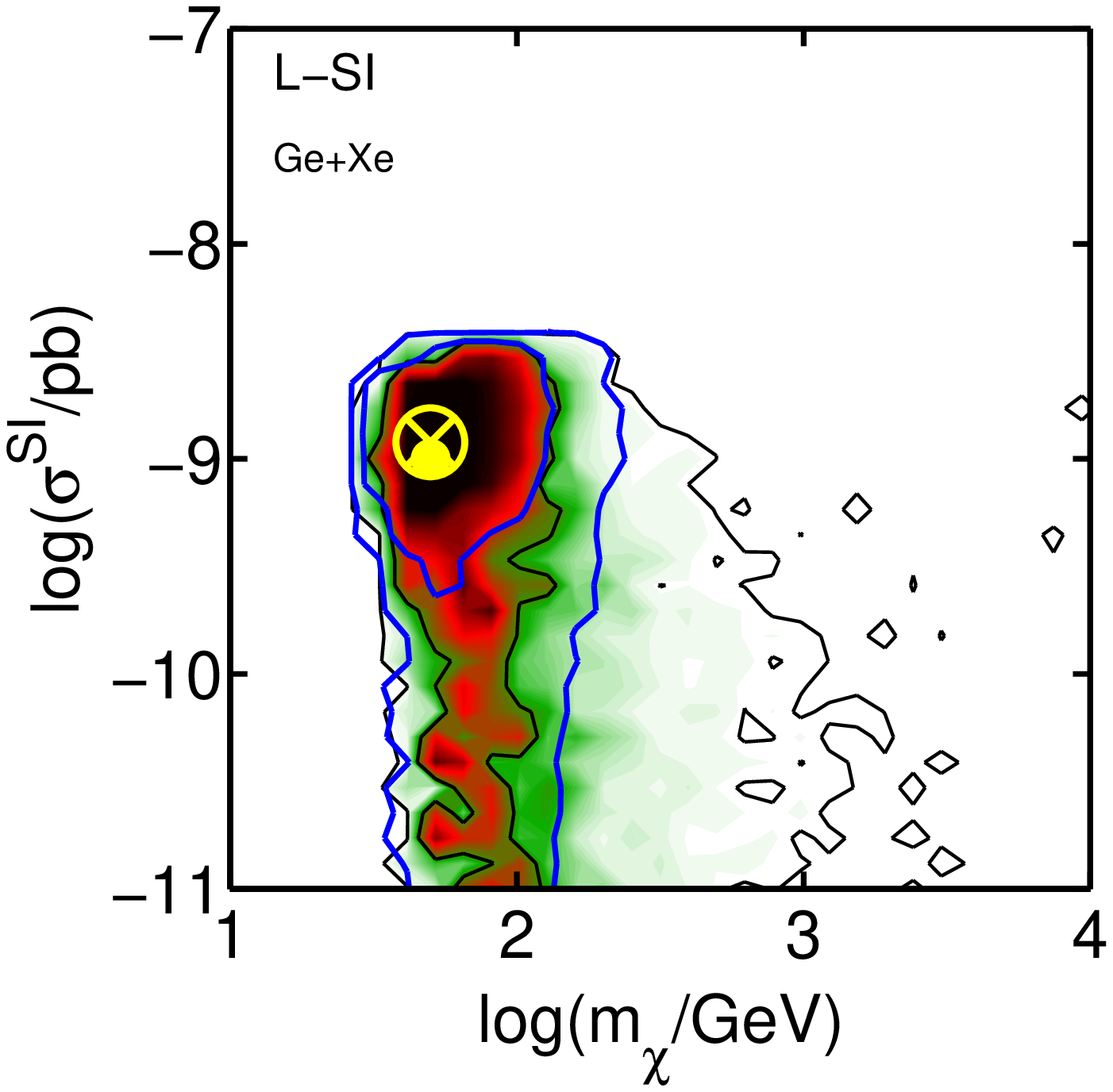,width=5.9cm}\hspace*{-0.6cm}
	\epsfig{file=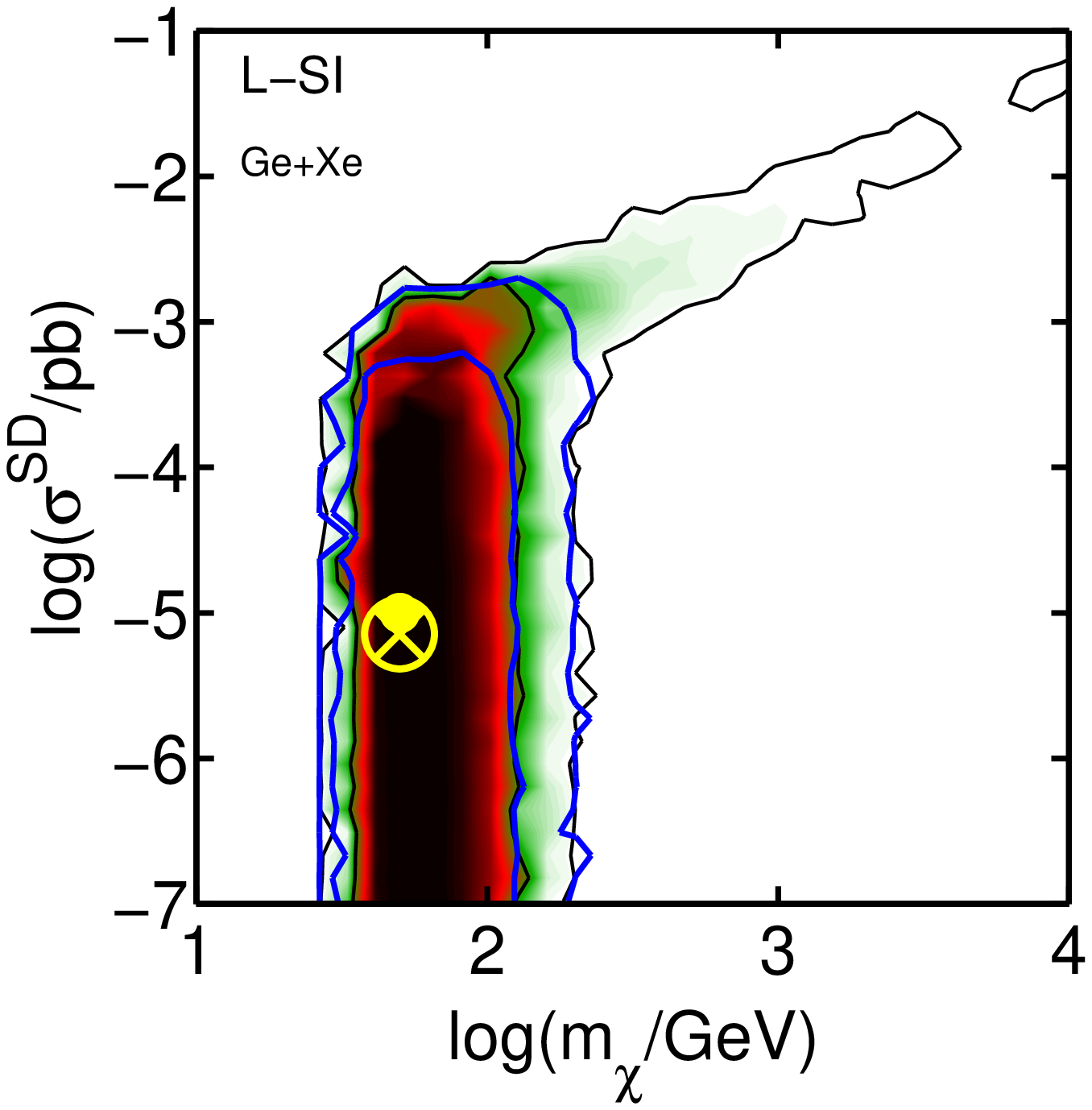,width=5.9cm}\hspace*{-0.6cm}
	\epsfig{file=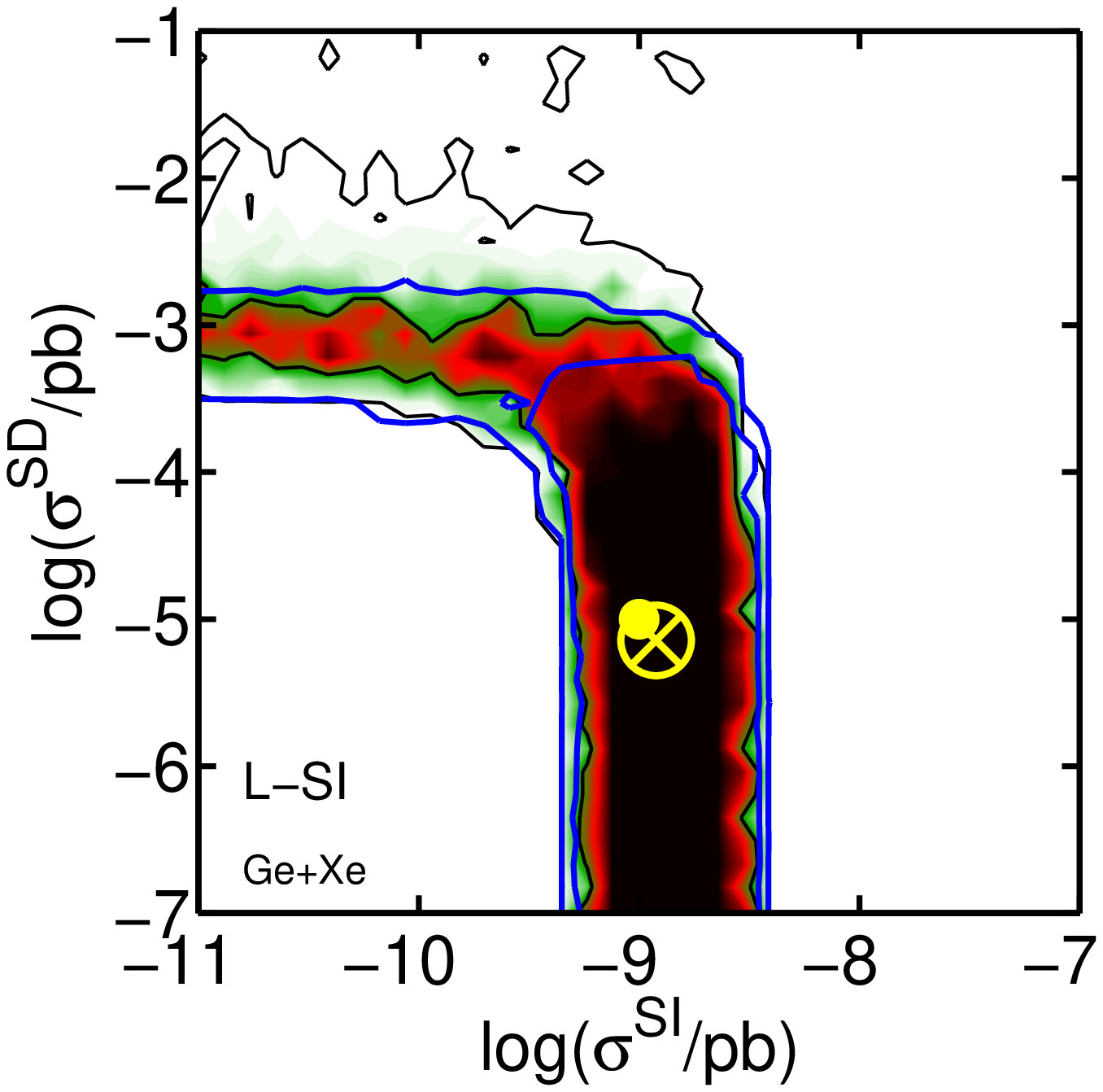,width=5.9cm}\\[-0.5cm]
	\hspace*{-1cm}
	\epsfig{file=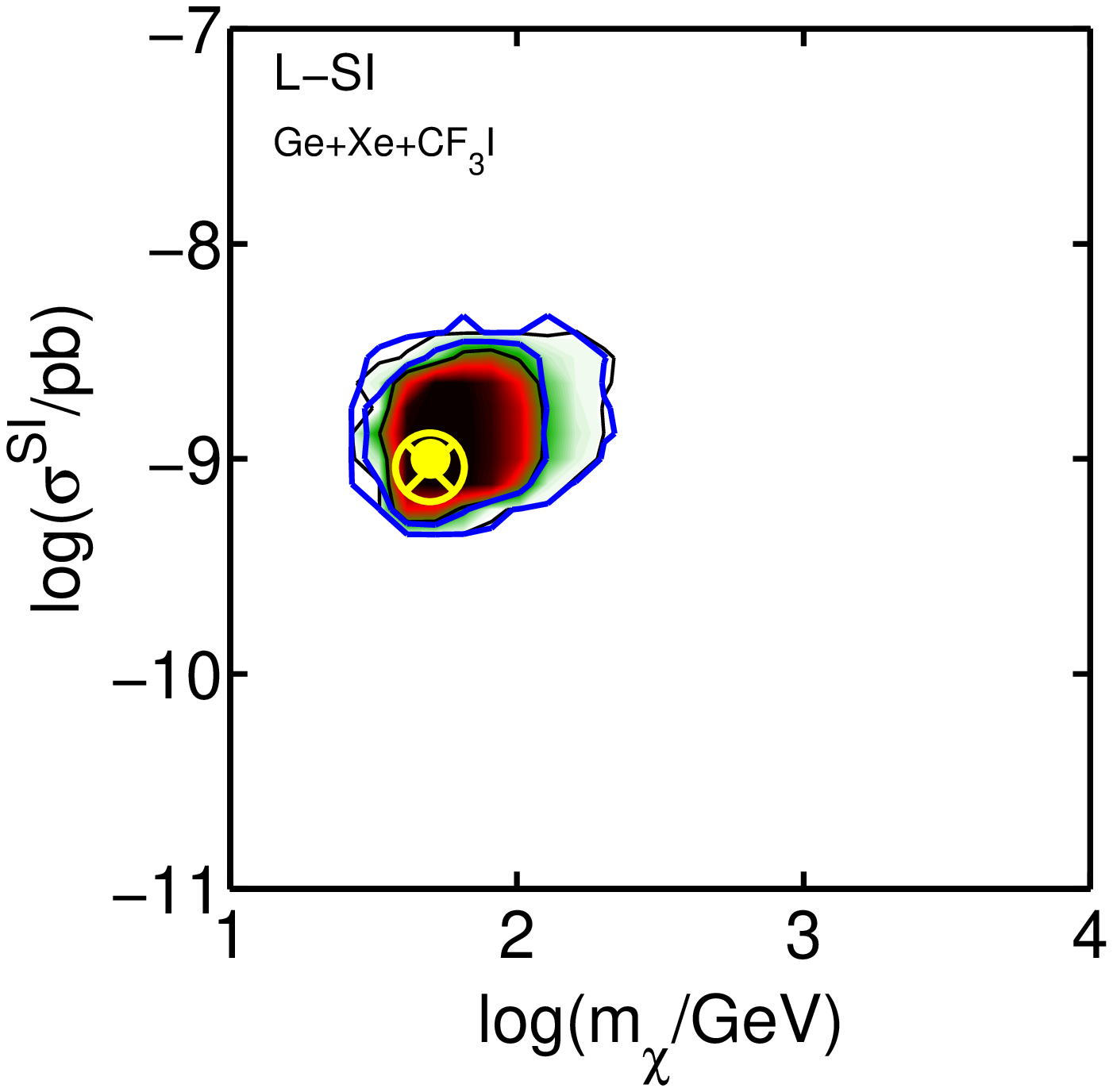,width=5.9cm}\hspace*{-0.6cm}
	\epsfig{file=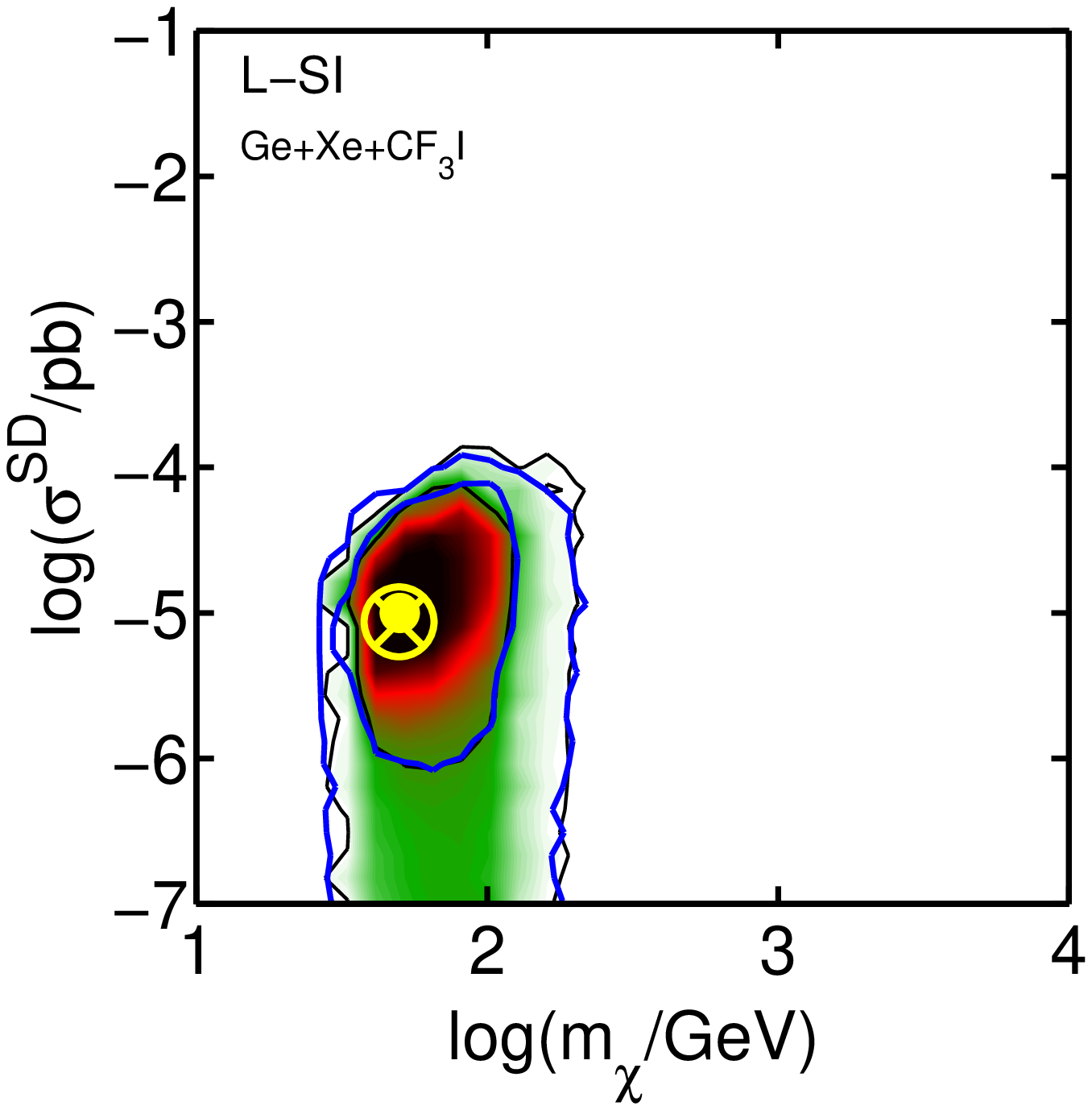,width=5.9cm}\hspace*{-0.6cm}
	\epsfig{file=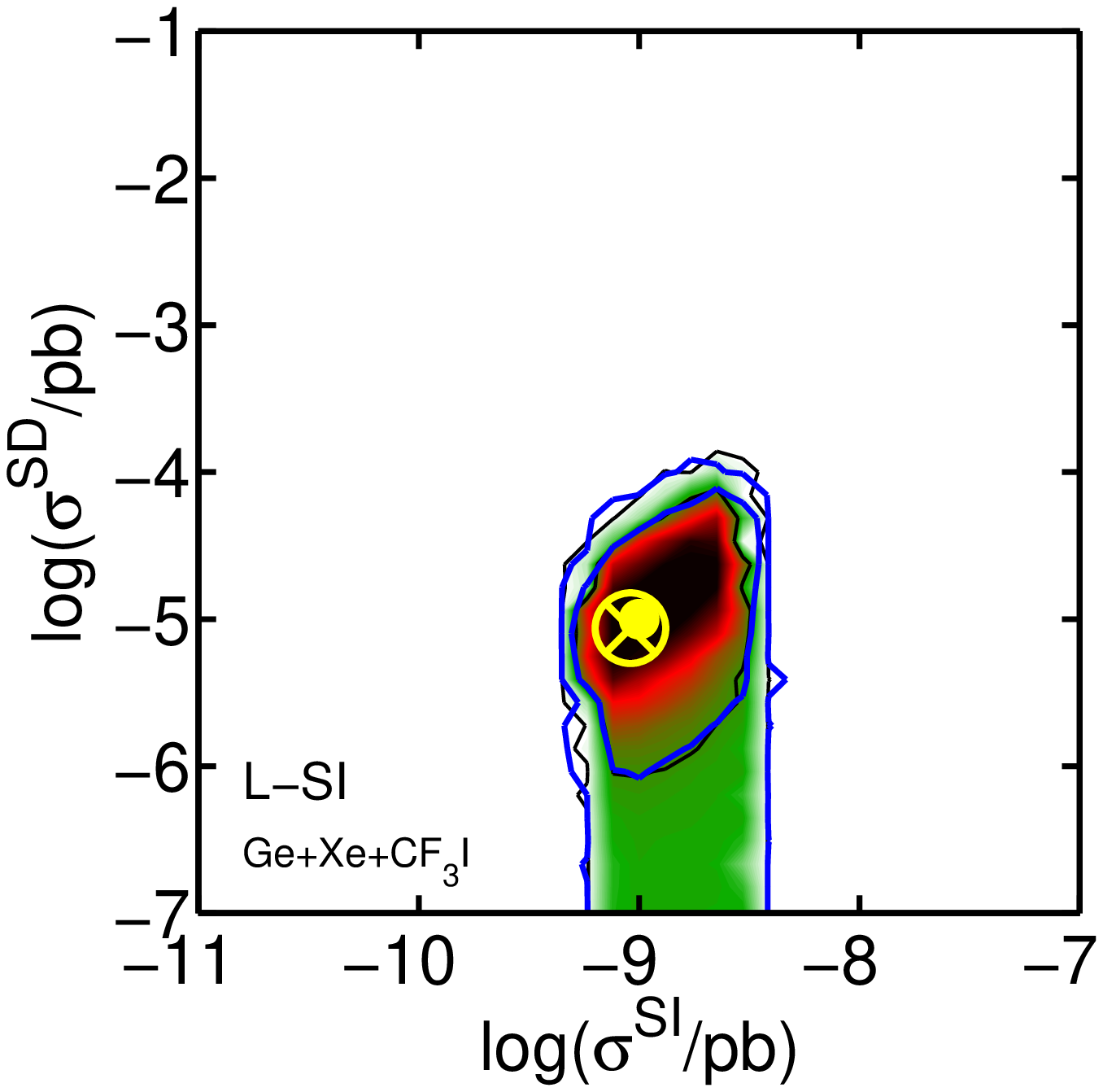,width=5.9cm}
	  \caption{\small The same as in Fig.\,\ref{fig:cdmsxenon-bmmsi-pl}, but for the benchmark point \bmlsi.}
  \label{fig:cdmsxenon-bmlsi}
\end{figure}

These results depend on the benchmark point, but the general conclusion on how different experiments combine remains valid.
When the number of events increases, either because the scattering cross section is larger or because the WIMP mass is closer to the optimal sensitivity for these targets, the determination of the DM parameters is better.
For example, Fig.\,\ref{fig:cdmsxenon-bmlsi} represents the hypothetical detection of a WIMP with properties determined by case \bmlsi\ in Table\,\ref{tab:bm}, in which $\mwimp=50$~GeV. Since the number of recoils is higher, the 68\% and 99\% contours in the PL are smaller. In particular, using only data from Ge and Xe detectors there is a good reconstruction of the WIMP mass, as well as a significant improvement in the determination of $\sigsi$. Notice that even then, the SD cross section is not properly determined, as only an upper bound is obtained.
As before, after the inclusion of C$_3$FI data, the SD cross section is better reconstructed (i.e. lower and upper limits are obtained) at the 68\% confidence level, with the upper limit being valid also at the 99\% confidence level. A good reconstruction of $\sigsd$ has also the consequence of providing a lower bound for $\sigsi$ and reducing the uncertainty on $\mwimp$. Both quantities are now determined with an uncertainty of approximately one order of magnitude, an extremely interesting scenario made possible by the nice interplay between the three experiments.

\begin{figure}[t!]
	\hspace*{-1cm}
	\epsfig{file=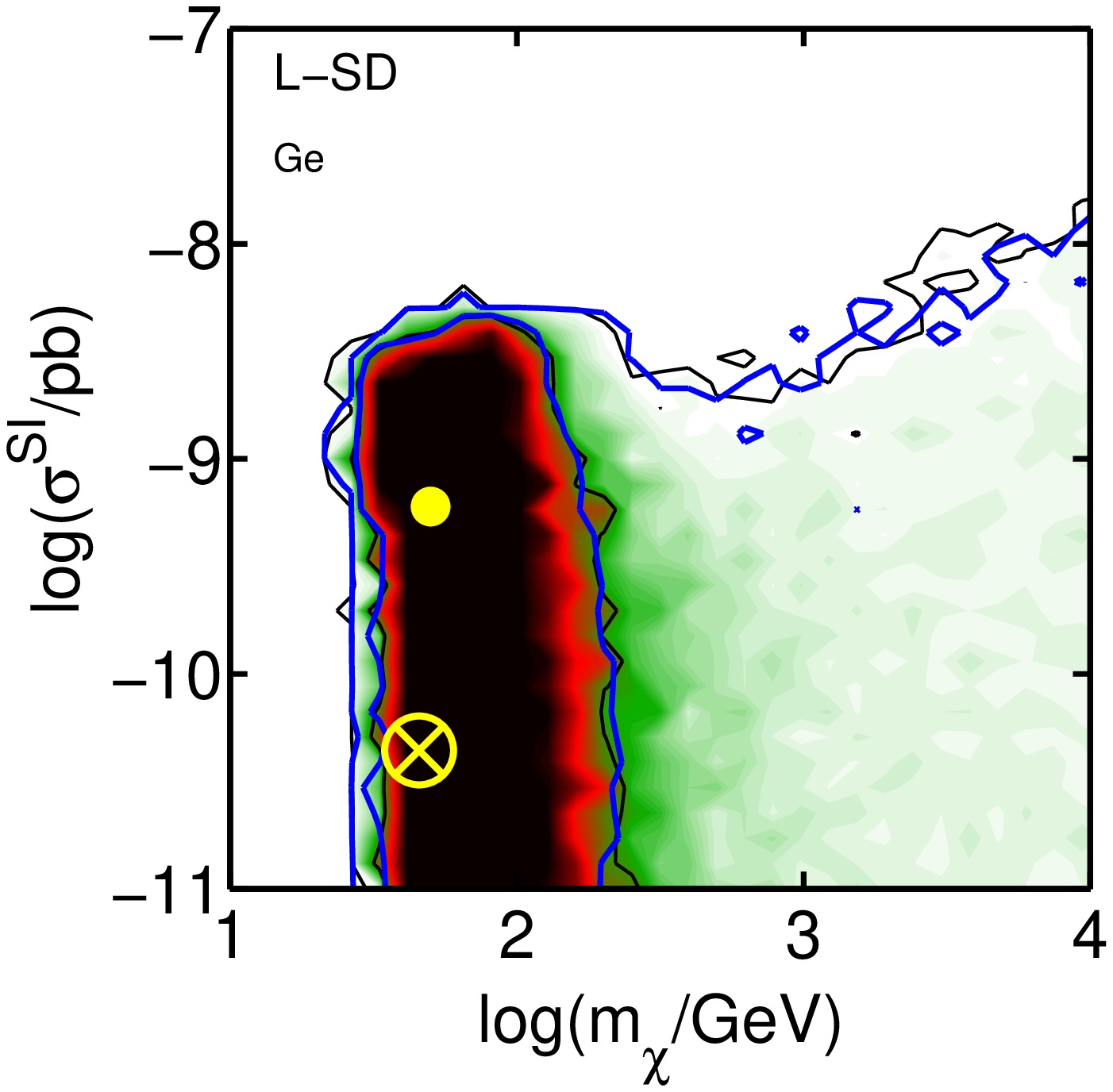,width=5.9cm}\hspace*{-0.6cm}
	\epsfig{file=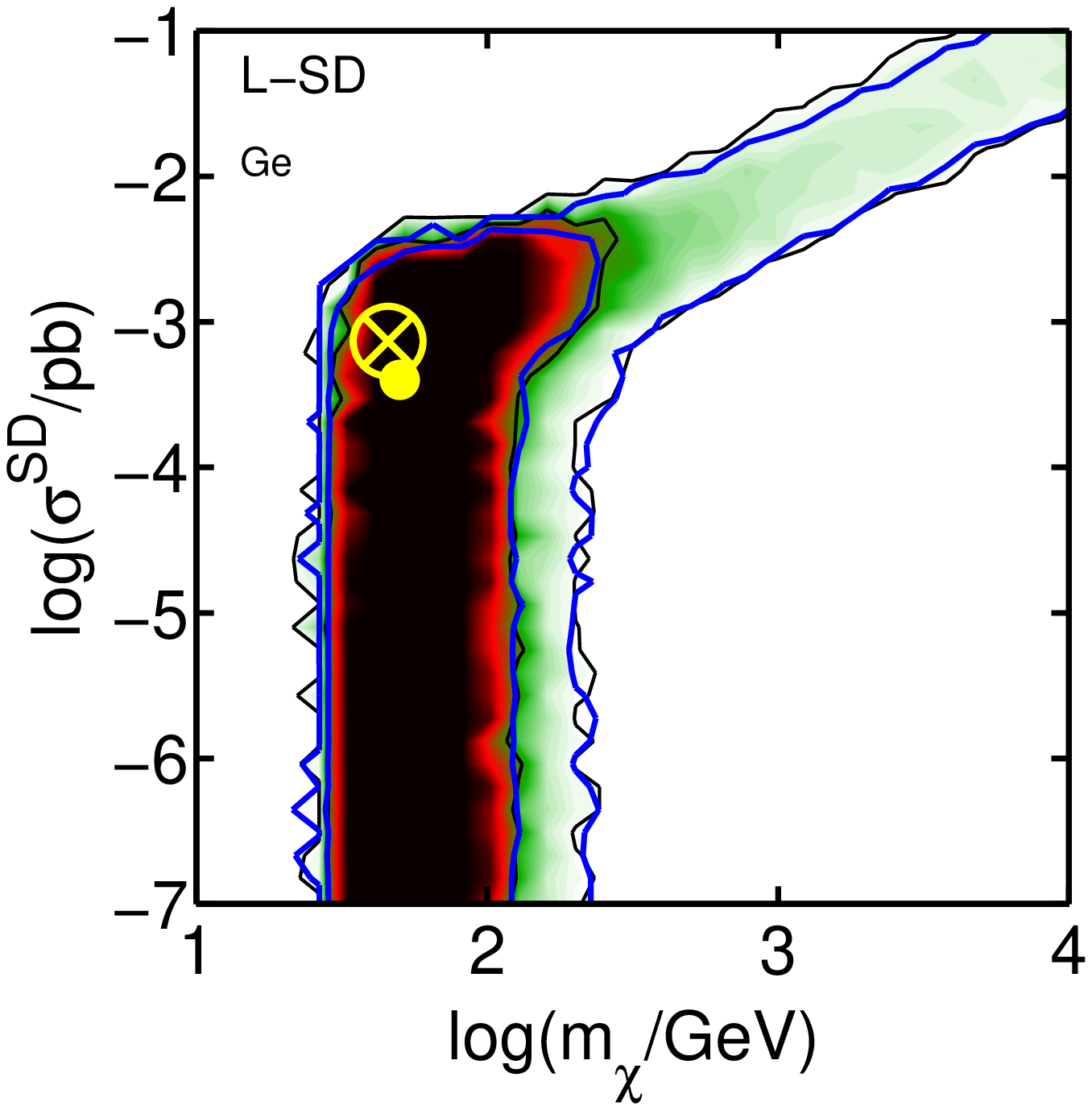,width=5.9cm}\hspace*{-0.6cm}
	\epsfig{file=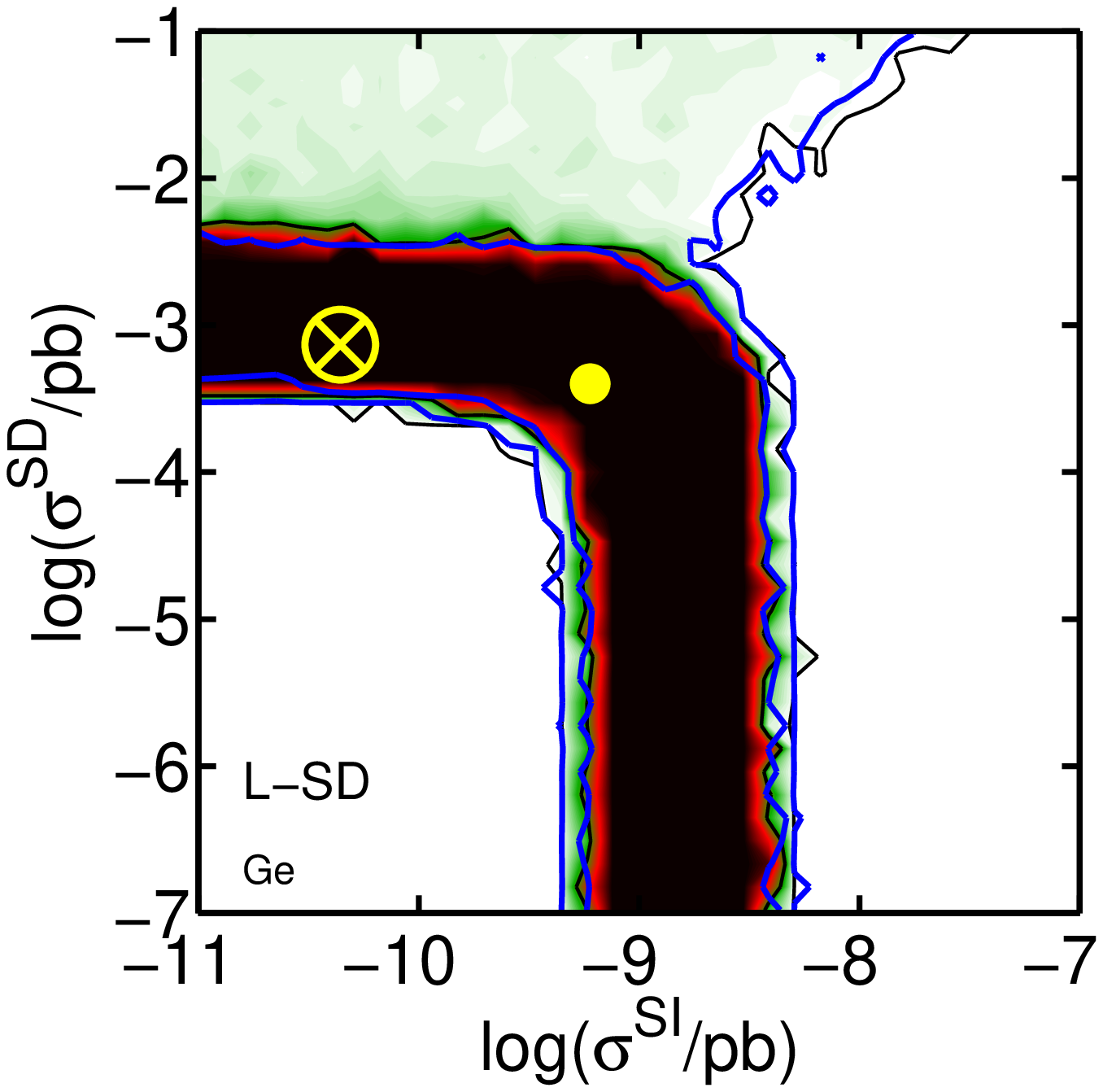,width=5.9cm}\\[-0.5cm]
	\hspace*{-1cm}
	\epsfig{file=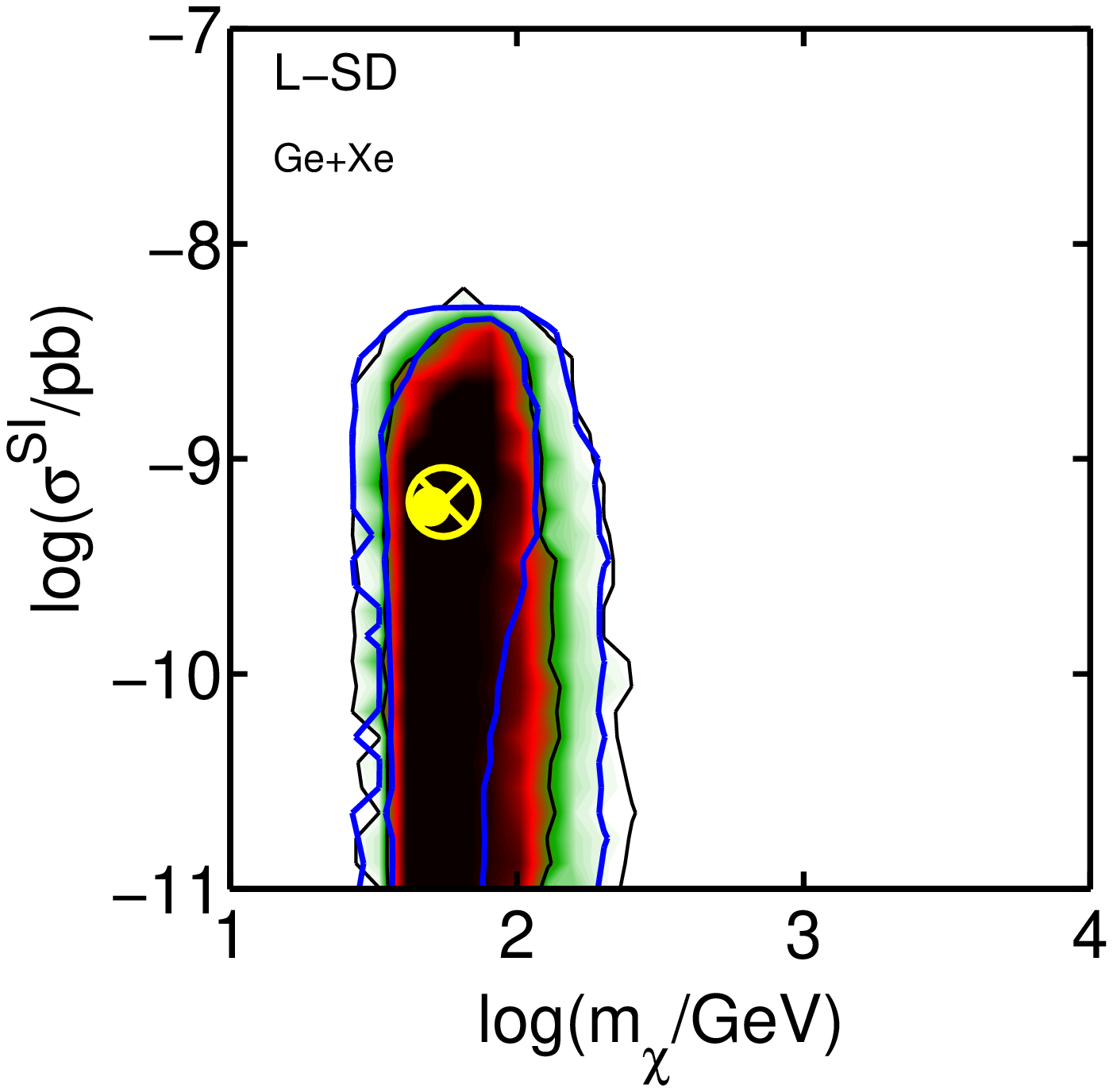,width=5.9cm}\hspace*{-0.6cm}
	\epsfig{file=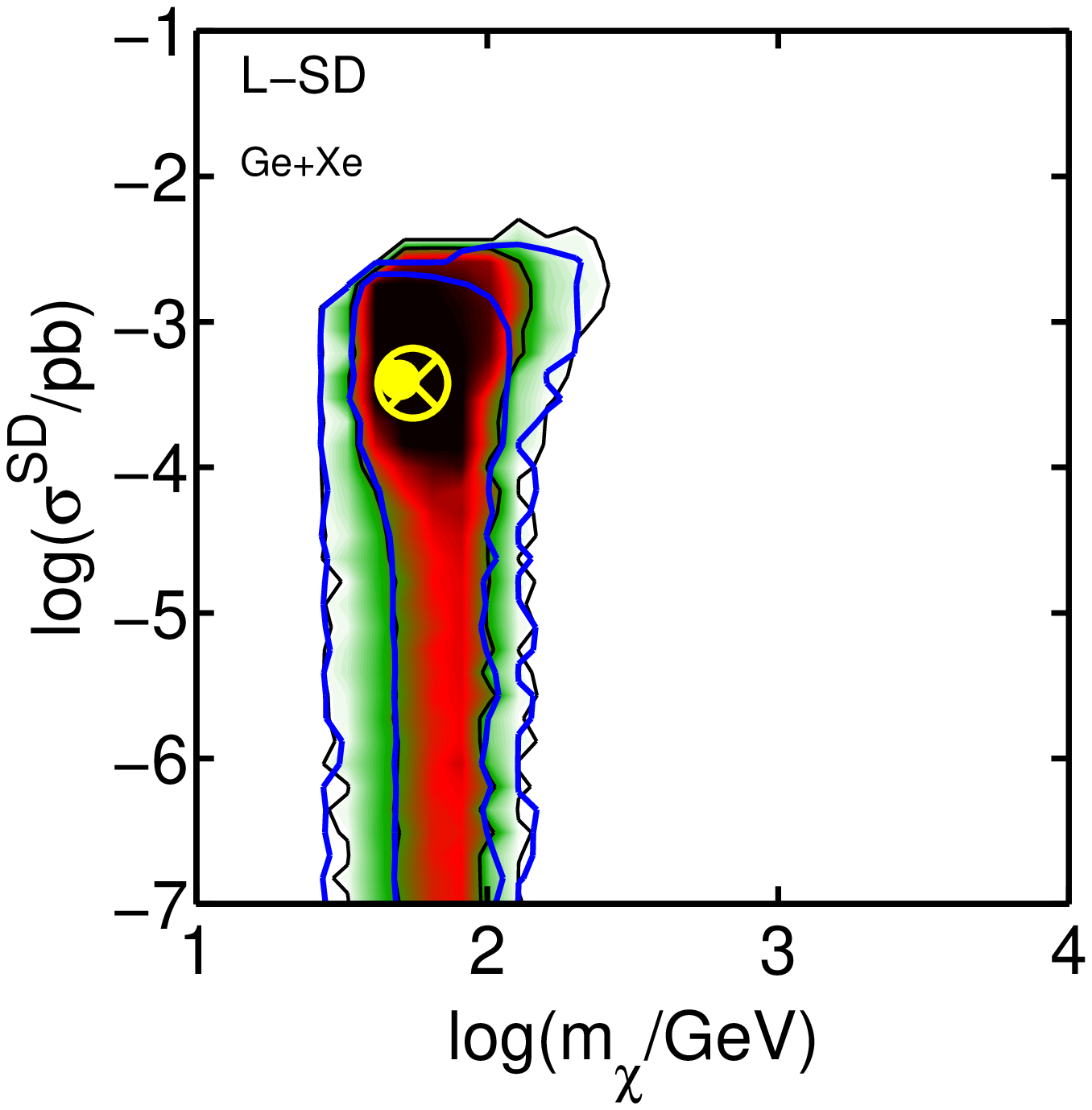,width=5.9cm}\hspace*{-0.6cm}
	\epsfig{file=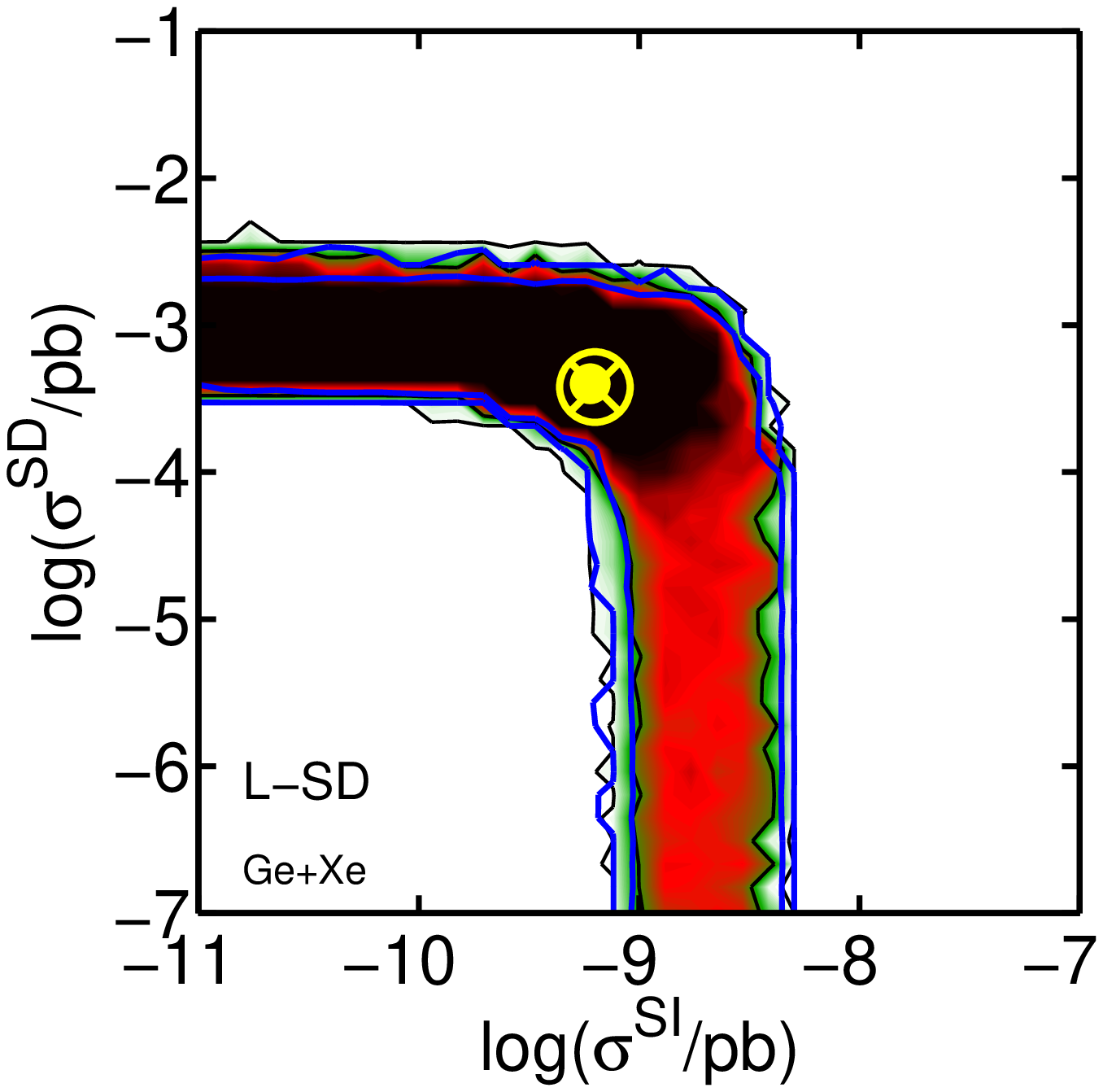,width=5.9cm}\\[-0.5cm]
	\hspace*{-1cm}
	\epsfig{file=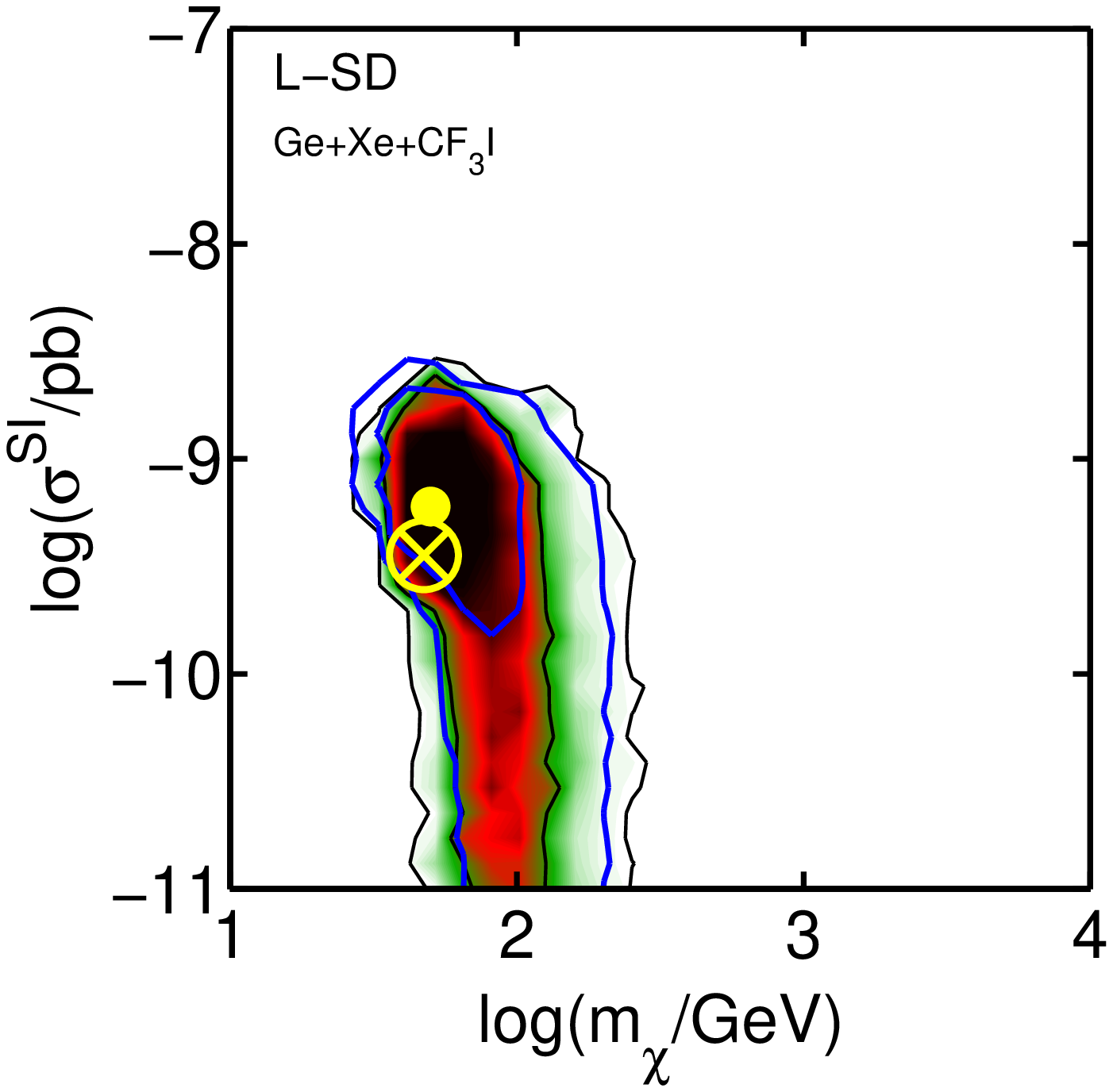,width=5.9cm}\hspace*{-0.6cm}
	\epsfig{file=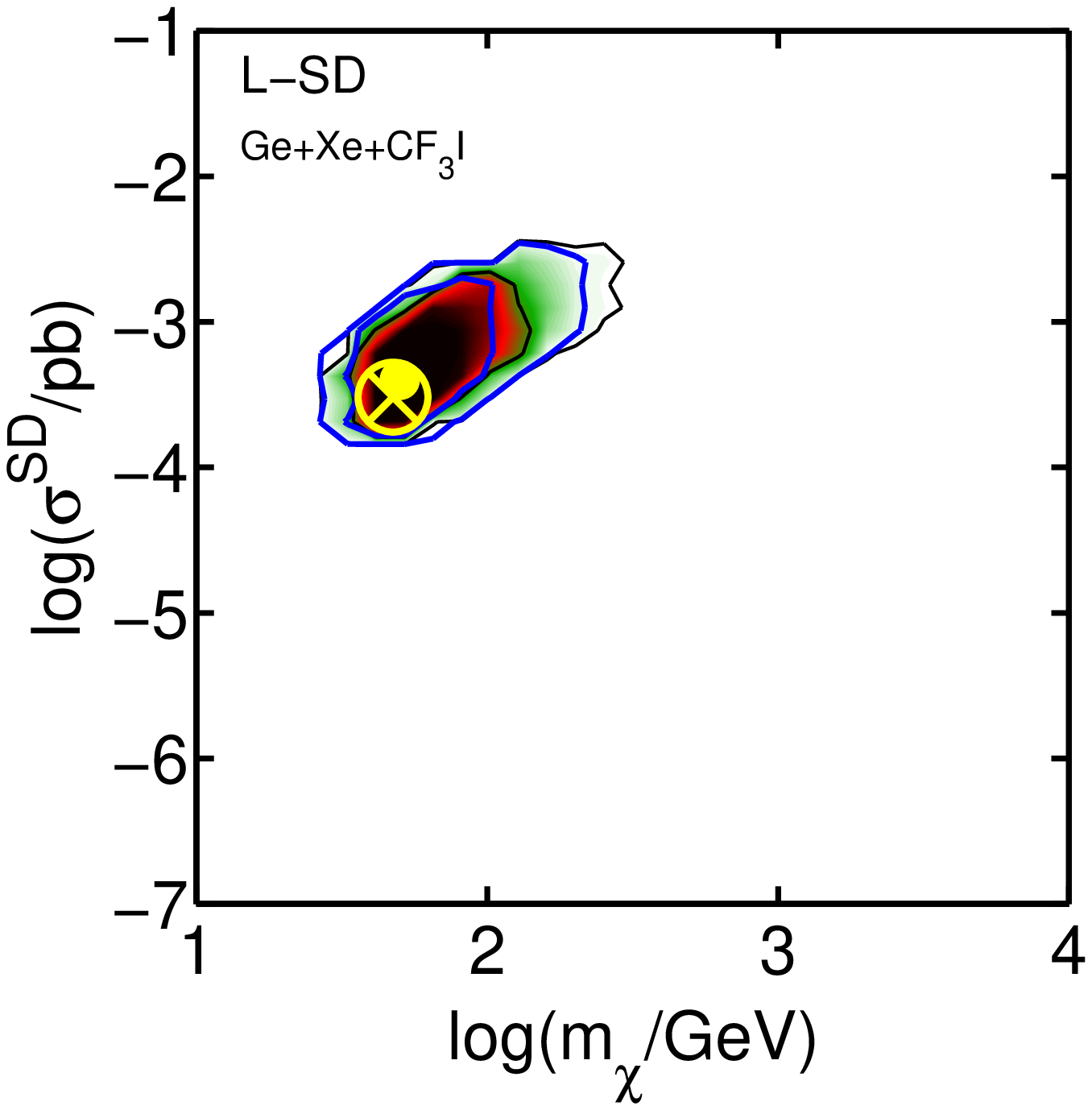,width=5.9cm}\hspace*{-0.6cm}
	\epsfig{file=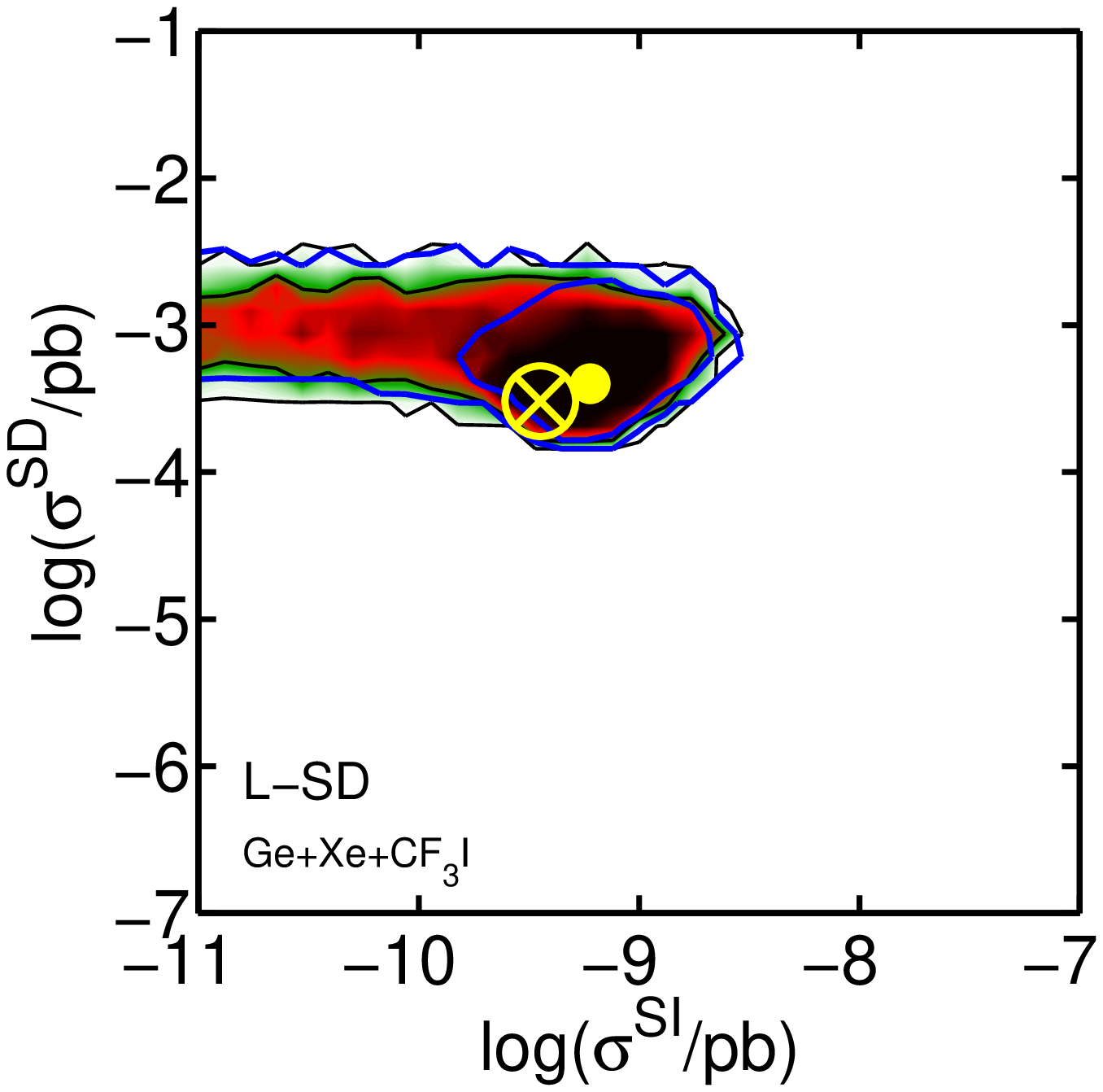,width=5.9cm}
  \caption{\small The same as in Fig.\,\ref{fig:cdmsxenon-bmmsi-pl}, but for the benchmark point \bmlsd.}
  \label{fig:cdmsxenon-bmmsd}
\end{figure}

The importance of nuclear uncertainties is evidenced when we consider the combination of the signal in Ge and Xe. For a fixed SD form factor it is not possible to provide a good fit to both signals with a SD-dominated energy spectrum. This produces an upper limit on $\sigsd$. On the other hand, when also the signal in CF$_3$I is included, selecting only points with a SI-dominated rate, the details of the SD form factor become irrelevant again and we see no difference between the blue and the black contours.

The last example considered is given in Fig.\,\ref{fig:cdmsxenon-bmmsd}, corresponding to the case \bmlsd, in which the SD contribution to the number of DM events dominates. As in the previous cases, the combination of data from Ge and Xe is not enough to determine the SI and SD contributions (only a combination of both).
Including CF$_3$I data has the effect of excluding small values of $\sigsd$ (contrary to what happened for \bmmsi\ and \bmlsi). Moreover, small SI cross sections also produce a slightly worse fit to the data now. Thus complementarity could be obtained for the 68\% confidence region but only if we do not include nuclear uncertainties.

To sum up, although 1 ton experiments based on germanium or xenon have an excellent discovery potential for WIMP DM, they are mostly sensitive to the SI part of the WIMP-nucleus interaction. As a consequence, the combination of data from both detectors would generally not allow the determination of the DM parameters in an unambiguous way. 
If we look for complementary information, the obvious alternative is a target that is sensitive to the SD component. 
CF$_3$I is certainly a good choice in this respect, since it incorporates fluorine $^{19}$F, which has an unpaired proton and a large nuclear nuclear spin $J$. However, COUPP would not provide information about the recoil spectrum, which limits its complementarity capability. Even considering a 1 ton phase the degeneracies in the reconstruction of DM parameters would not be completely removed.

\section{Complementarity of bolometric targets}
\label{sec:rosebud}

A much more appealing situation could come from an experiment which is sensitive to the SD WIMP-nucleon interaction and in which the energy spectrum can be accurately determined, as well. This can be the case of certain scintillating bolometric targets that have been developed and studied by the ROSEBUD  \cite{coron} and CRESST collaborations and are planned or could be used by the EURECA project. We will consider the following targets:
\begin{itemize} 
\item CaWO$_4$: This is a material which has been tested and used as a WIMP target by ROSEBUD \cite{cebrian} and it is the current target in CRESST \cite{Angloher:2011uu}. The sensitivity to SD interactions comes from tungsten, for which the isotope $^{183}$W (with a natural isotopic abundance of 14.3\%) has an unpaired neutron. 
However, tungsten is a very heavy material and therefore it is mostly sensitive to the SI component of the WIMP interactions. For this reason, we should expect compatibility with experiments based on Ge or Xe to be limited. Notice that our analysis does not include the possibility (shown by CRESST) of disentangling O, Ca and W nuclear recoils.

\item Al$_2$O$_3$: Sapphire is a very promising DM target because it is sensitive to low mass WIMPs (Al and O are both light nuclei). It is also sensitive to SD interactions ($^{27}$Al has 100\% isotopic abundance and $J=5/2$) and recent tests indicate very high light yields \cite{amare,Luca:2009zz} and a particle discrimination threshold below $10$~keV seems to be possible \cite{amare2006}. 

\item LiF: This target is also sensitive to low mass and SD interactions (Li and F are light nuclei, $^{19}$F has 100\% isotopic abundance and $J=1/2$, $^7$Li has 92.5\% isotopic abundance and $J=3/2$). However, up to now, low particle discrimination thresholds have not been obtained and more R{\&}D is needed on this target to be used for DM searches.
\end{itemize}

In this section we determine the conditions under which the different scintillating bolometric targets can provide good complementary information when combined with germanium and xenon detectors.
As in the previous section, for each benchmark and for each detector target, we simulate the number of recoils predicted in the different energy bins. This constitutes our set of observables. Then, we perform a scan on the three-dimensional parameter space ($\mwimp$, $\sigsi$, $\sigsd$) and apply the Bayesian inference method to determine the reconstruction of these quantities. Astrophysical and nuclear uncertainties are included as described above.

\begin{table}

\begin{center}

\begin{tabular}{|c||c|c|c||c|c|c|}
\hline
Benchmark Point & $\mwimp$ (GeV) & $\sigsi$ (pb) & $\sigsd$ (pb) & $N_{\rm CaWO_4}$ & $N_{\rm Al_2O_3}$ & $N_{\rm LiF}$\\
\hline
\hline
\bmmsi & $100$ & $10^{-9}$ & $10^{-5}$ &27 (27) &14 (4) &25 (2)\\
\bmlsi & $50$ & $10^{-9}$ & $10^{-5}$ &30 (30) &17 (5) &35 (2)\\
\bmlsd & $50$ & $6\times10^{-10}$ & $4\times10^{-4}$ &20 (18) &505 (3) &1295 (1)\\
\bmvlsi & $15$ & $10^{-8}$ & $10^{-5}$ &8 (8) &23 (18) &29 (11)\\
\hline
\end{tabular}
\caption{\small Total recoil events for the set of benchmark points expected on each of the bolometric targets considered in this work for an exposure $\epsilon=300$~kg\,yr.}
\label{tab:rosebud}
\end{center}
\end{table}

\begin{figure}
\hspace*{-1cm}
	\epsfig{file=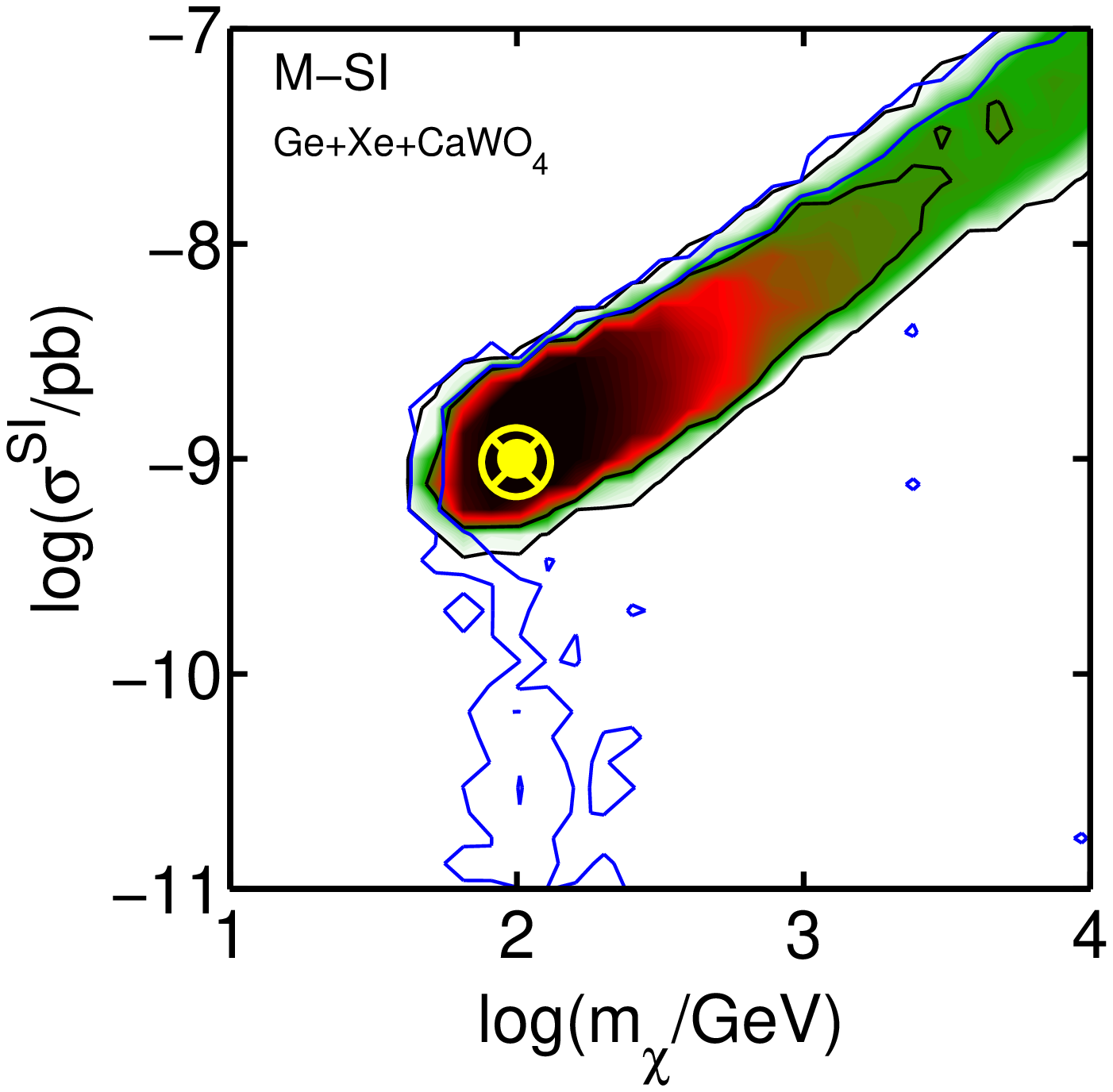,width=5.9cm}\hspace*{-0.6cm}
	\epsfig{file=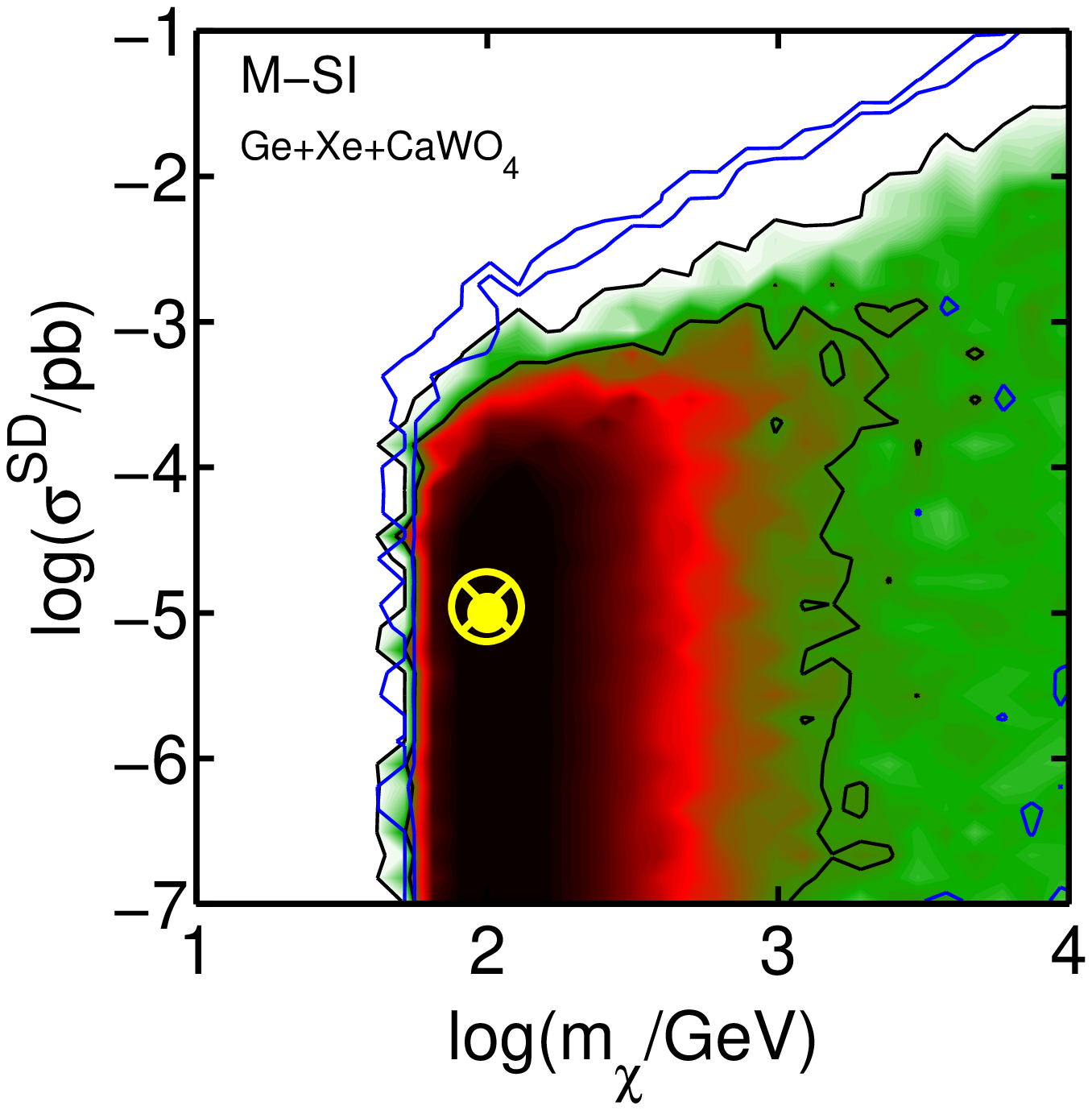,width=5.9cm}\hspace*{-0.6cm}
	\epsfig{file=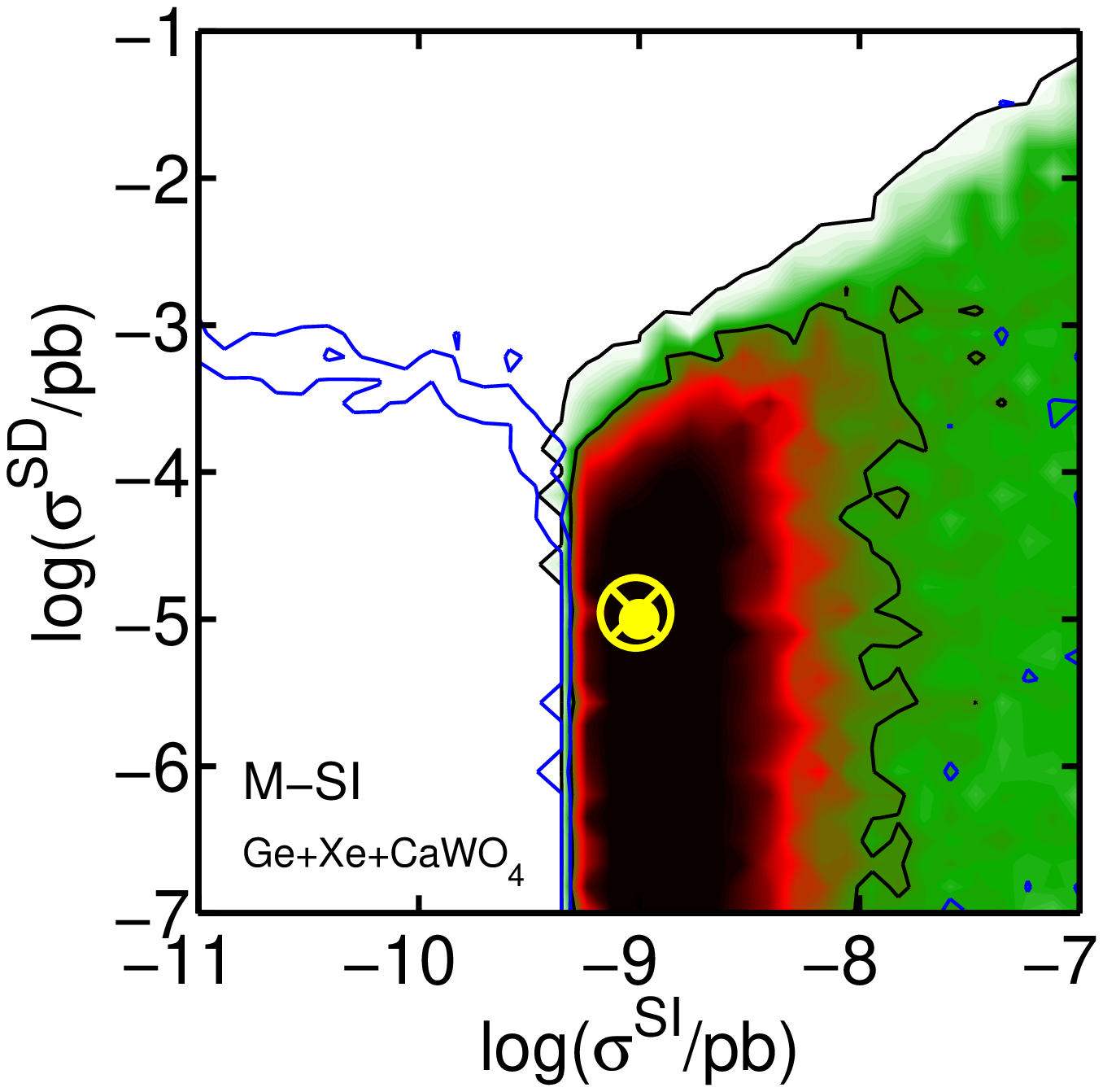,width=5.9cm}\\[-0.5cm]
	\hspace*{-1cm}
	\epsfig{file=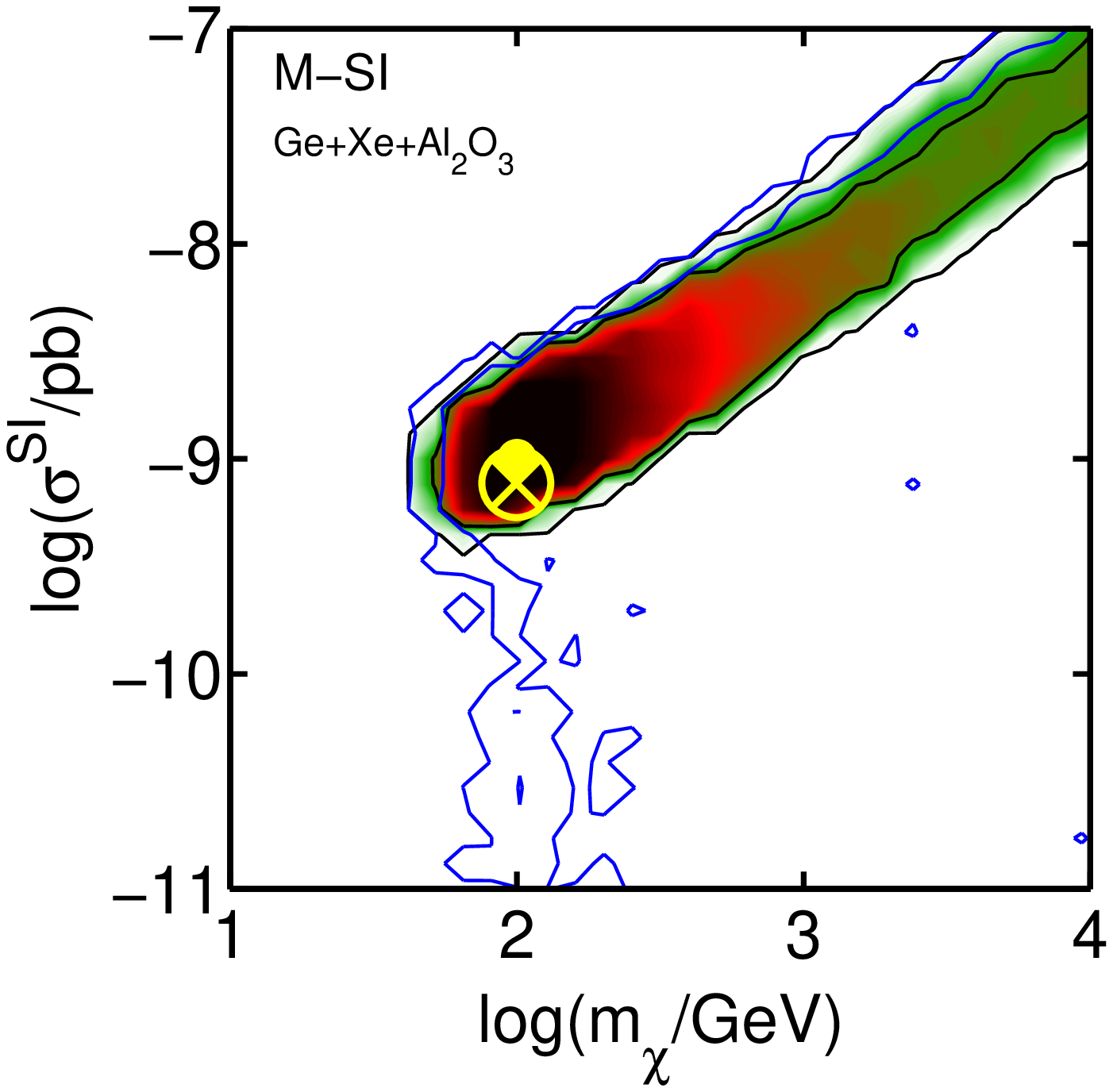,width=5.9cm}\hspace*{-0.6cm}
	\epsfig{file=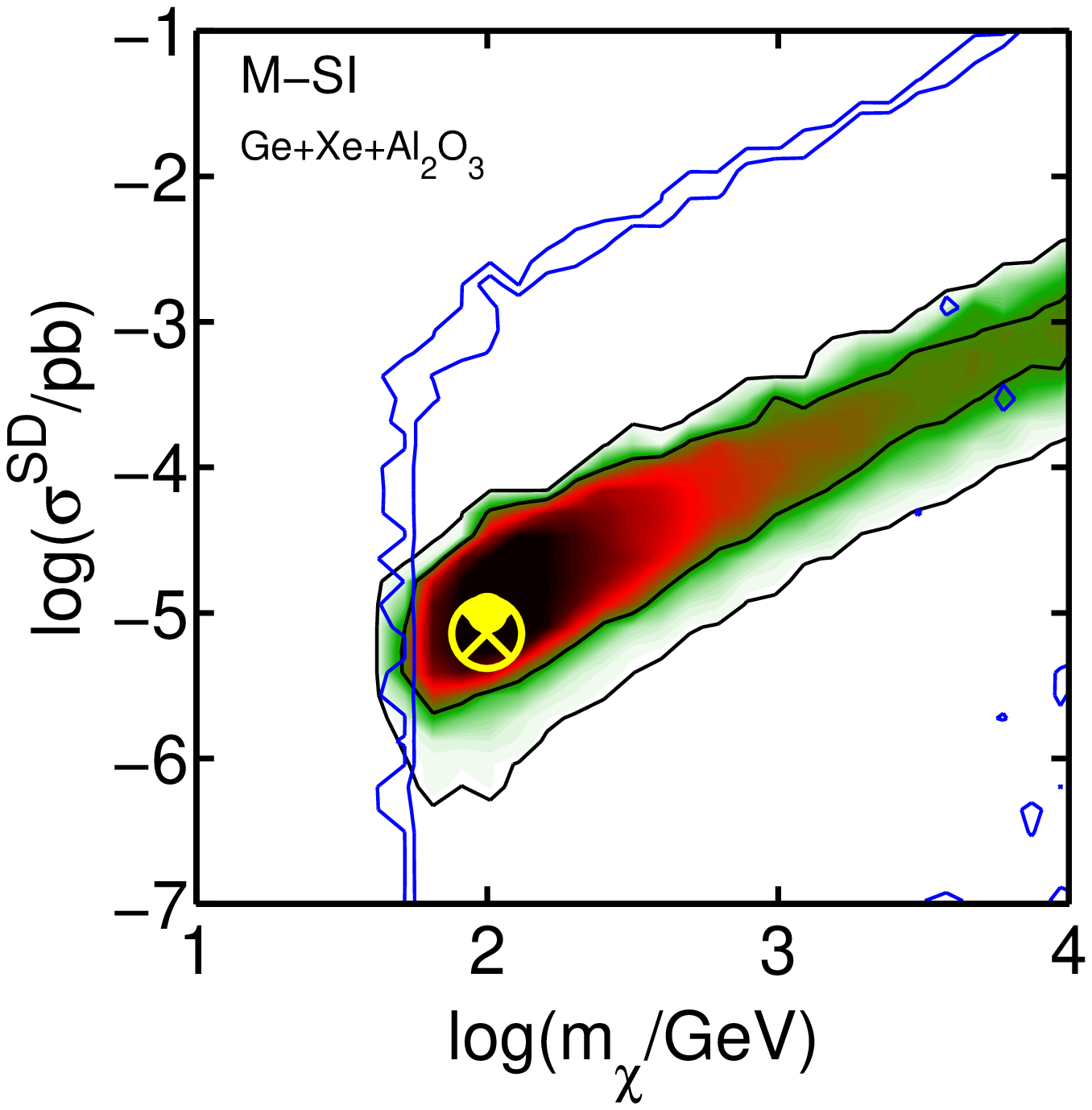,width=5.9cm}\hspace*{-0.6cm}
	\epsfig{file=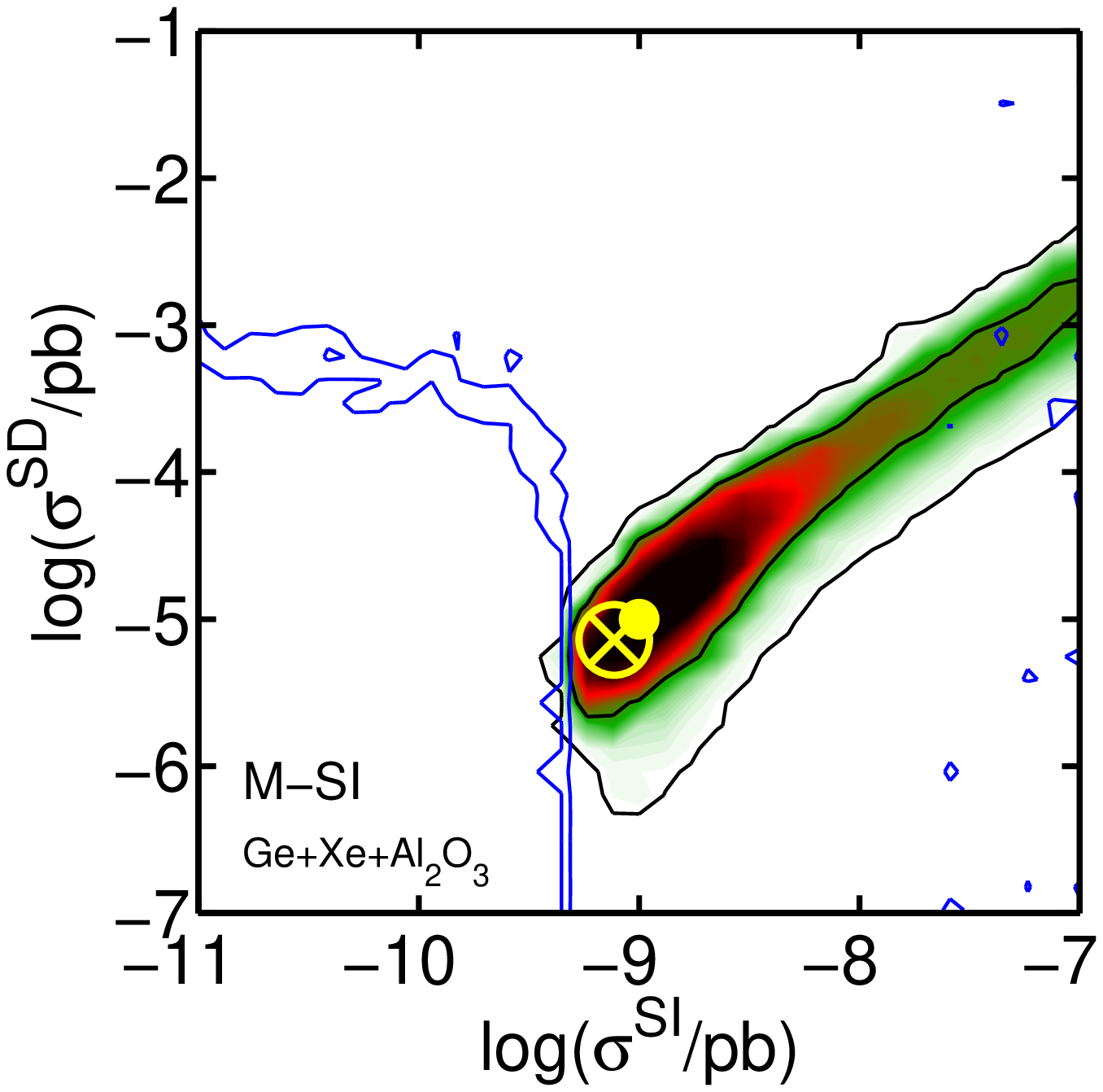,width=5.9cm}\\[-0.5cm]
	\hspace*{-1cm}
	\epsfig{file=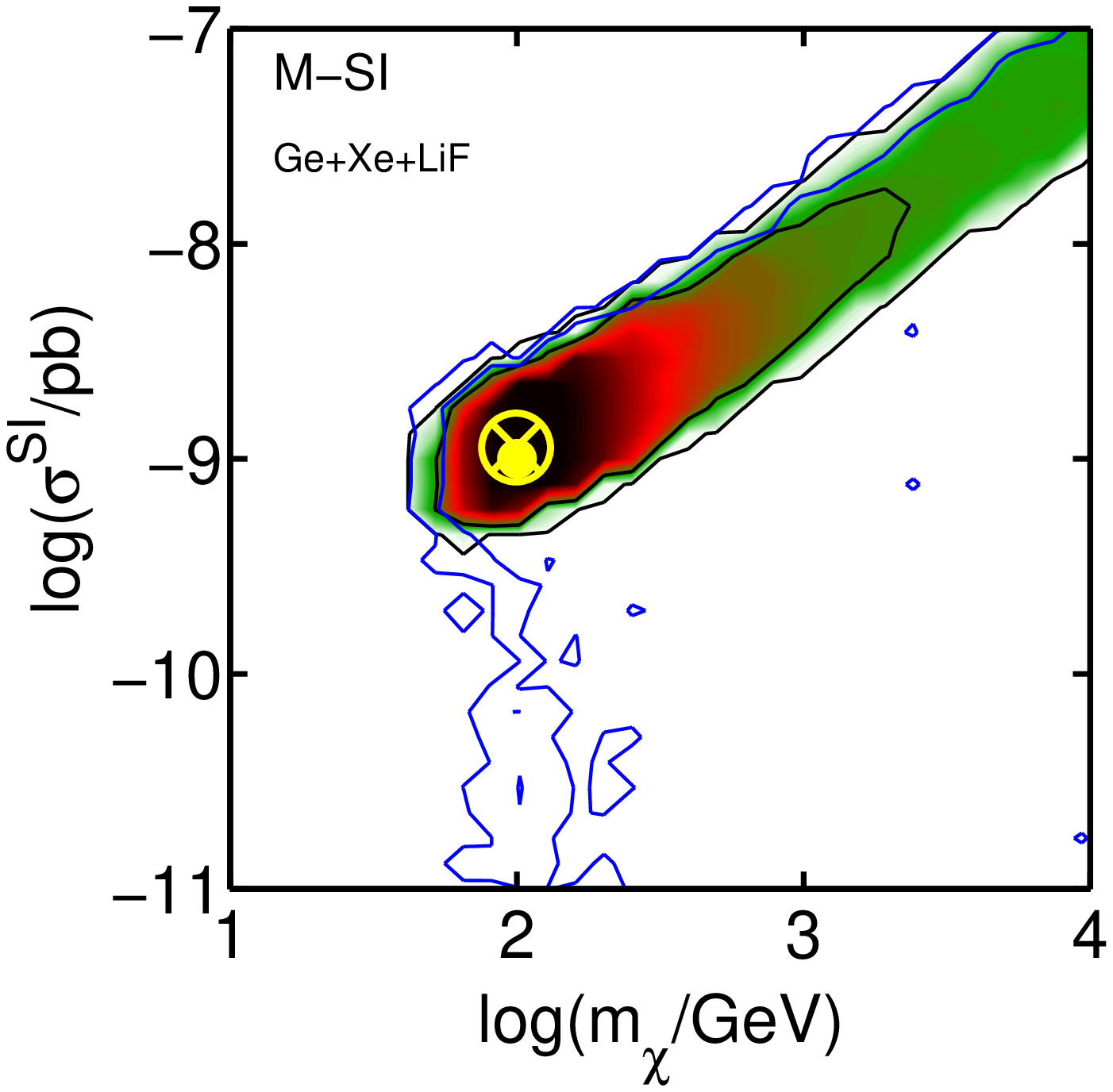,width=5.9cm}\hspace*{-0.6cm}
	\epsfig{file=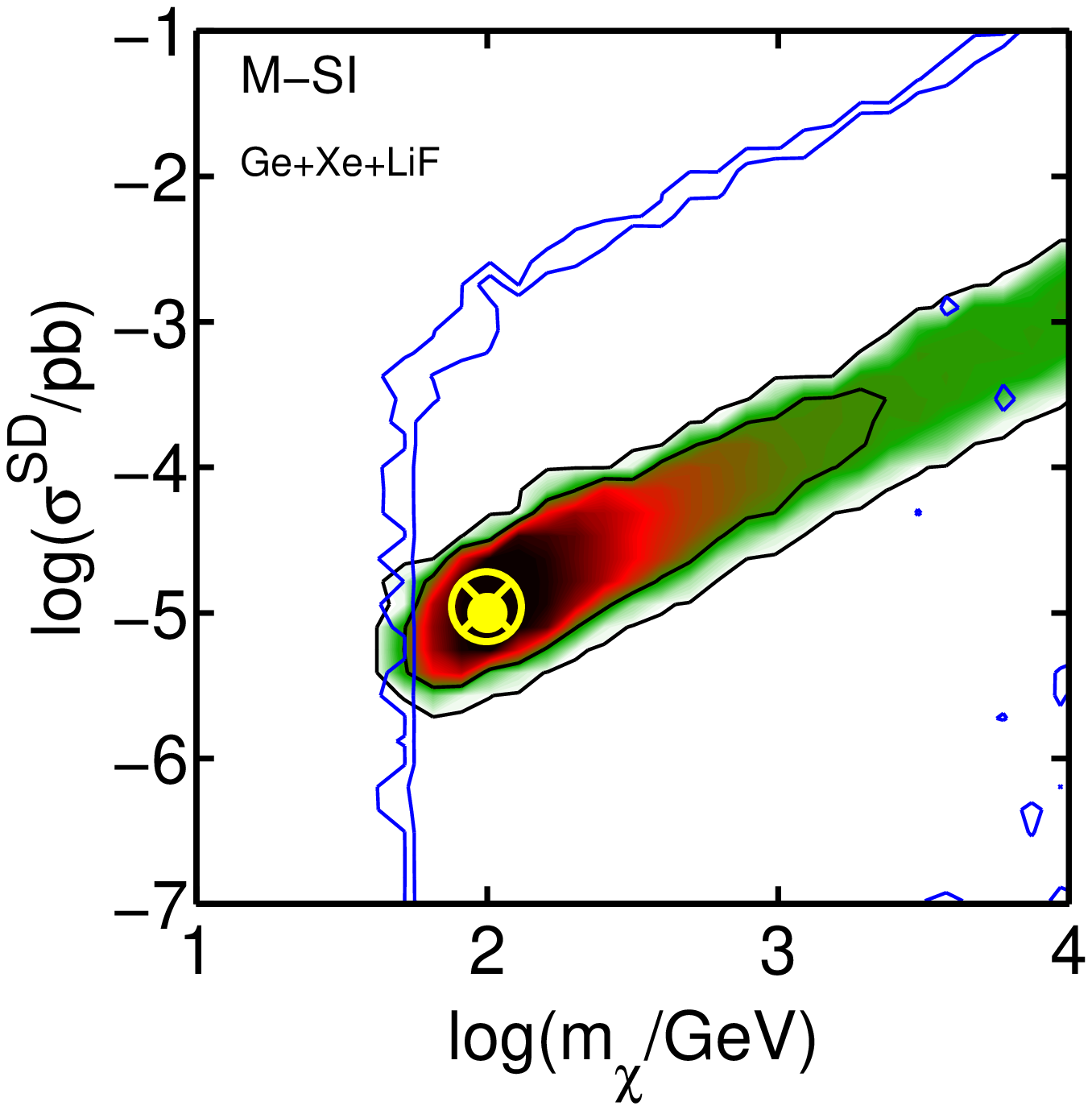,width=5.9cm}\hspace*{-0.6cm}
	\epsfig{file=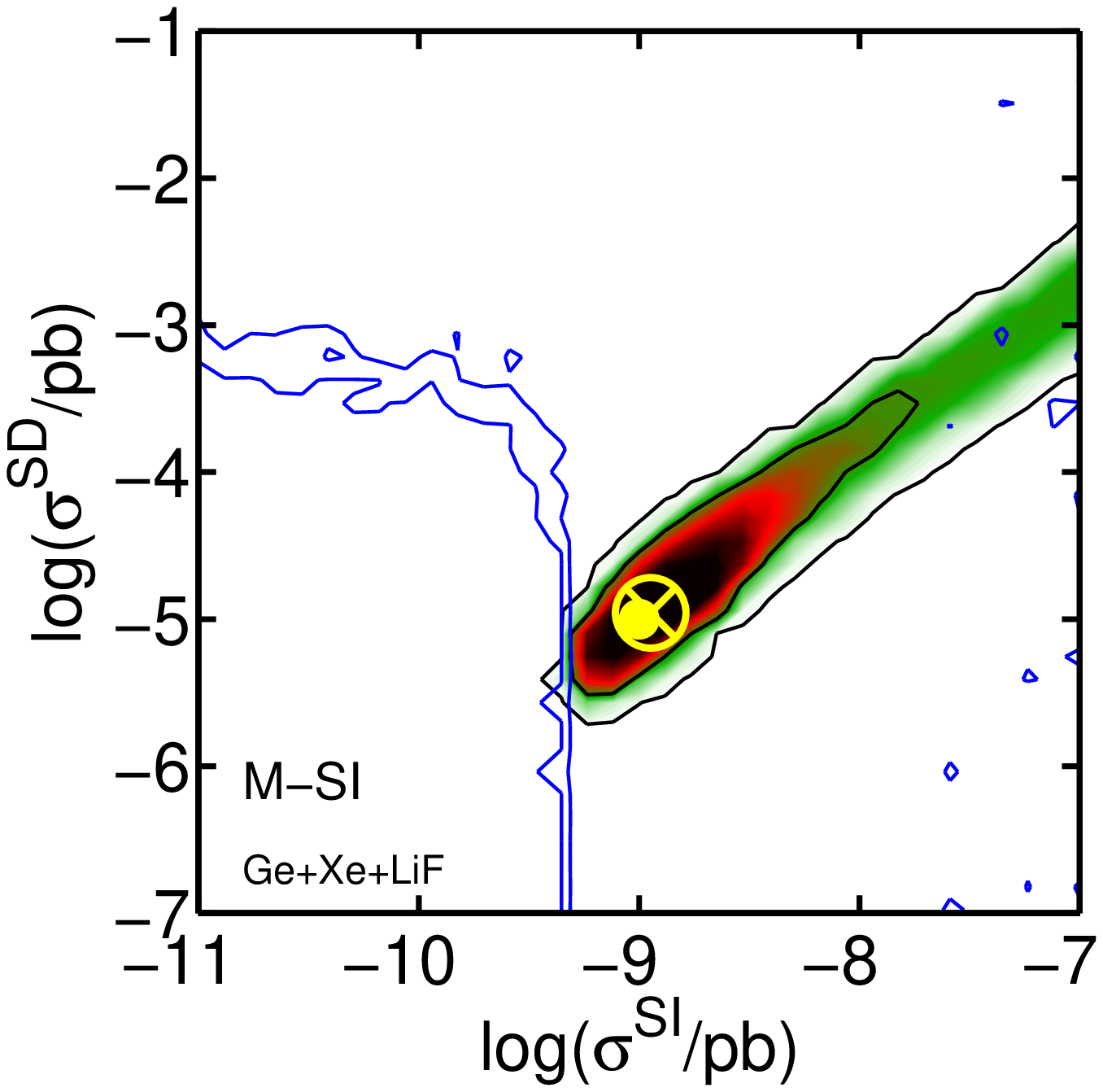,width=5.9cm}
\caption{\small Profile likelihood for the DM parameters in the $\msi$, $\msd$, and $\sisd$ planes for the benchmark point \bmmsi\   after the combination of data from a Ge detector, a Xe detector, and a bolometric target (CaWO$_4$, Al$_2$O$_3$ and LiF from top to bottom, respectively). The exposure is $\epsilon=300$~kg~yr for all the experiments. From the inside out, contours enclose 68\% and 99\% of confidence interval. The yellow dot represents the nominal point and the yellow cross the posterior mean. The blue lines correspond to the case when only Ge and Xe are used.
}
\label{fig:bmmsi}
\end{figure}

\begin{figure}
\hspace*{-1cm}
	\epsfig{file=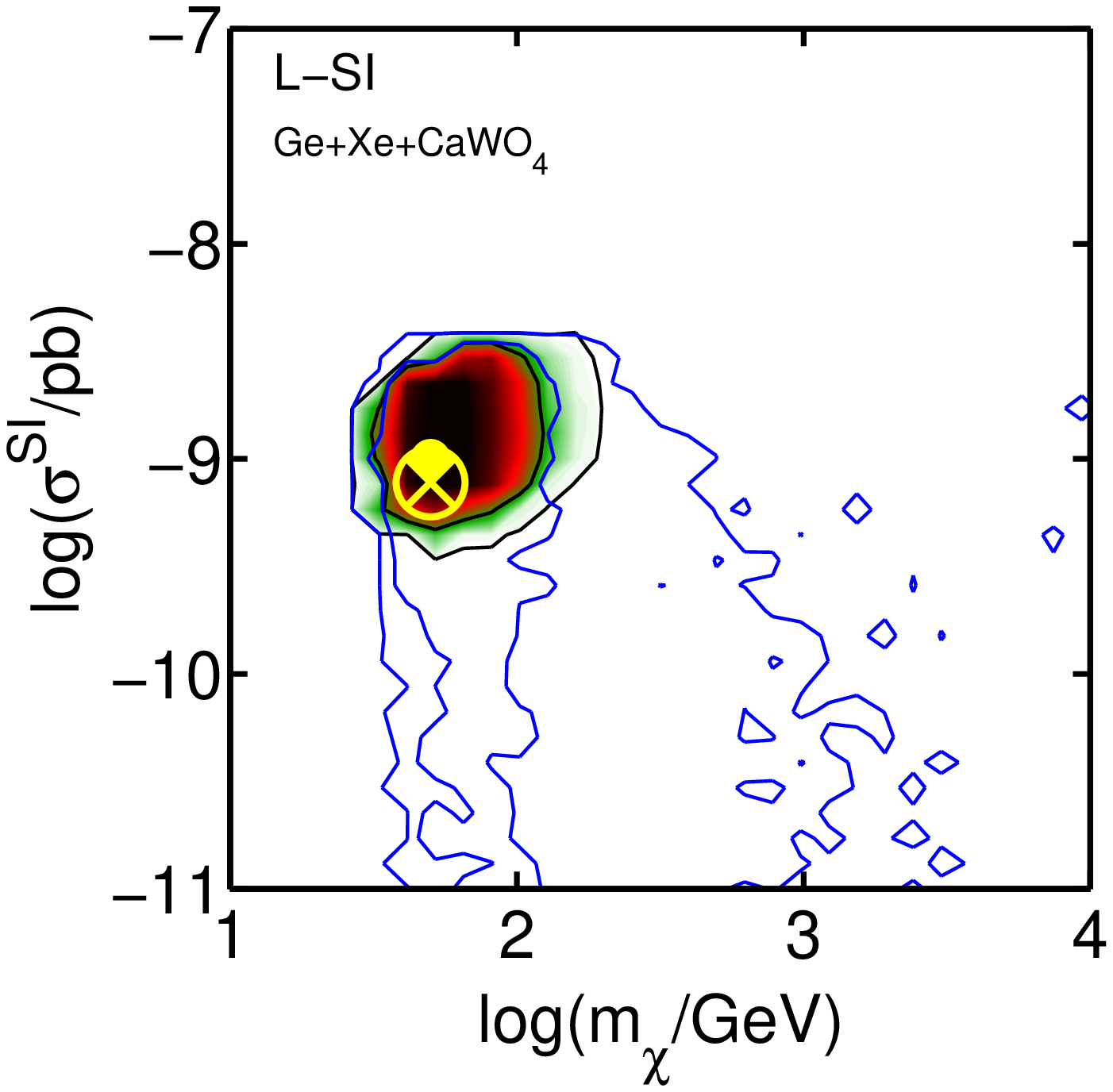,width=5.9cm}\hspace*{-0.6cm}
	\epsfig{file=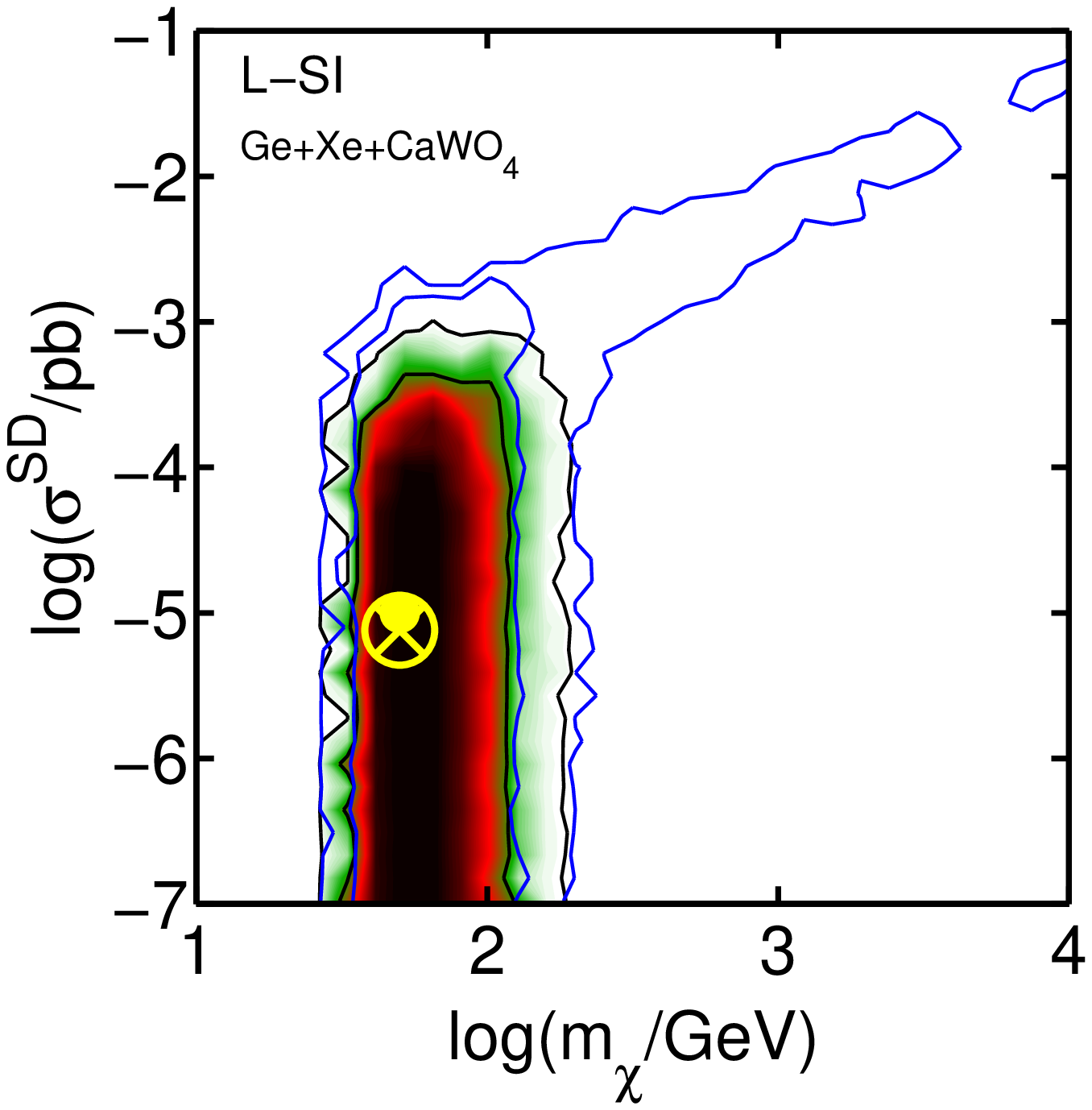,width=5.9cm}\hspace*{-0.6cm}
	\epsfig{file=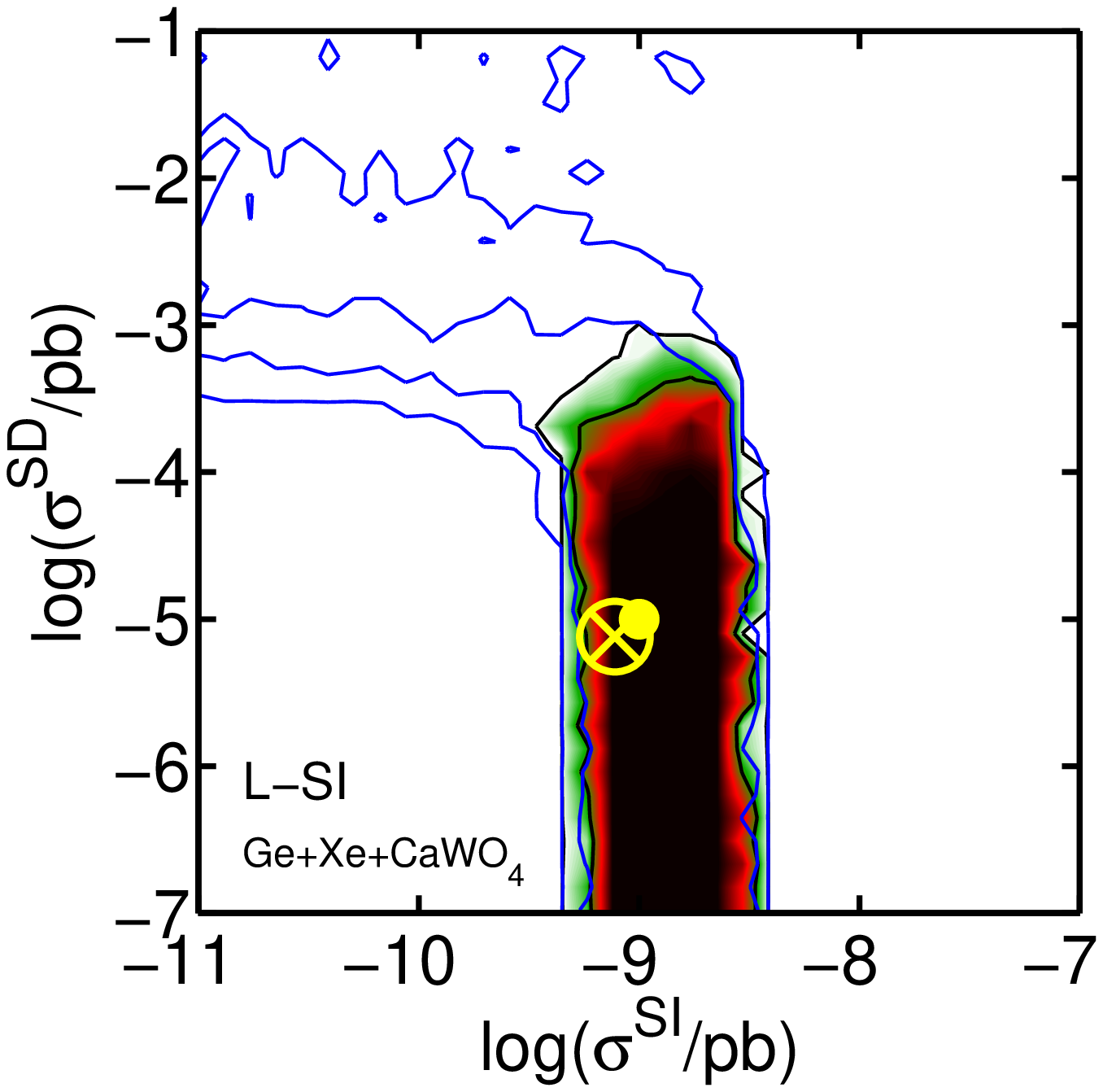,width=5.9cm}\\[-0.5cm]
	\hspace*{-1cm}
	\epsfig{file=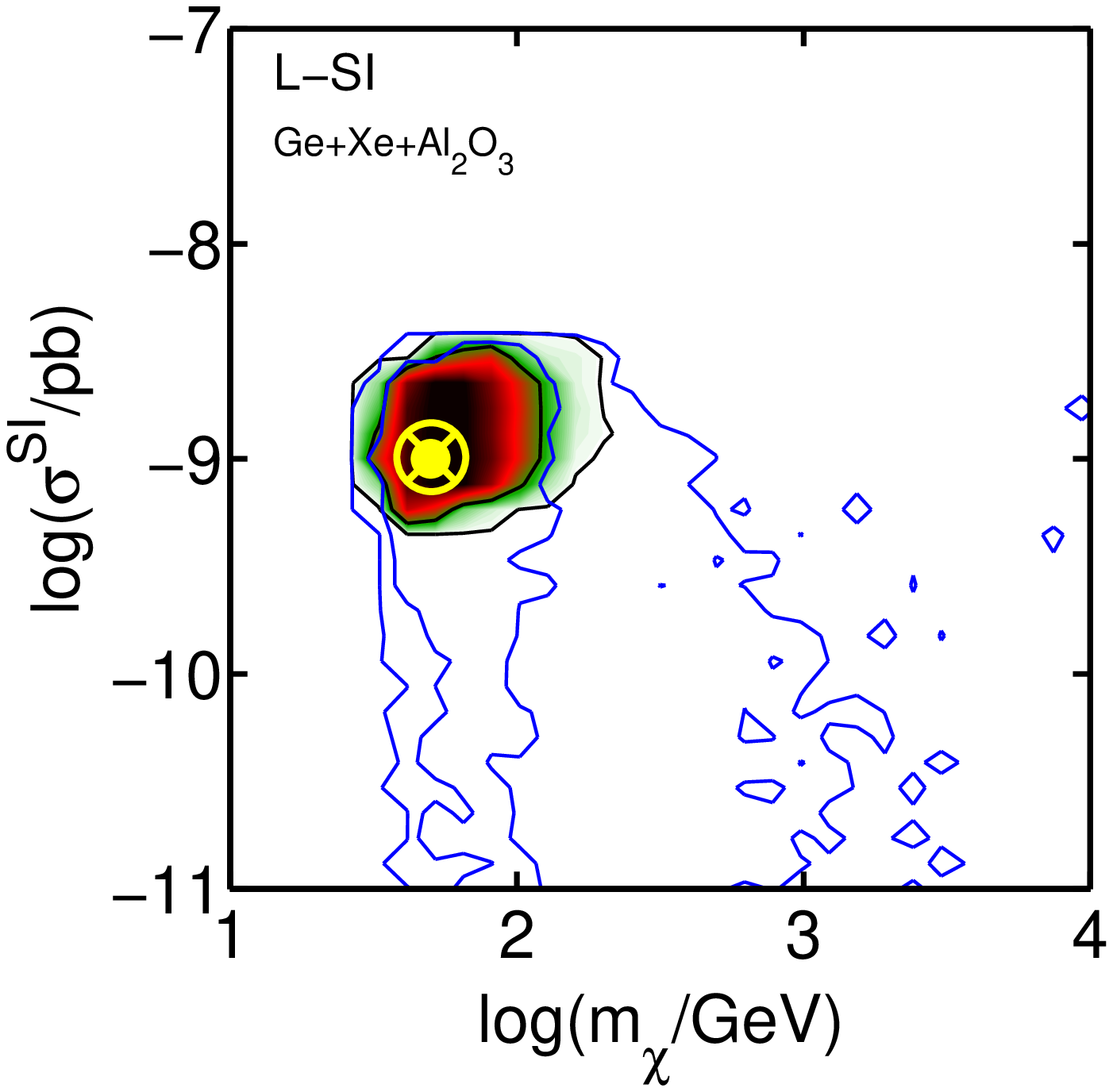,width=5.9cm}\hspace*{-0.6cm}
	\epsfig{file=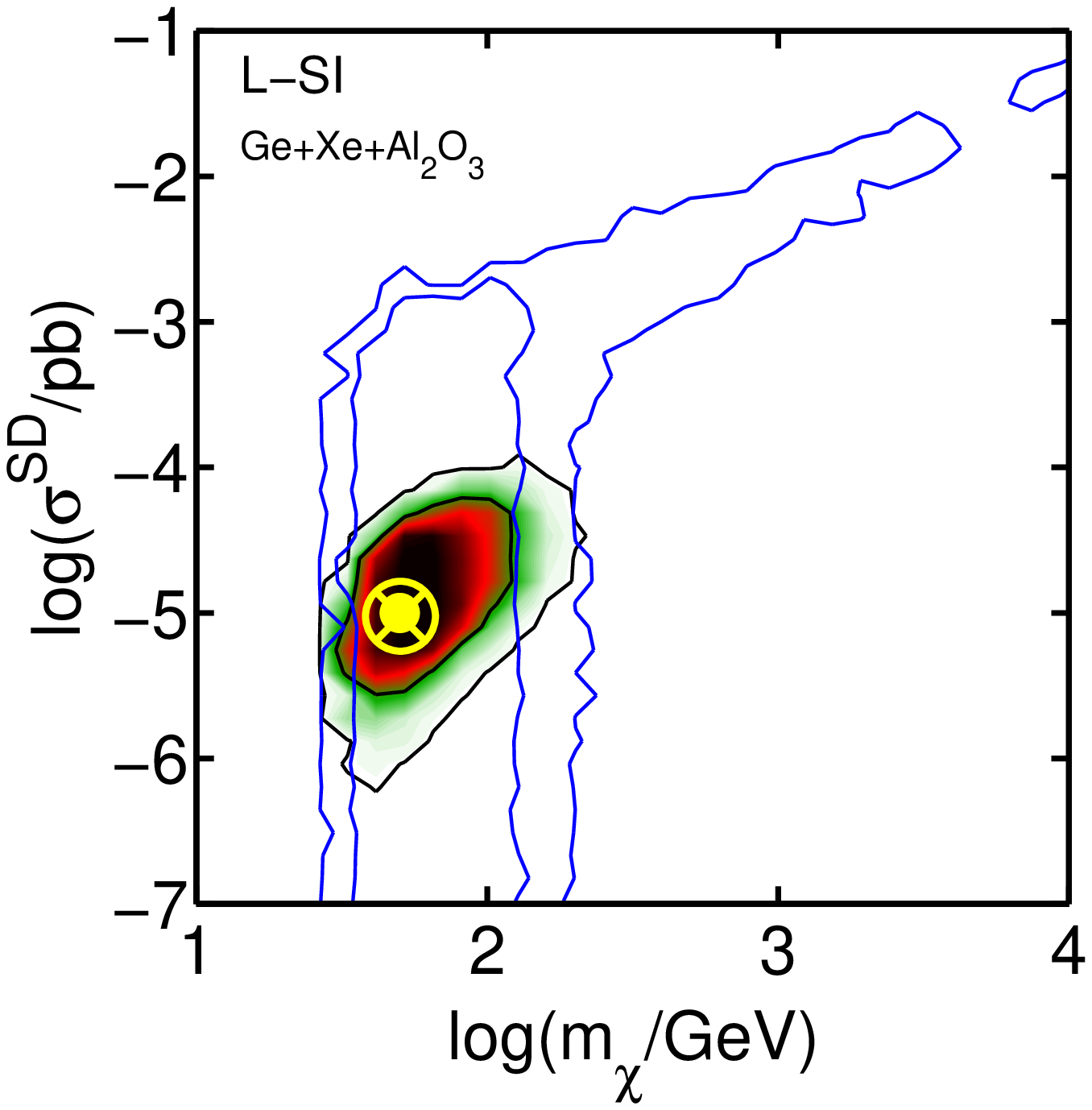,width=5.9cm}\hspace*{-0.6cm}
	\epsfig{file=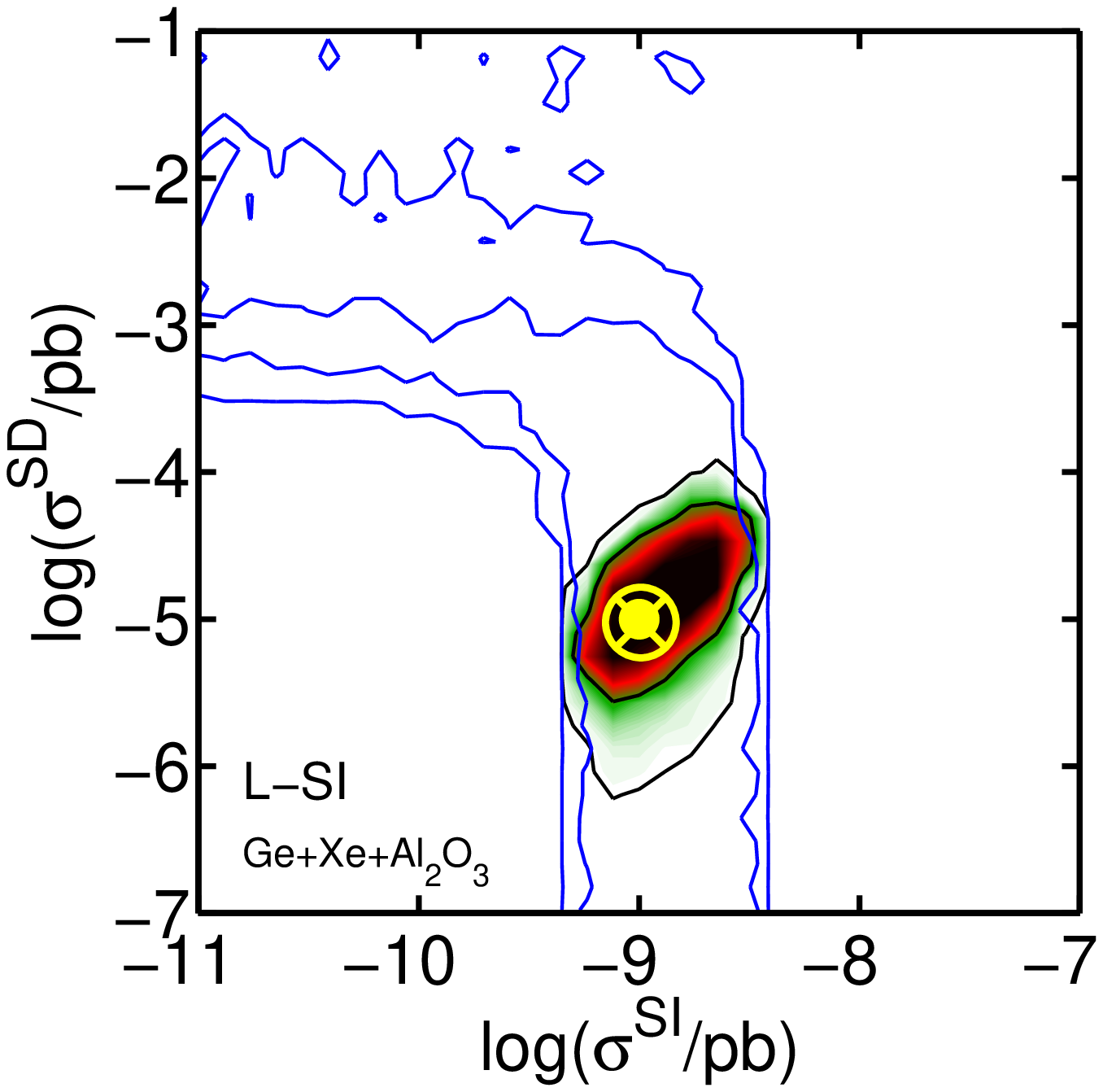,width=5.9cm}\\[-0.5cm]
	\hspace*{-1cm}
	\epsfig{file=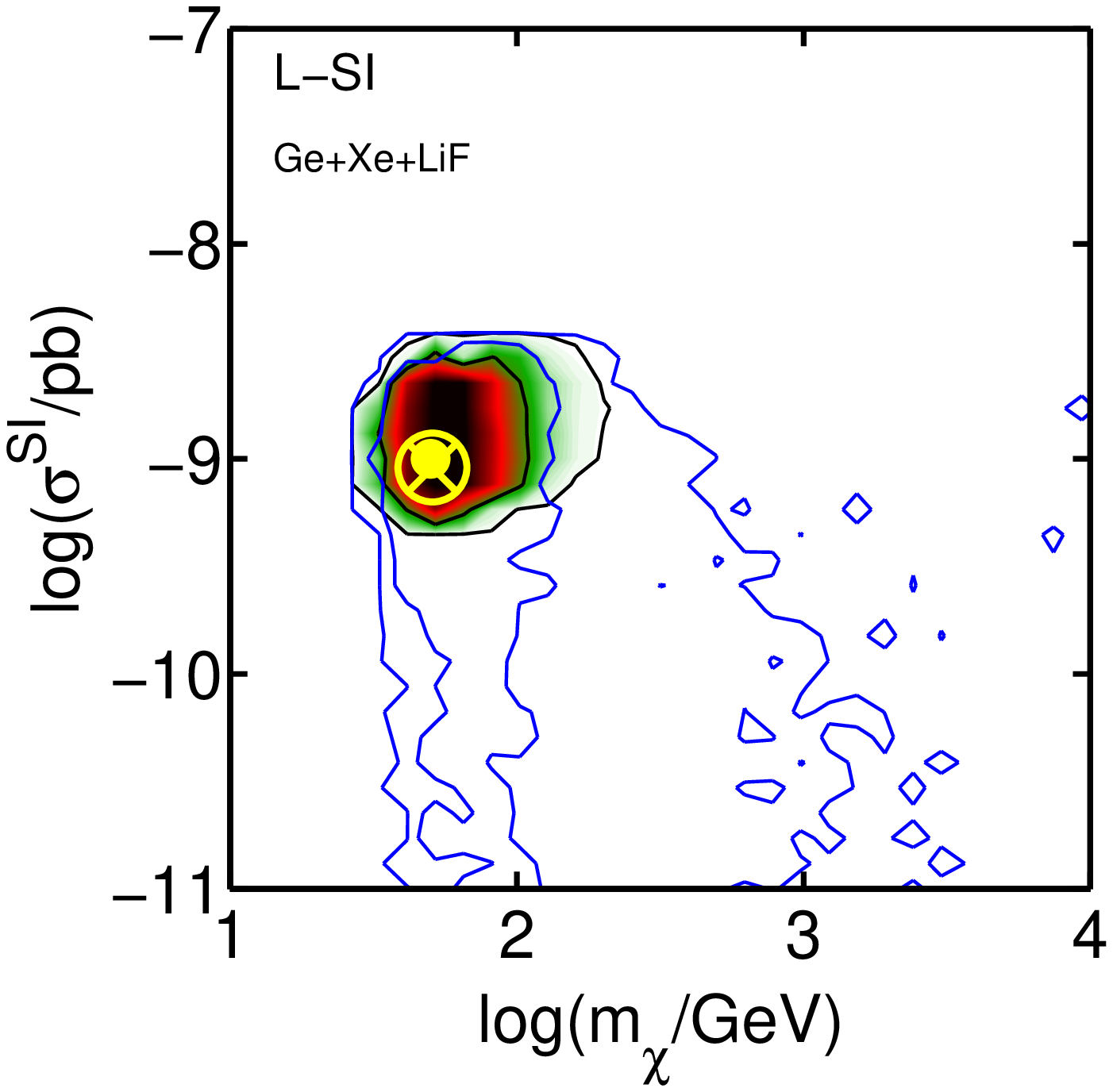,width=5.9cm}\hspace*{-0.6cm}
	\epsfig{file=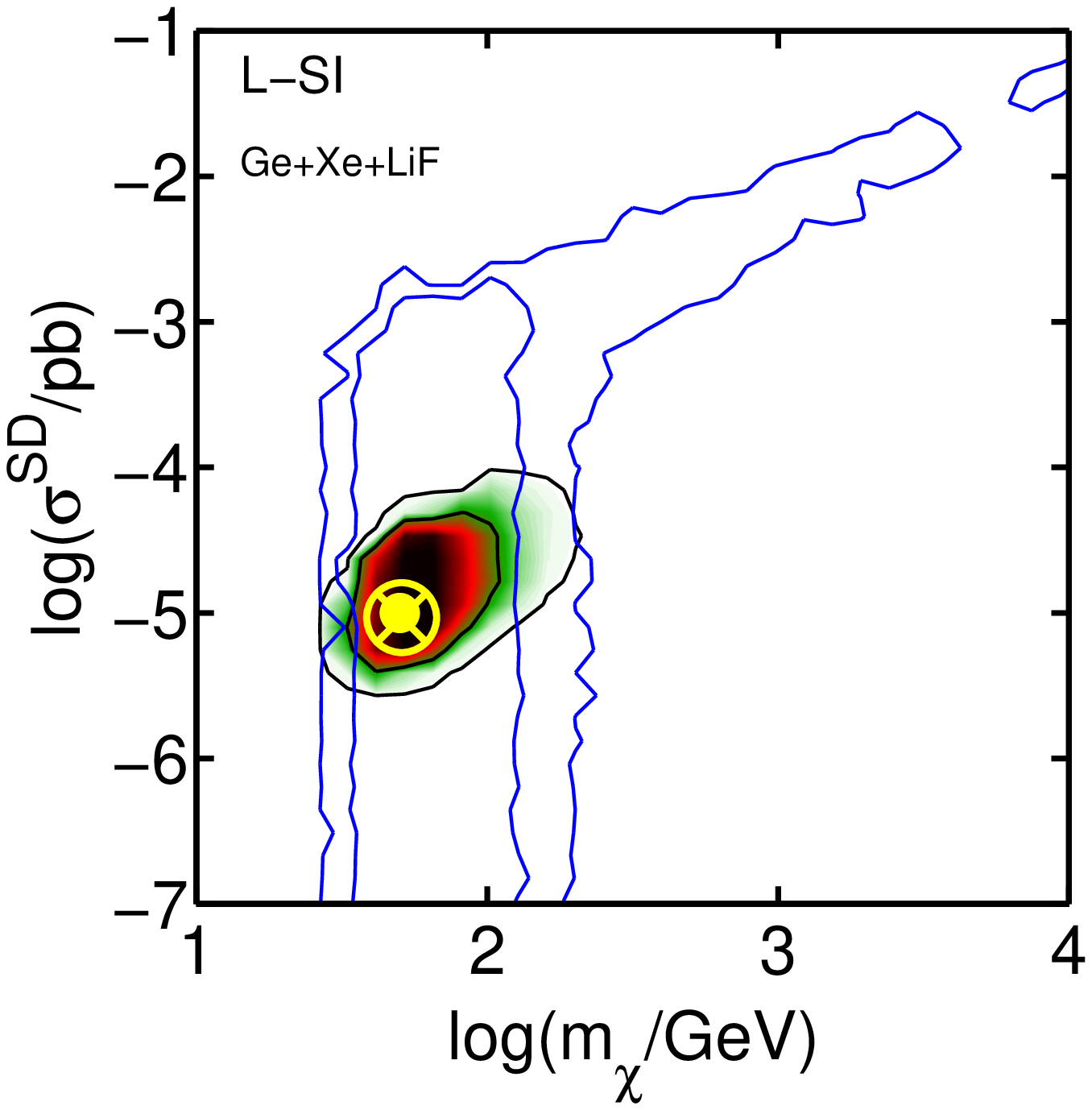,width=5.9cm}\hspace*{-0.6cm}
	\epsfig{file=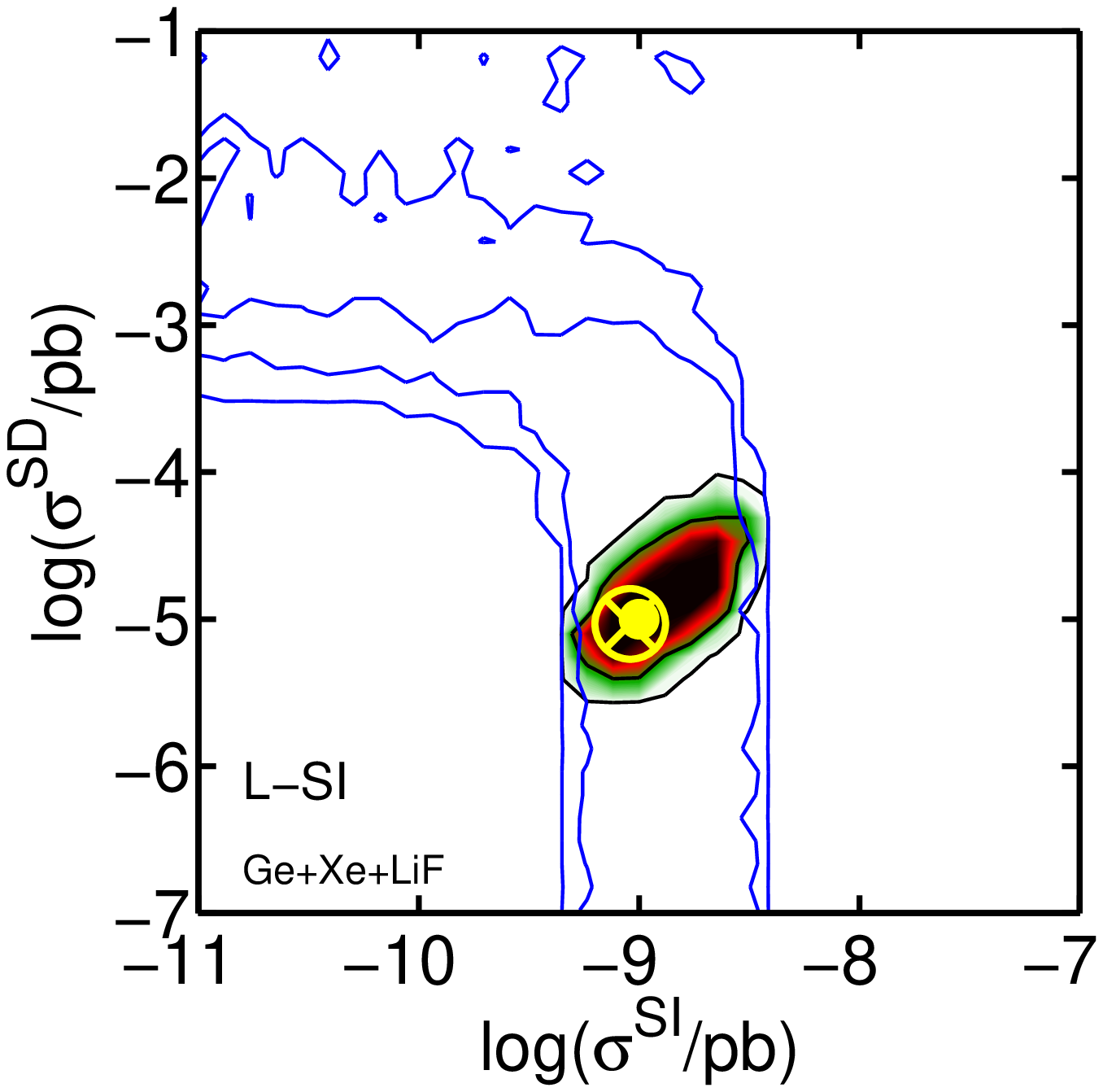,width=5.9cm}
\caption{\small The same as in Fig.\,\ref{fig:bmmsi} but for the case of \bmlsi.}
\label{fig:bmlsi}
\end{figure}

\begin{figure}
\hspace*{-1cm}
	\epsfig{file=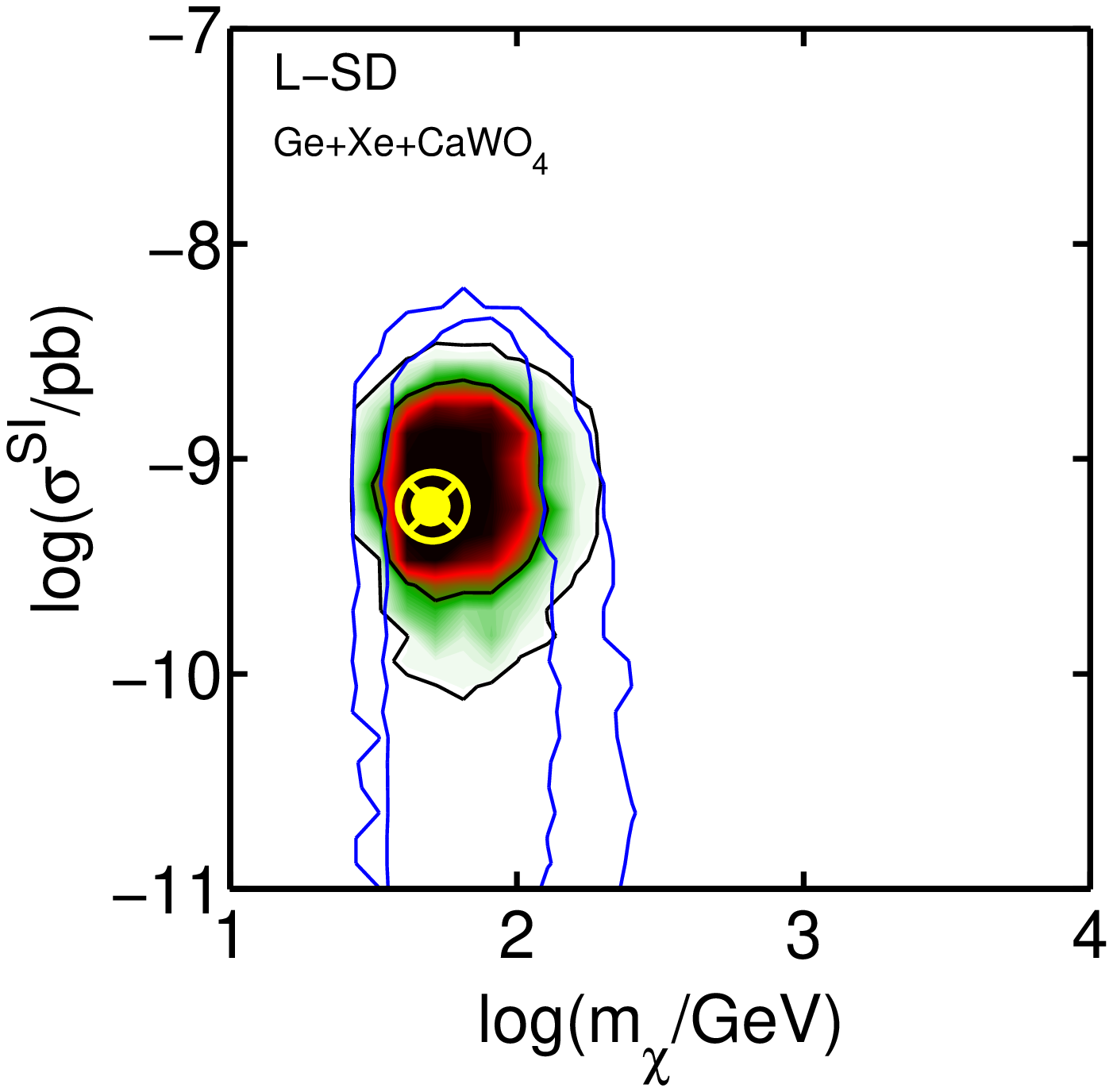,width=5.9cm}\hspace*{-0.6cm}
	\epsfig{file=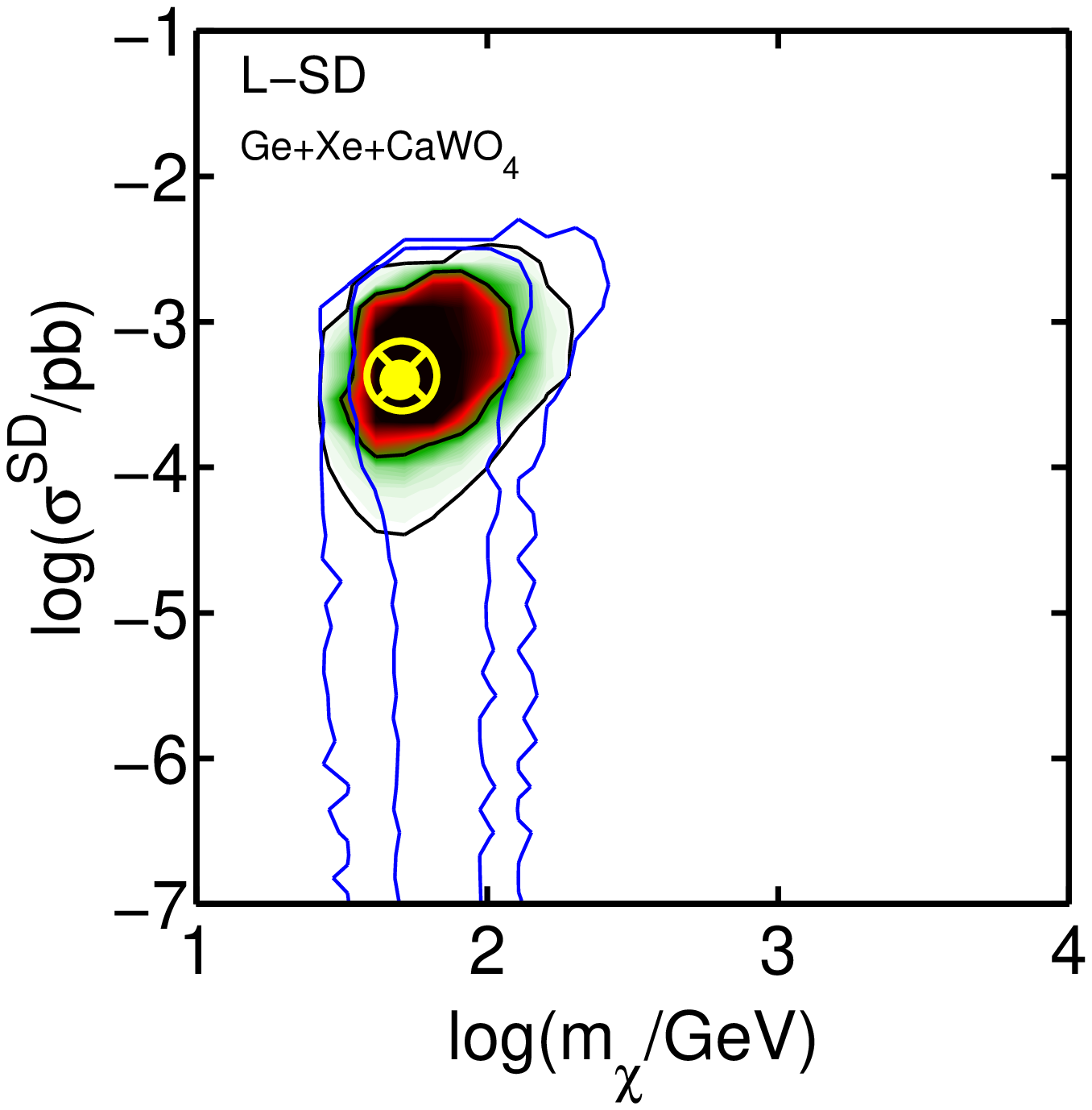,width=5.9cm}\hspace*{-0.6cm}
	\epsfig{file=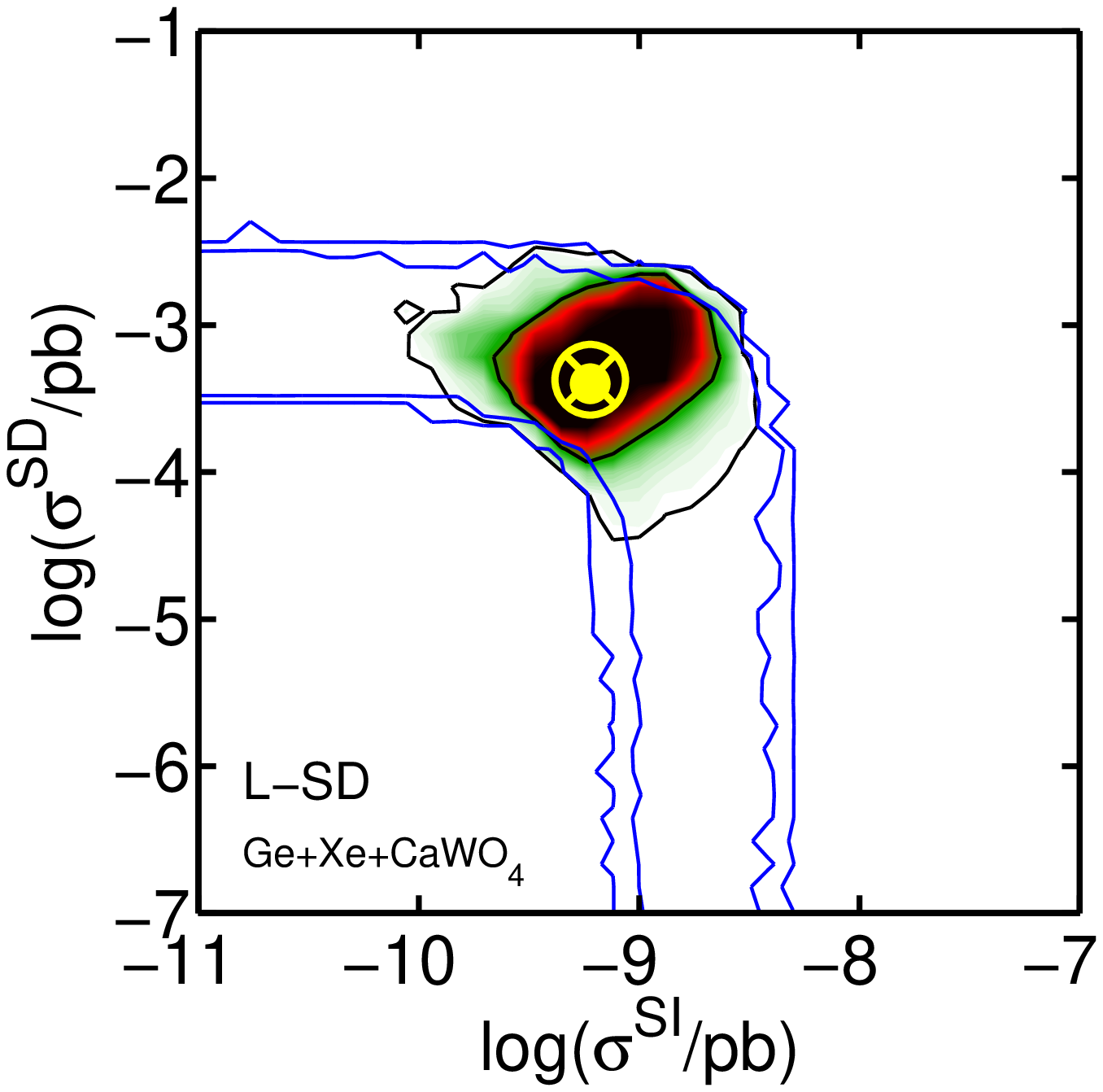,width=5.9cm}\\[-0.5cm]
	\hspace*{-1cm}
	\epsfig{file=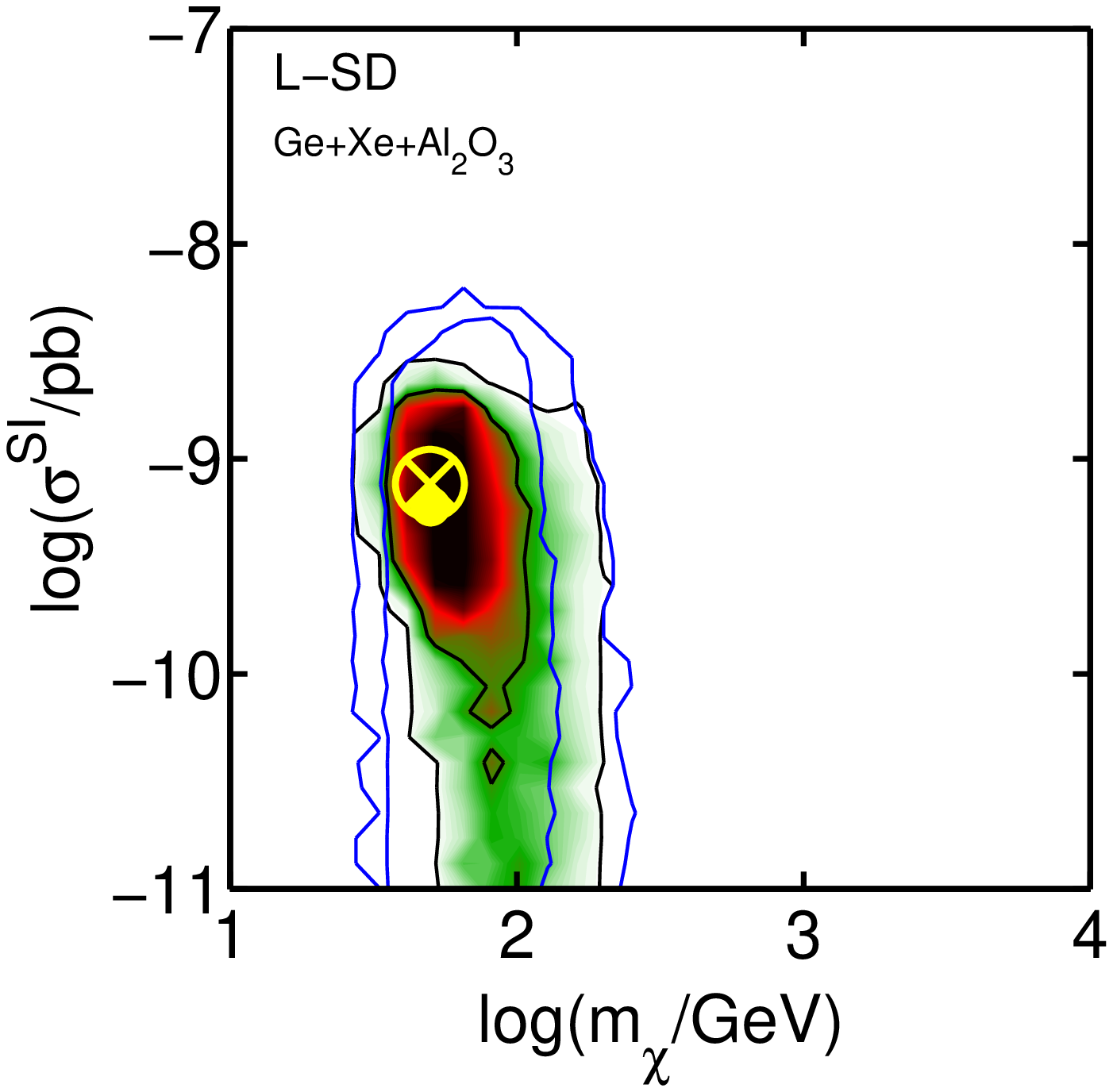,width=5.9cm}\hspace*{-0.6cm}
	\epsfig{file=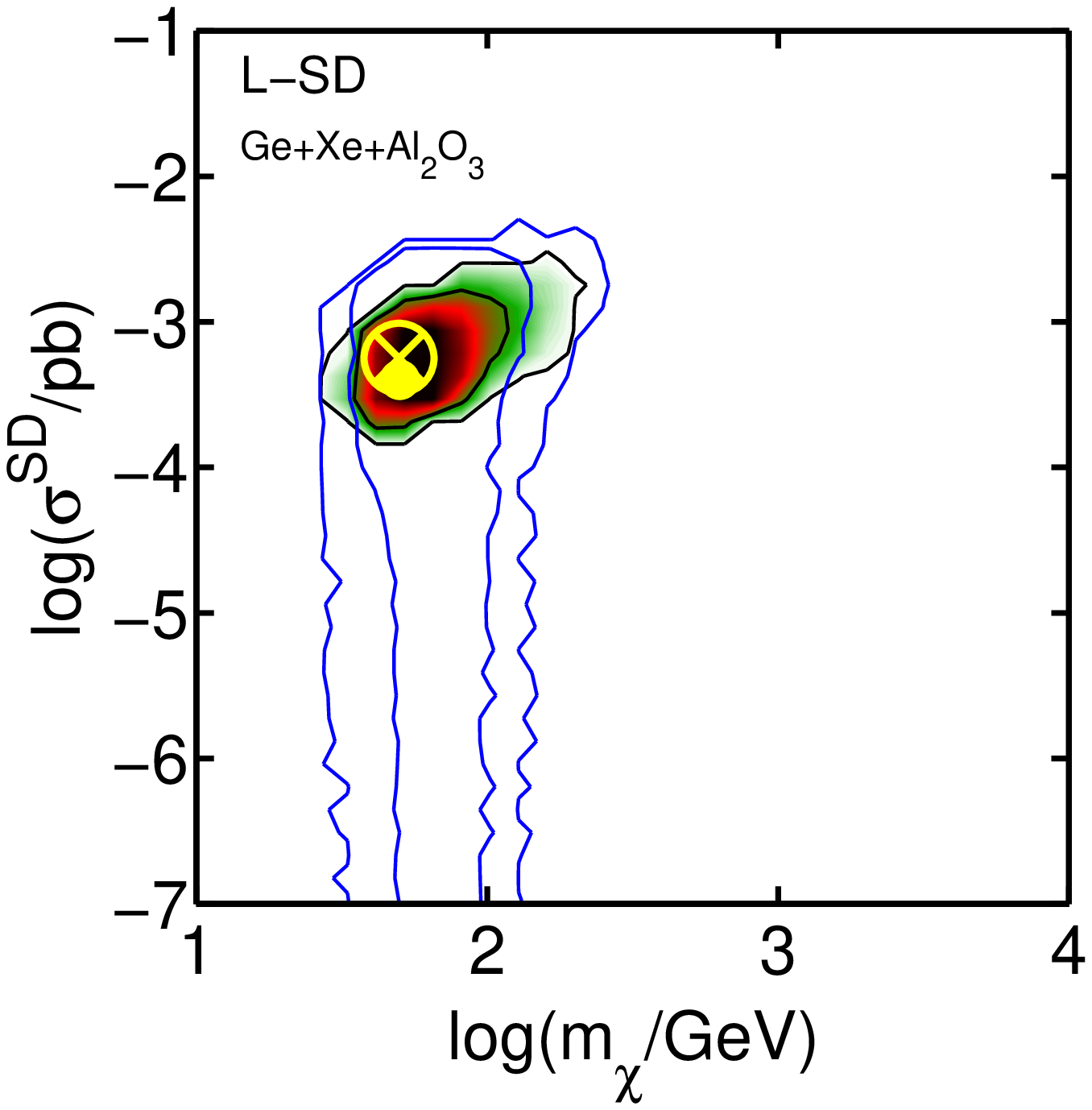,width=5.9cm}\hspace*{-0.6cm}
	\epsfig{file=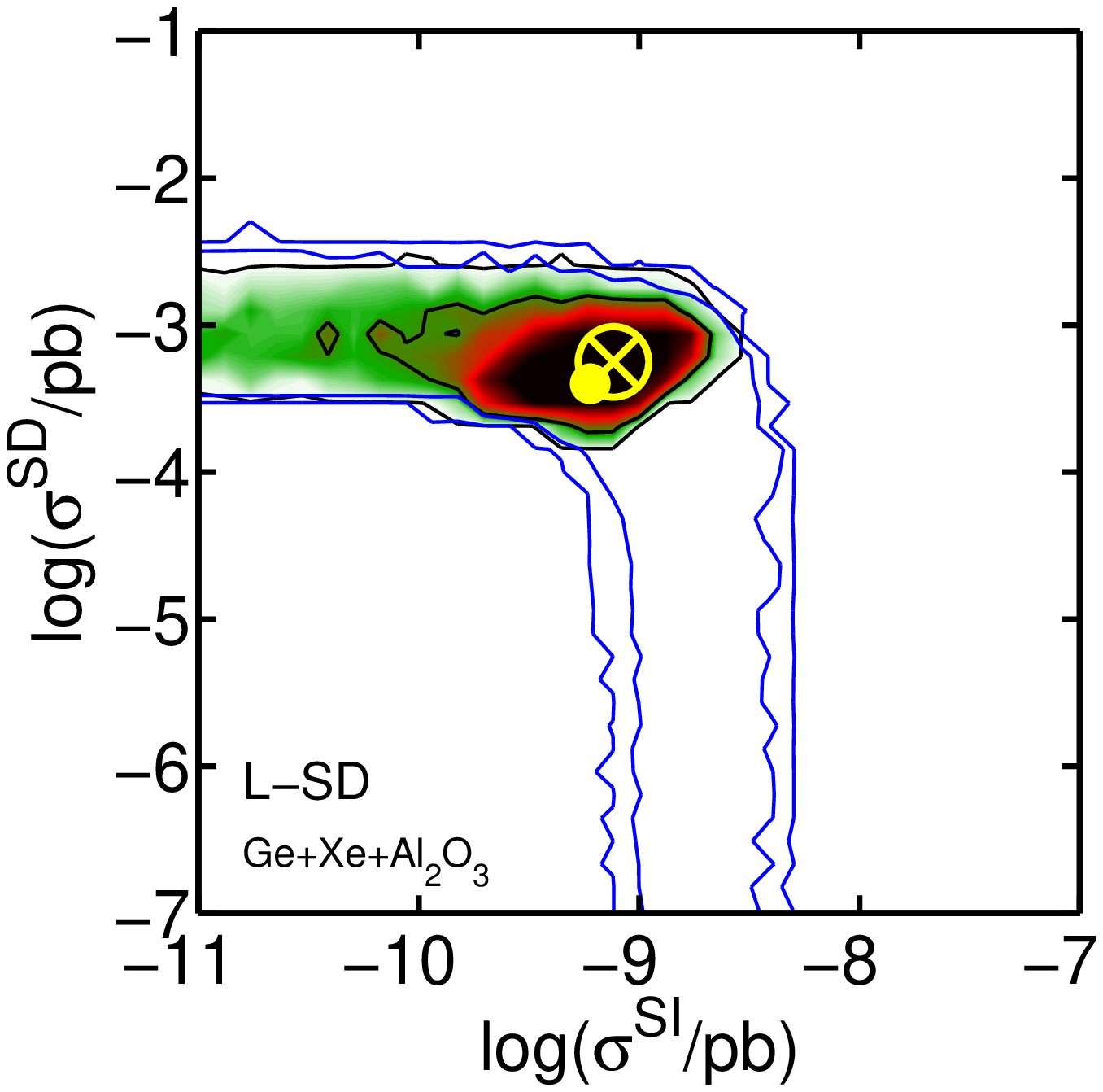,width=5.9cm}\\[-0.5cm]
	\hspace*{-1cm}
	\epsfig{file=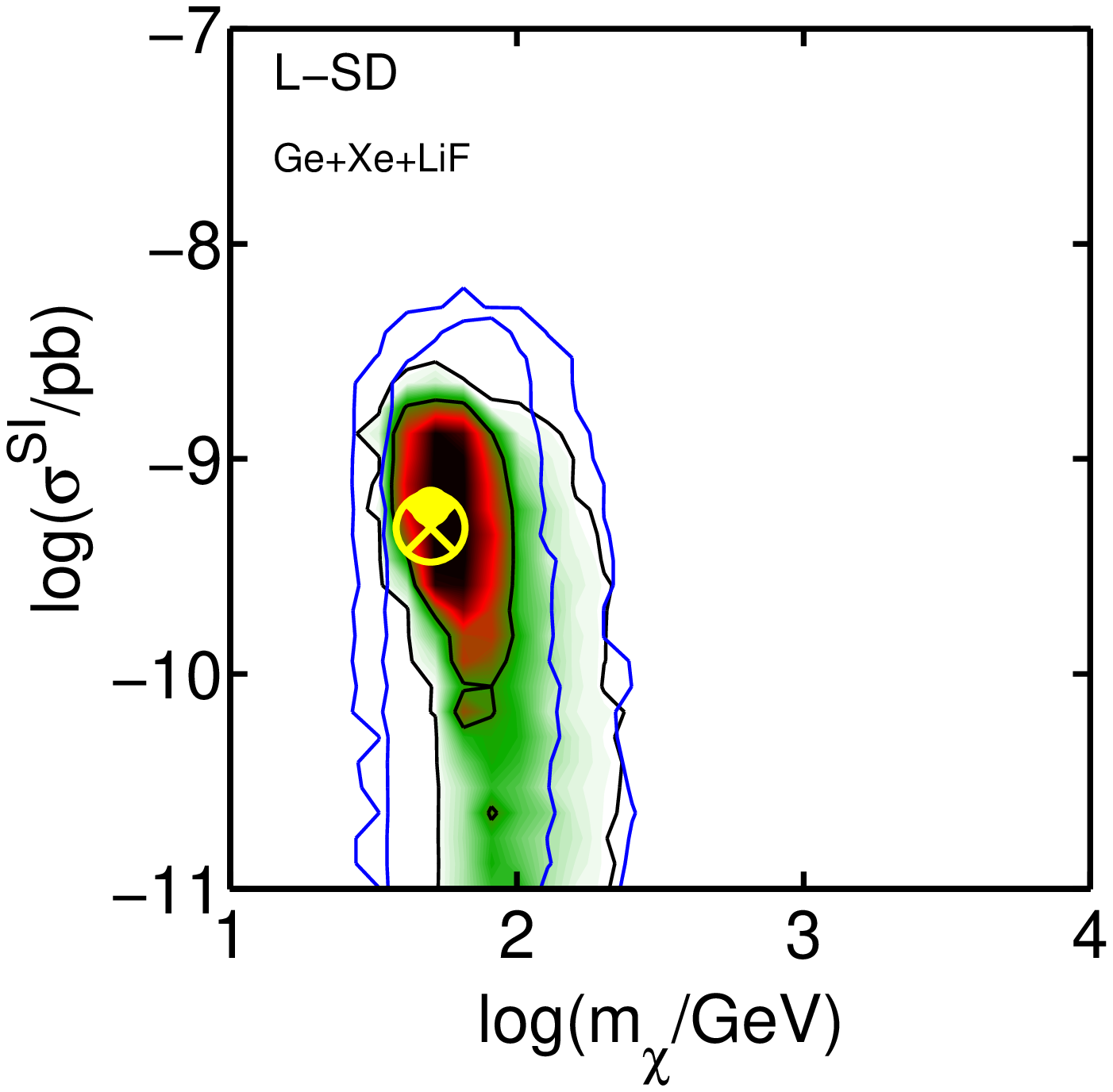,width=5.9cm}\hspace*{-0.6cm}
	\epsfig{file=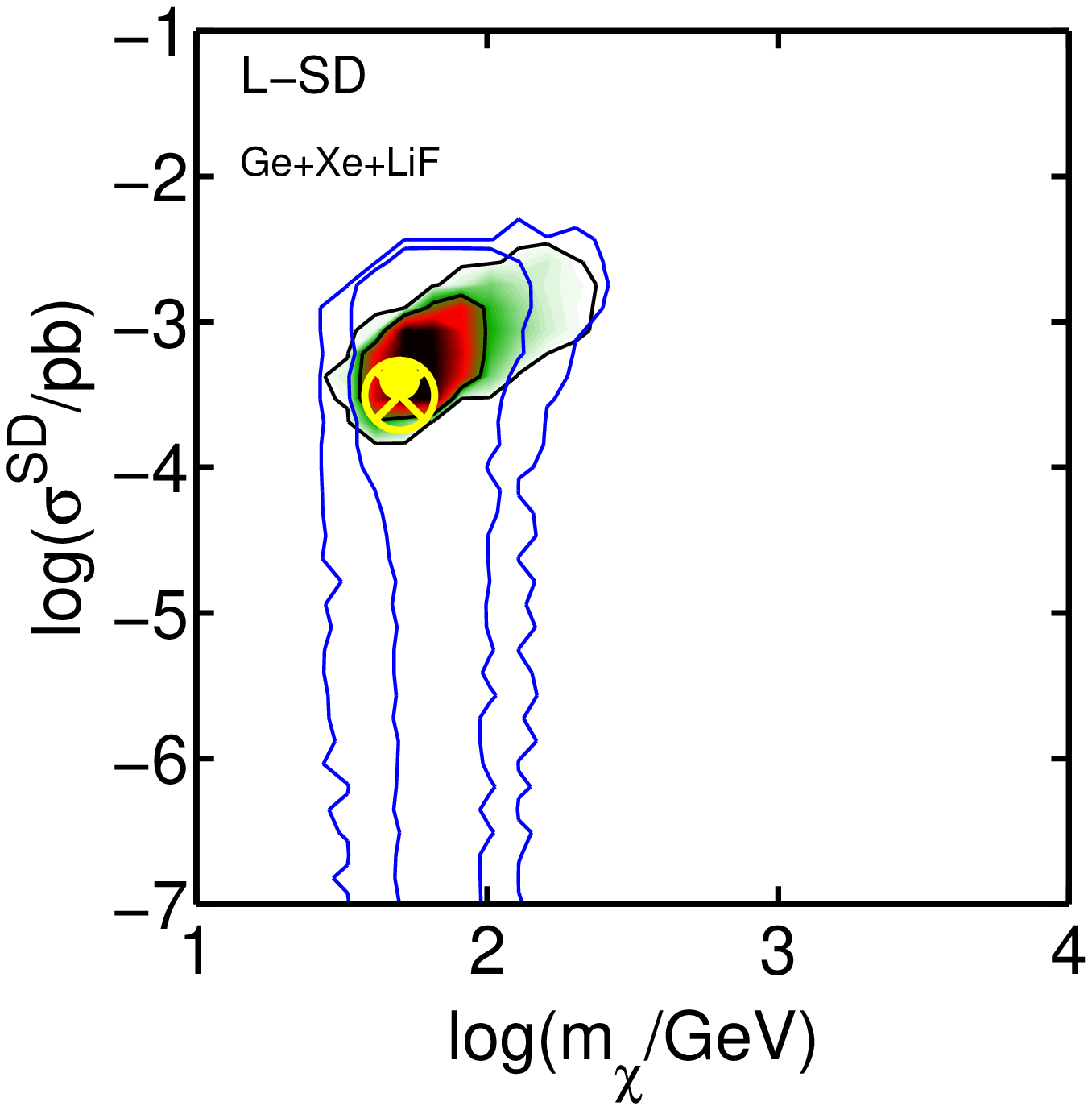,width=5.9cm}\hspace*{-0.6cm}
	\epsfig{file=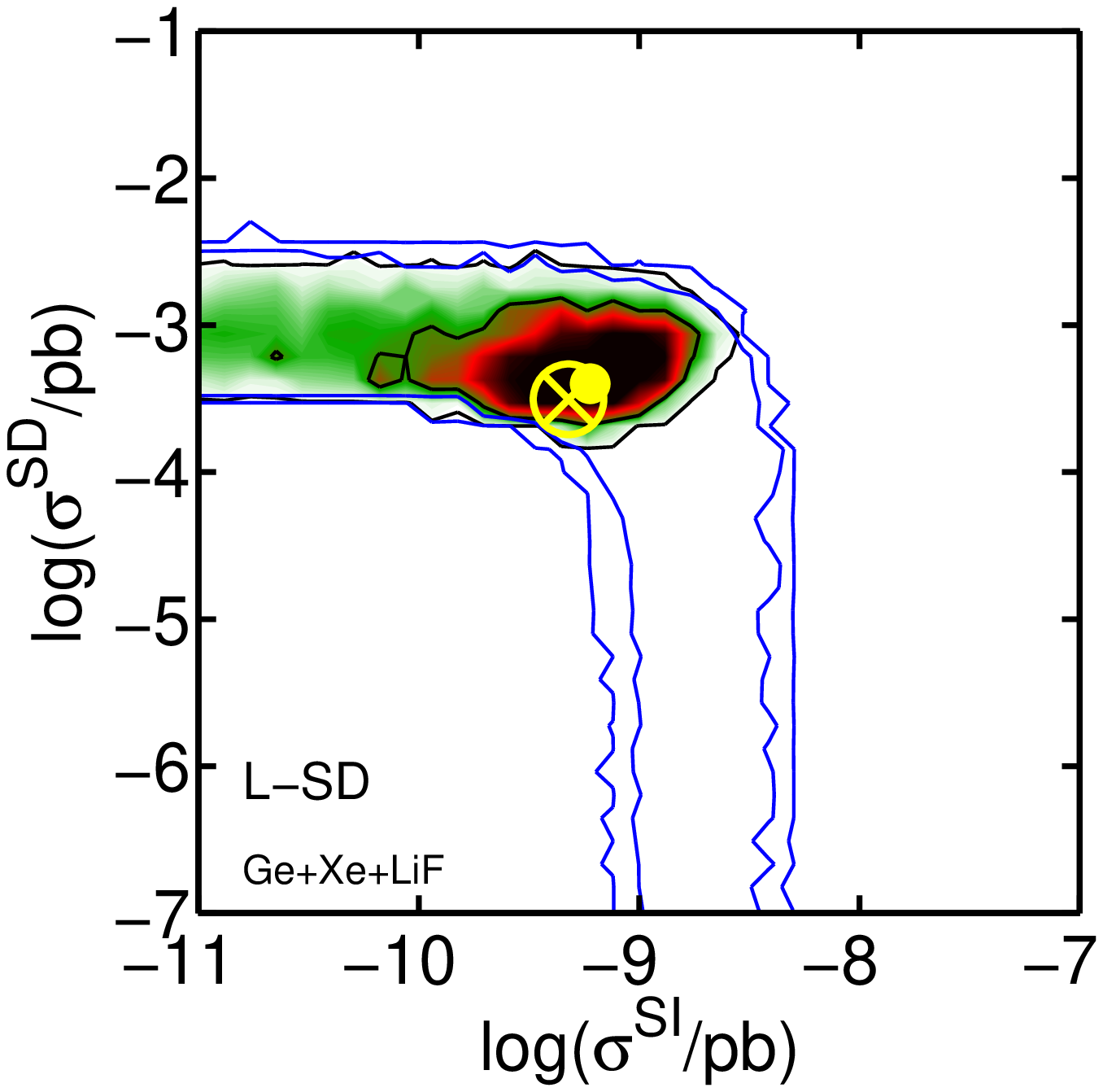,width=5.9cm}
\caption{\small The same as in Fig.\,\ref{fig:bmmsi} but for the case of \bmlsd.}
\label{fig:bmlsd}
\end{figure}

\begin{figure}
\hspace*{-1cm}
	\epsfig{file=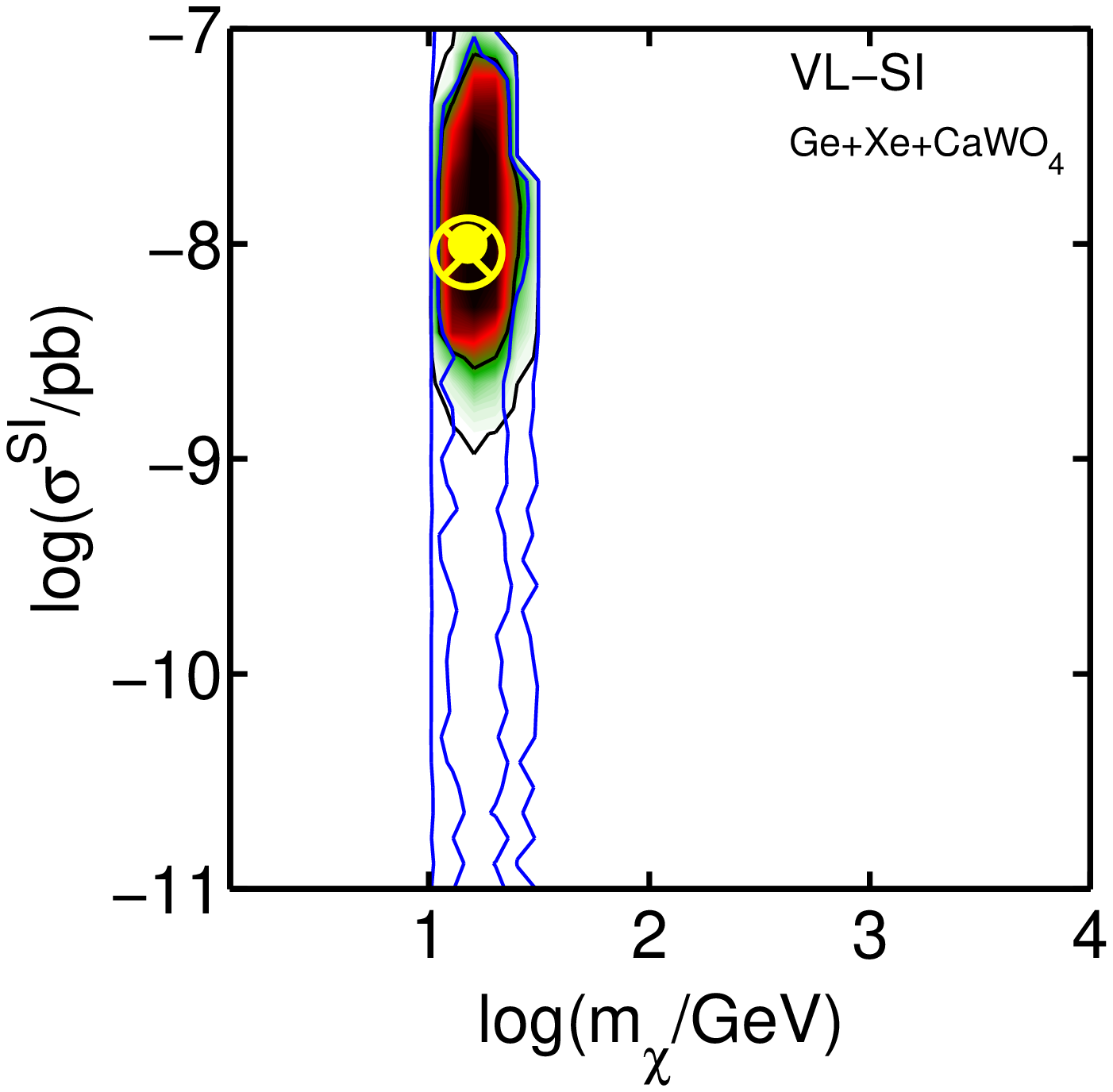,width=5.9cm}\hspace*{-0.6cm}
	\epsfig{file=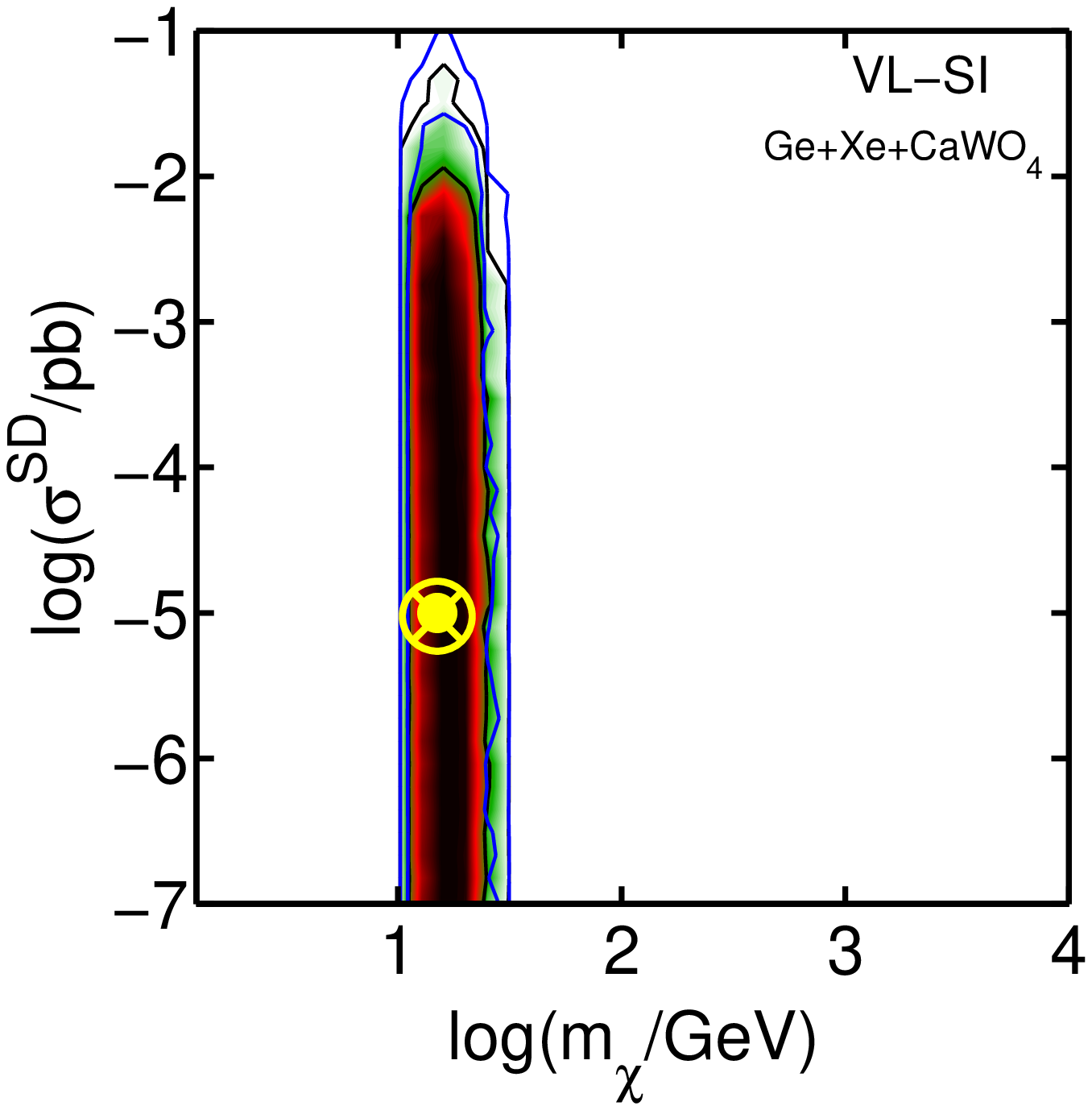,width=5.9cm}\hspace*{-0.6cm}
	\epsfig{file=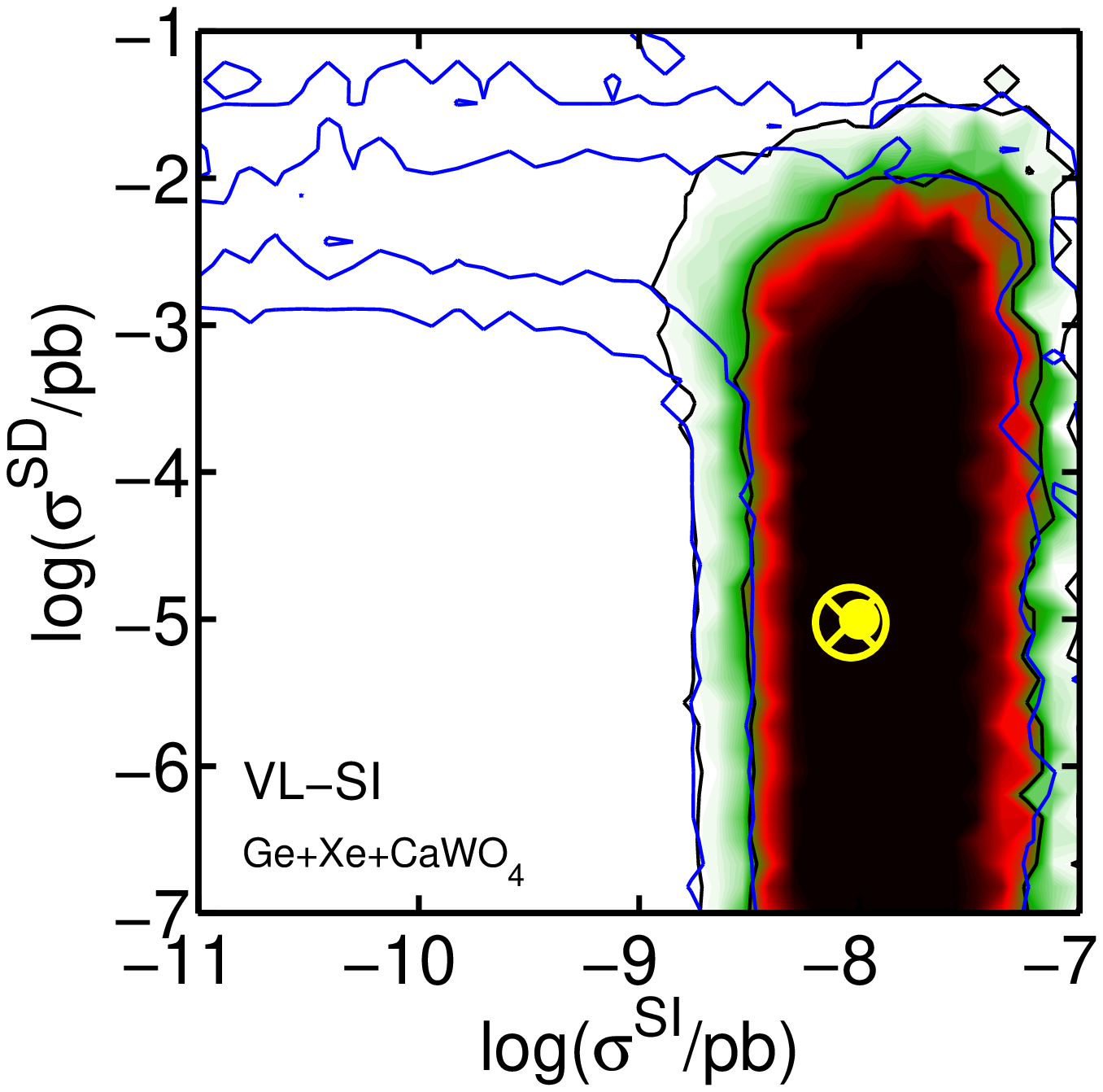,width=5.9cm}\\[-0.5cm]
	\hspace*{-1cm}
	\epsfig{file=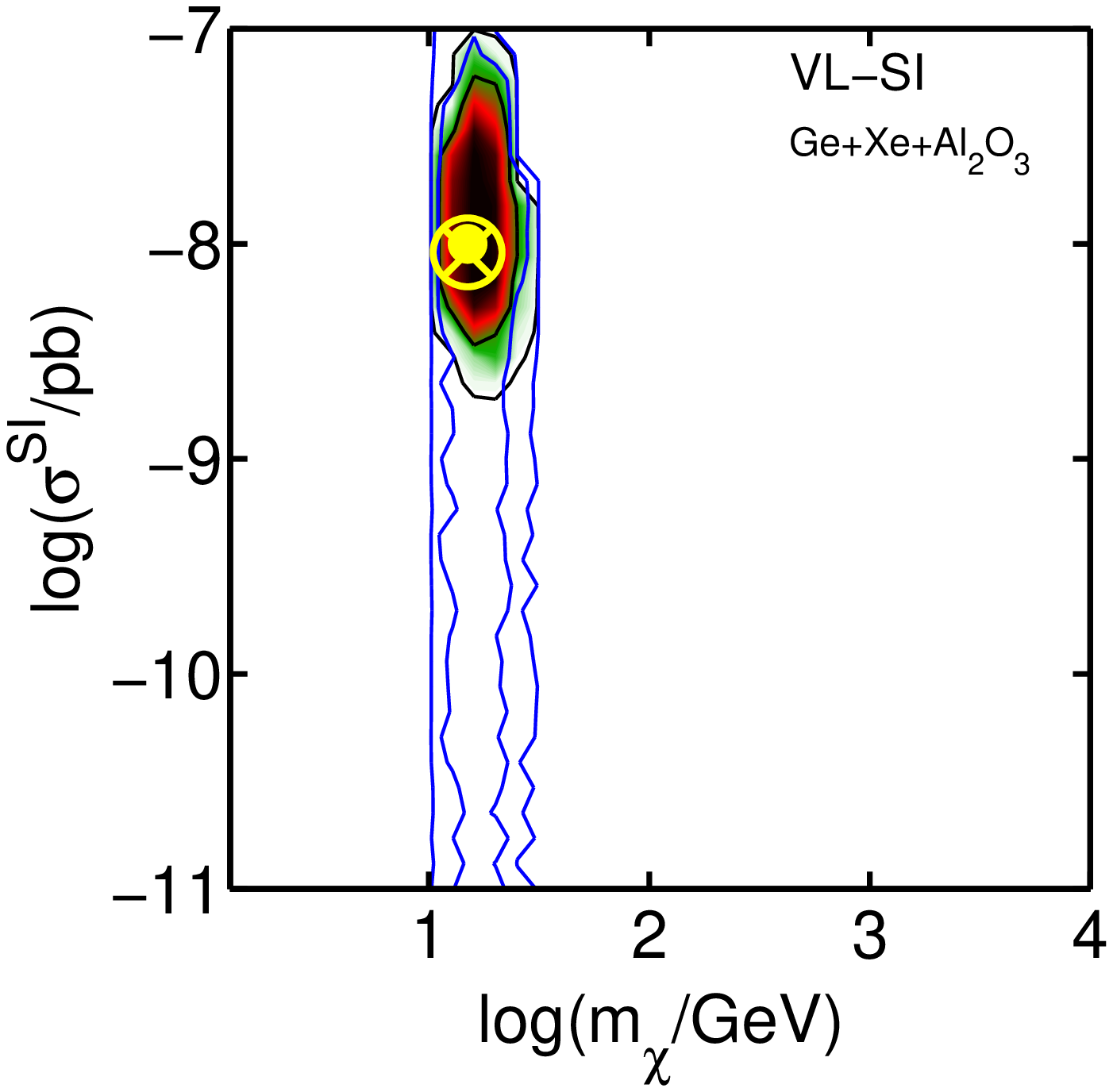,width=5.9cm}\hspace*{-0.6cm}
	\epsfig{file=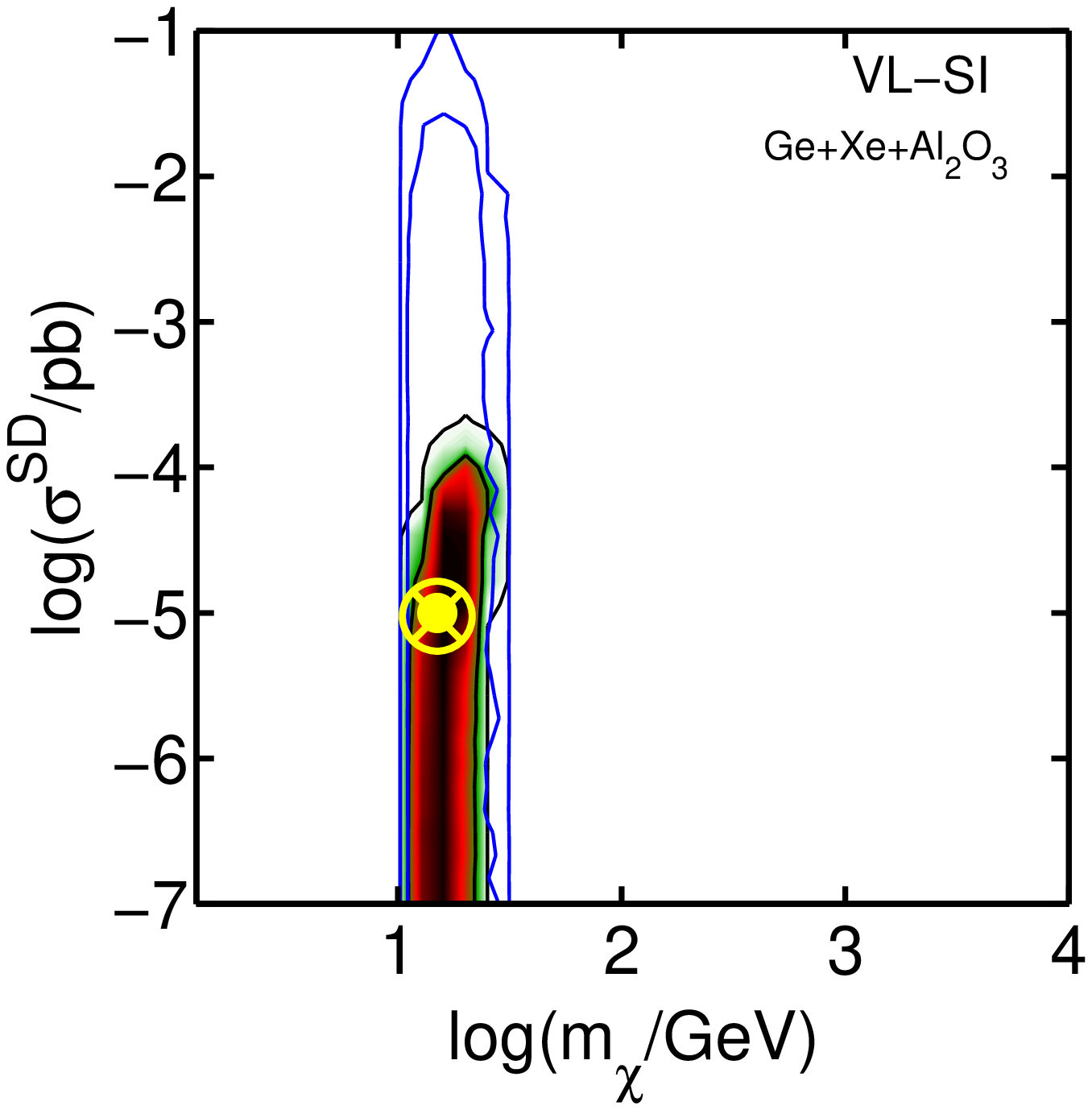,width=5.9cm}\hspace*{-0.6cm}
	\epsfig{file=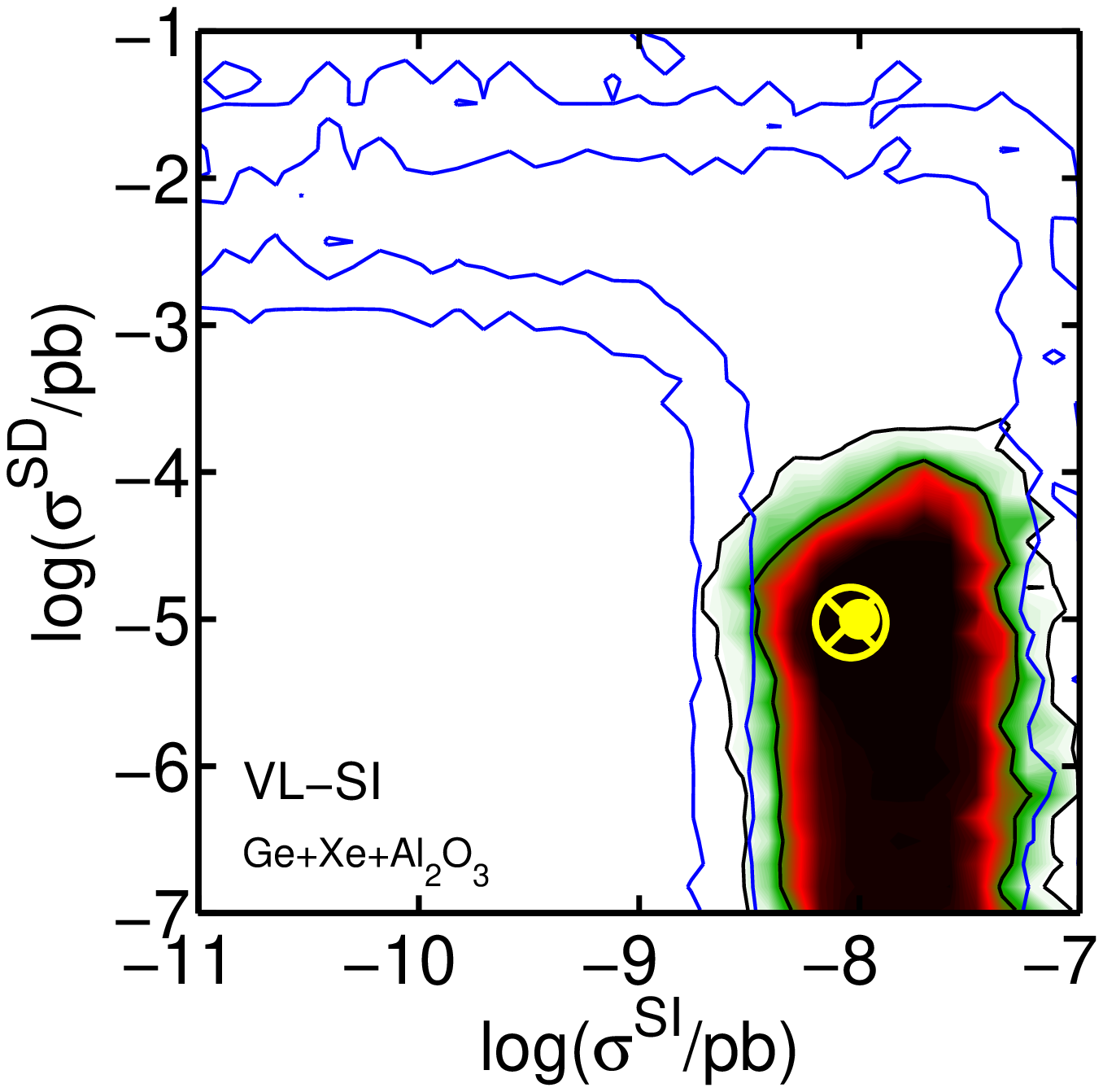,width=5.9cm}\\[-0.5cm]
	\hspace*{-1cm}
	\epsfig{file=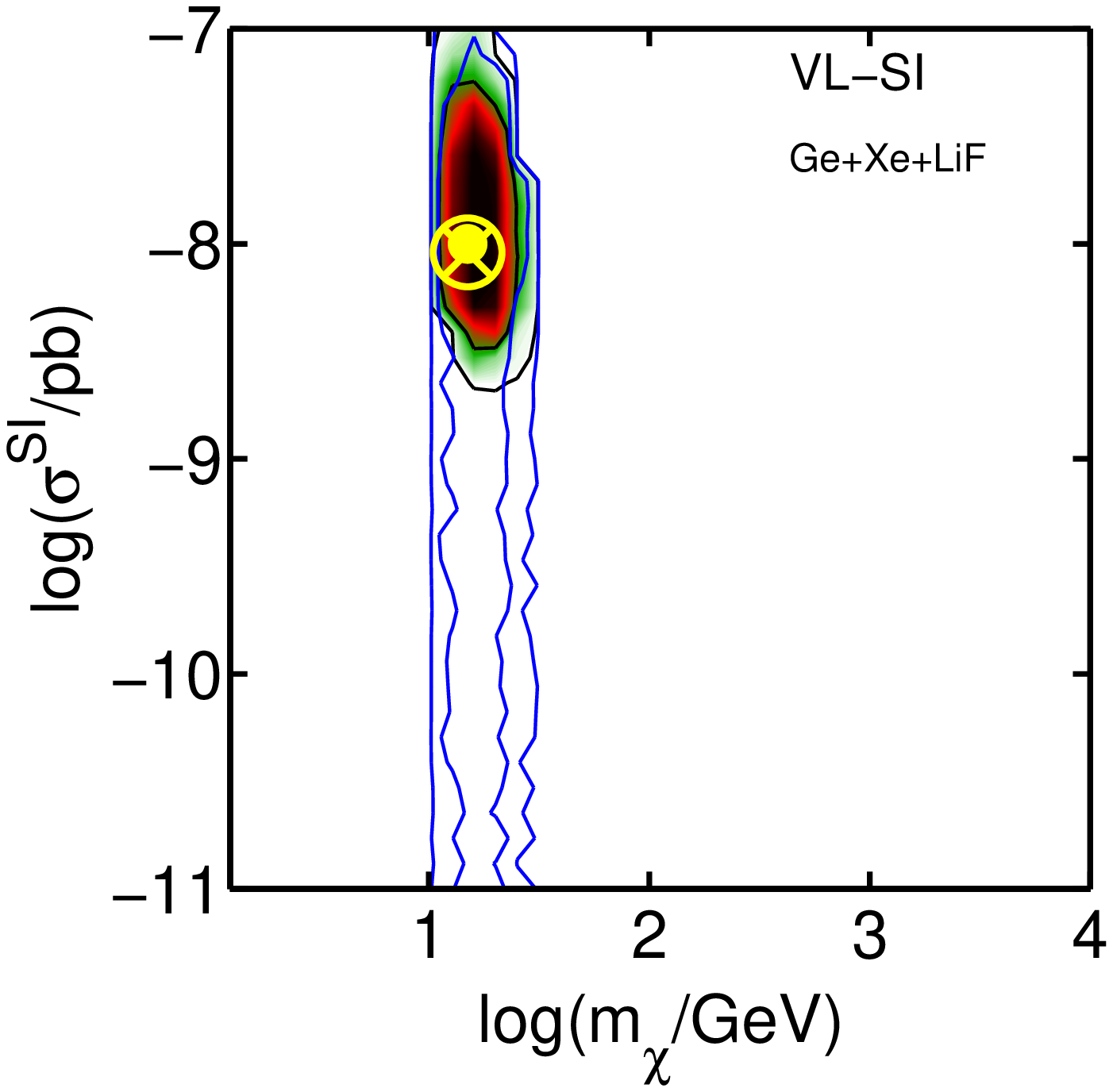,width=5.9cm}\hspace*{-0.6cm}
	\epsfig{file=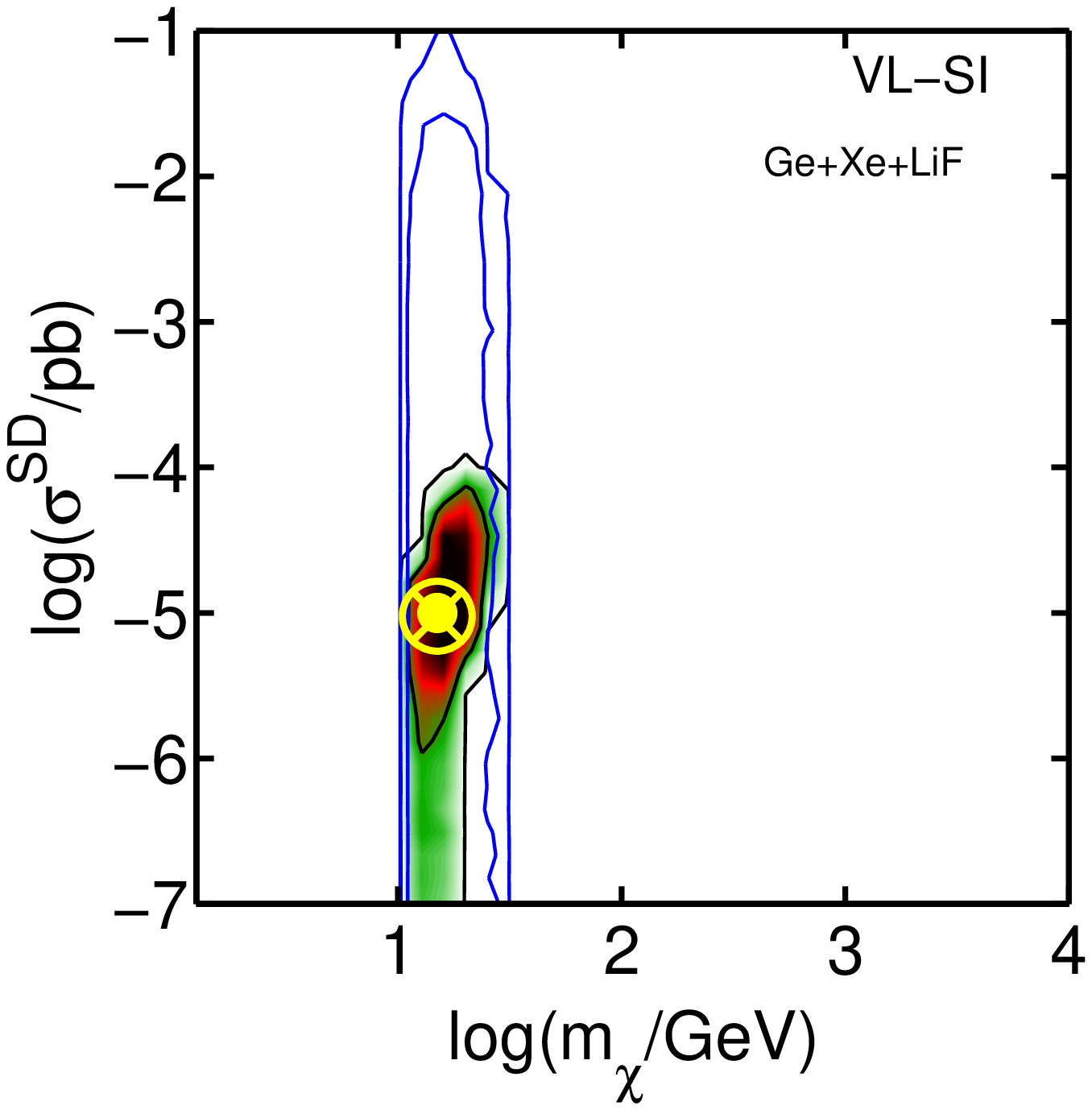,width=5.9cm}\hspace*{-0.6cm}
	\epsfig{file=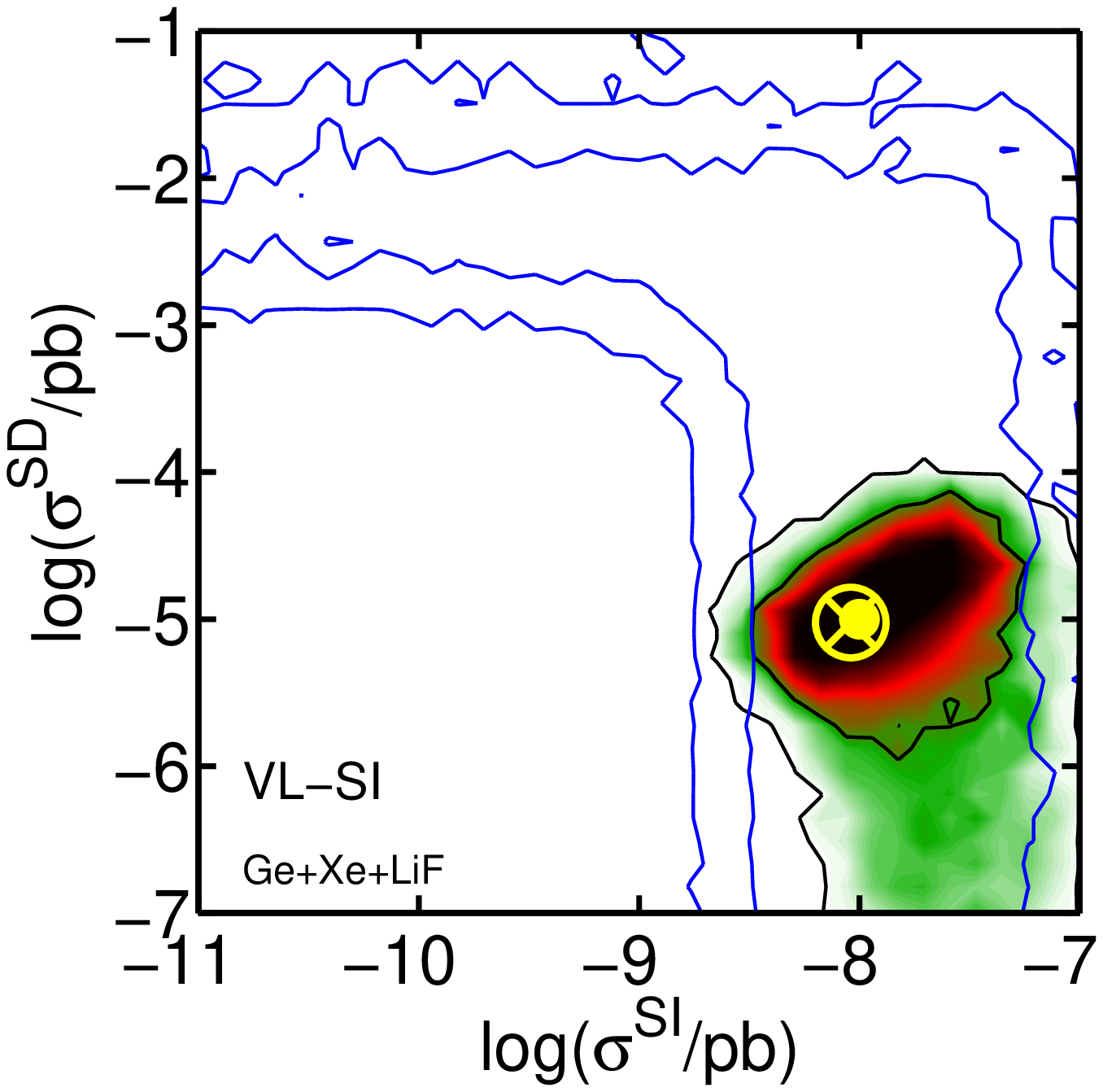,width=5.9cm}
\caption{\small The same as in Fig.\,\ref{fig:bmmsi} but for the case of \bmvlsi.}
\label{fig:bmvl}
\end{figure}

We start by considering the same exposure for all the experiments, $\epsilon=300$~kg~yr, and study the same benchmark scenarios as in the previous section. In Table\,\ref{tab:rosebud} we indicate the number of recoil events for each of the bolometric targets. Initially we assume zero background for the three targets and postpone the discussion about the influence of the background level to the end of this section. In Figs.\,\ref{fig:bmmsi}, \ref{fig:bmlsi}, \ref{fig:bmlsd} and \ref{fig:bmvl} we represent the PL for reconstructed DM parameters in benchmark points \bmmsi, \bmlsi, \bmlsd, and \bmvlsi, respectively. 
The black contours correspond to the combination of Ge and Xe data with a bolometric target: CaWO$_4$ for the upper row, Al$_2$O$_3$ in the middle and LiF for the lower row. For comparison, the blue contours illustrate the results when only germanium and xenon detectors are used.

We observe how in general Al$_2$O$_3$ and LiF provide good complementarity with germanium and xenon in points for which the detection rate in the latter is dominated by SI contributions. This is the case of benchmark points \bmmsi\ and \bmlsi. The area in the parameter space that is compatible with the simultaneous observation of a WIMP in the three experiments becomes much narrower than the case with only Ge and Xe. Although the WIMP mass cannot be completely reconstructed in some of these cases (e.g., in \bmmsi), a lower constraint is generally obtained for $\sigsd$ (which would suffice, e.g., to discriminate the observation from the case $\sigsd=0$ of scalar DM). In the examples for which the number of DM events is larger, such as in \bmlsi, the extra information on the recoil spectrum results in a better determination of the WIMP mass and consequently leads to closed contours in the three quantities. This is the situation that we describe as complementary. Although in these cases data from CaWO$_4$ does not allow a complete complementarity, an upper bound can be obtained in the value of $\sigsd$ which, in turn, leads to a lower bound on $\sigsi$ and a good determination of the latter (see, e.g., the case of \bmlsi\ in the top row of Fig.\,\ref{fig:bmlsi}).

On the contrary, CaWO$_4$ performs better for benchmark points where SD contributions dominate the rate for germanium and xenon. Since tungsten is a heavier nucleus than both germanium and xenon, it is more sensitive to $\sigsi$. This is, e.g., what happens in \bmlsd, where the reconstructed contours are closed and the reduction in the best fit areas is very significant. 
In this kind of points, Al$_2$O$_3$ and LiF do not perform that well since they are mostly sensitive to the SD component. In any case, they can be used to obtain a lower bound on $\sigsd$ which clearly rules out the possibility $\sigsd=0$.

It should also be noticed that in the case of the benchmark point \bmmsi\ represented in Fig.\,\ref{fig:bmmsi} the reconstruction of the DM mass has no upper bound. This is a generic feature for heavy DM particles, due to the fact that the spectrum becomes flatter and the fit to the mass is more inaccurate. The combination with bolometric targets does not improve this situation significantly. For this reason, we expect a worse reconstruction for DM particles heavier than in \bmmsi.

Very light WIMPs, on the other hand, constitute an interesting possibility that can also be explored with the aid of bolometric targets. In particular, as we said above, both Al$_2$O$_3$ and LiF can be sensitive to low-mass DM particles. 
In Fig.\,\ref{fig:bmvl} we show the case of a WIMP with $\mwimp=15$~GeV and scattering cross section as in \bmvlsi. 
Data from germanium and xenon are enough to determine the WIMP mass rather accurately, but a large uncertainty remains in both $\sigsi$ and $\sigsd$ as can be observed in the blue contours. 
Both Al$_2$O$_3$ and LiF remove significantly this degeneracy and in the case of LiF we even obtain an inner contour around the correct value of $\sigsd$. 
In the case of CaWO$_4$ only the SI component can be determined but no further information on $\sigsd$ is obtained.

\begin{figure}[!t]
	\epsfig{file=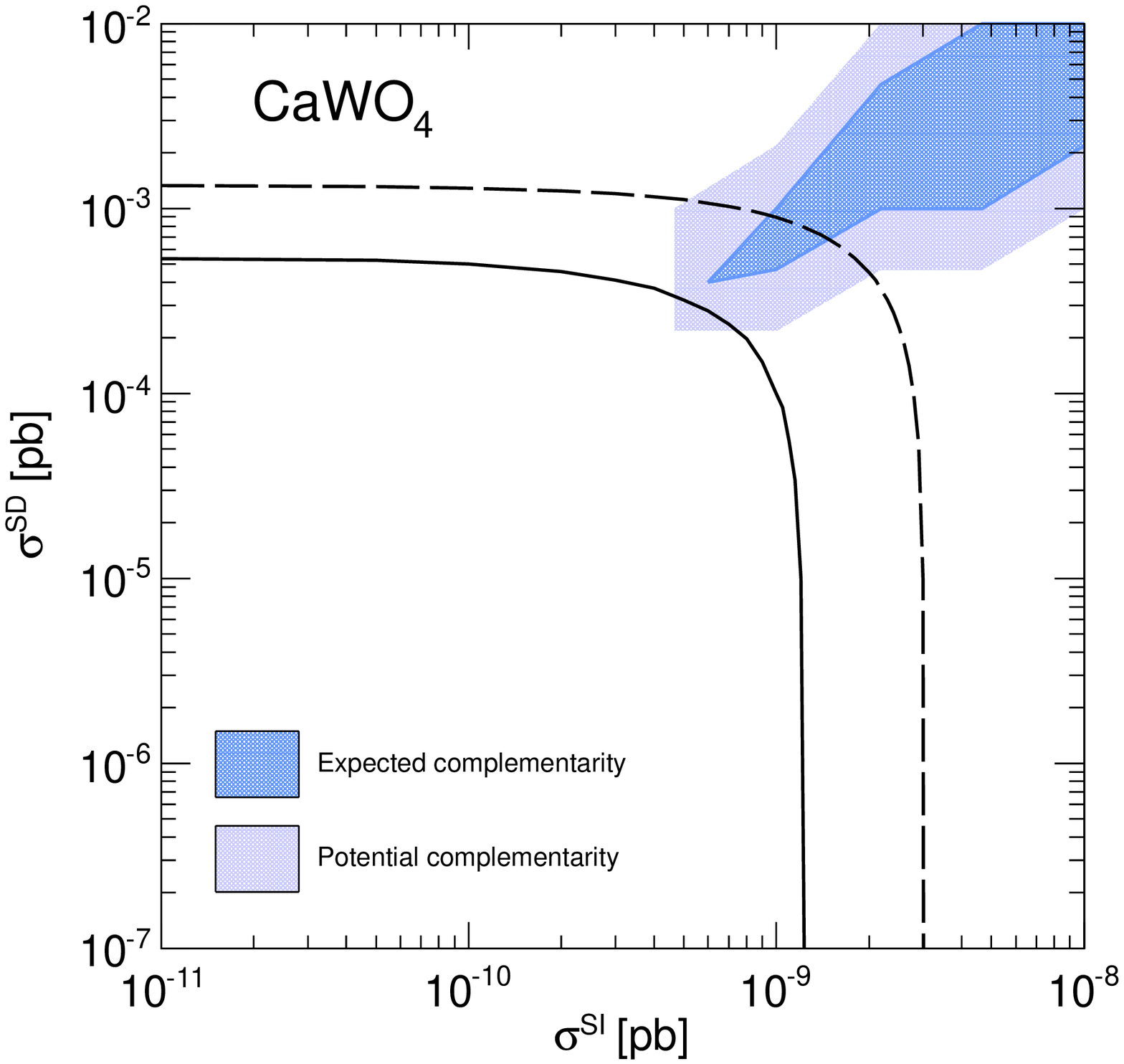,width=5.4cm}\hspace*{-0.3cm}
	\epsfig{file=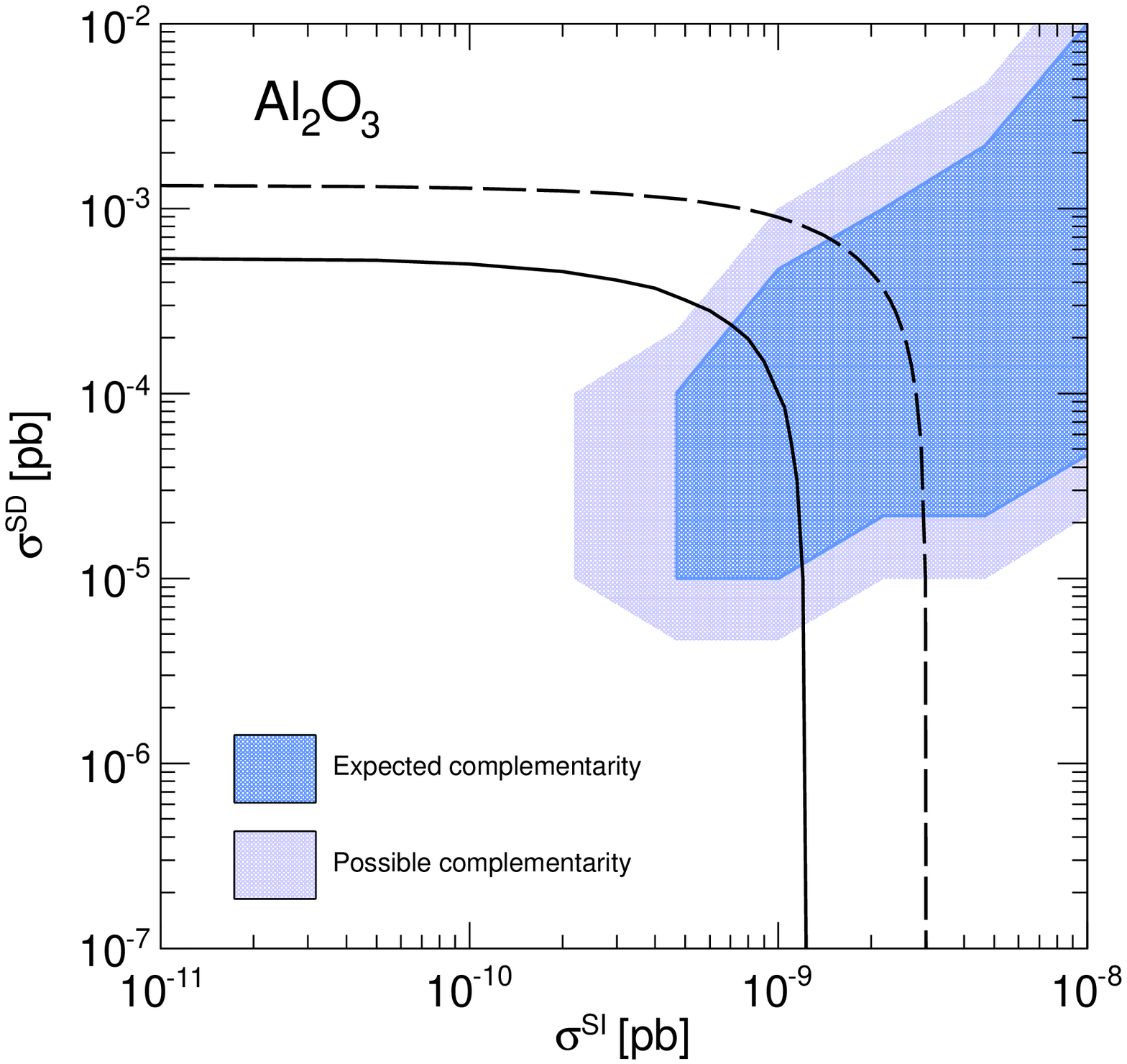,width=5.4cm}\hspace*{-0.3cm}
	\epsfig{file=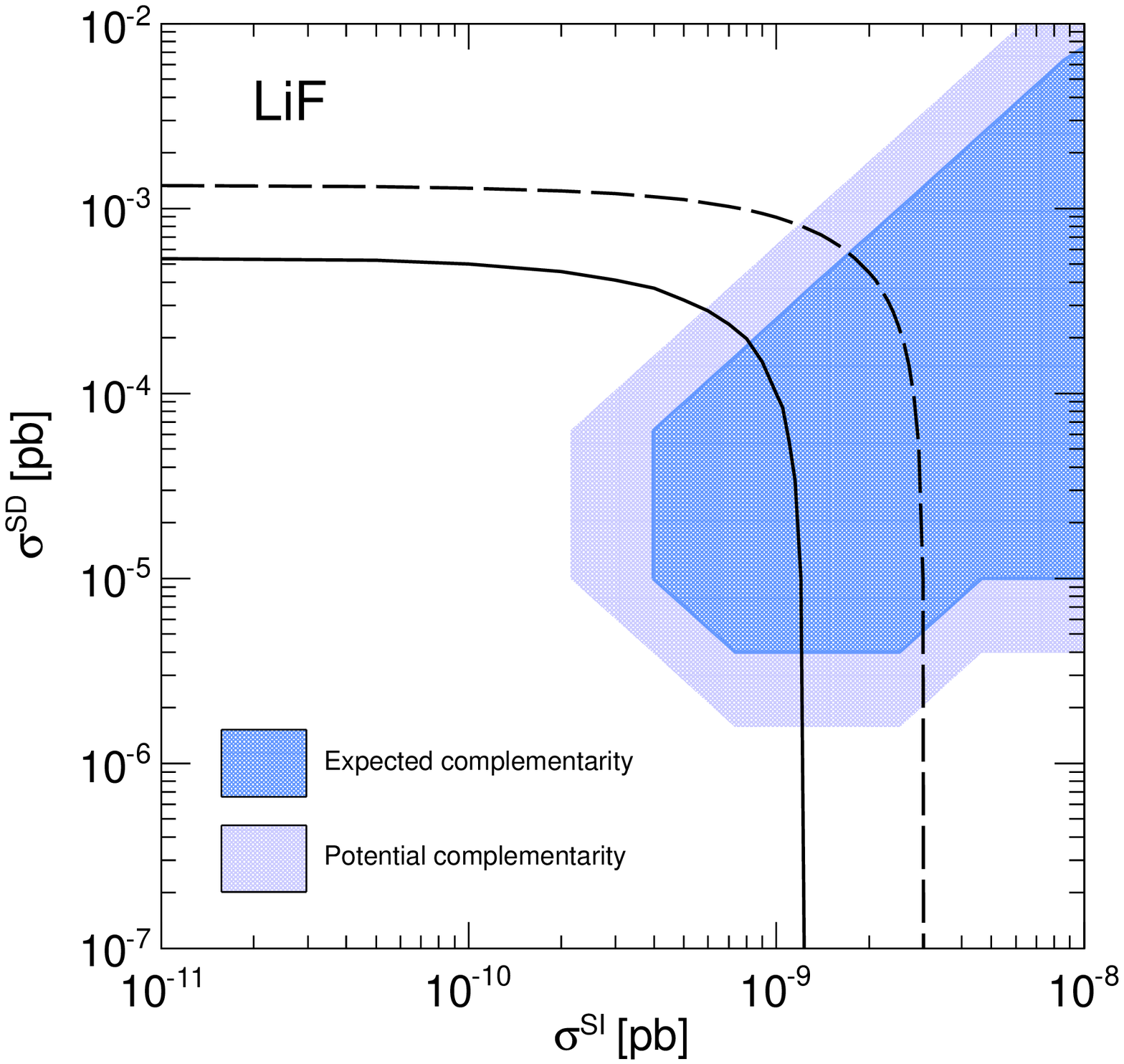,width=5.4cm}
\caption{\small The dark blue region represents the area of the $(\sigsi,\ \sigsd)$ plane for which complementarity is obtained for CaWO$_4$, Al$_2$O$_3$ and LiF in the case of a WIMP mass of $\mwimp=50$~GeV. 
The light blue region is a measure of the uncertainty of our grid scan (see text).
The black line corresponds to the upper constraint obtained from XENON100 data (using the BonnA result for the SD structure function \cite{HjorthJensen:1995ap}). 
Points above the dashed line predict more than 4.9 events in XENON100 and are therefore excluded by the recent experimental result \cite{Aprile:2012nq} following a Feldmans-Cousin method \cite{Garny:2012it}. For reference, points along the solid lines predict 2 events in XENON100.}
\label{fig:compl}
\end{figure}

So far we have observed that some targets perform better than others (in terms of complementarity) in certain regions of the parameter space. In particular, for each of the targets we can determine the regions in the parameter space for which complementarity is obtained. In Fig.\,\ref{fig:compl} we indicate the area (dark blue) of the $(\sigsi,\ \sigsd)$ plane for which we obtain closed contours in the reconstructed PL of the three DM parameters $(\mwimp,\ \sigsi,\ \sigsd)$  for each of the bolometric targets and a WIMP mass of $\mwimp=50$~GeV. 
In order to obtain this region we have performed a grid scan in the SD-SI plane for which the separation among points is limited by computing time. 
The light blue region separates points leading to complementarity from the nearest which do not, hence indicating the resolution of our grid scan.
The areas are different for the three bolometers studied. Consistently with the individual examples that were analysed previously, we observe that the areas for Al$_2$O$_3$ and LiF are larger as compared with the area for CaWO$_4$, and shifted towards smaller values of $\sigsd$. For CaWO$_4$ total complementarity only occurs for a small region, very close to the current upper constraint by XENON100. However, as we showed in various examples, it helps in the determination of the WIMP mass and $\sigsi$. 
Needless to say, a reduction in the exposure would result in a shift of the complementarity areas towards larger values of both $\sigsi$ and $\sigsd$. 

\begin{figure}
	\epsfig{file=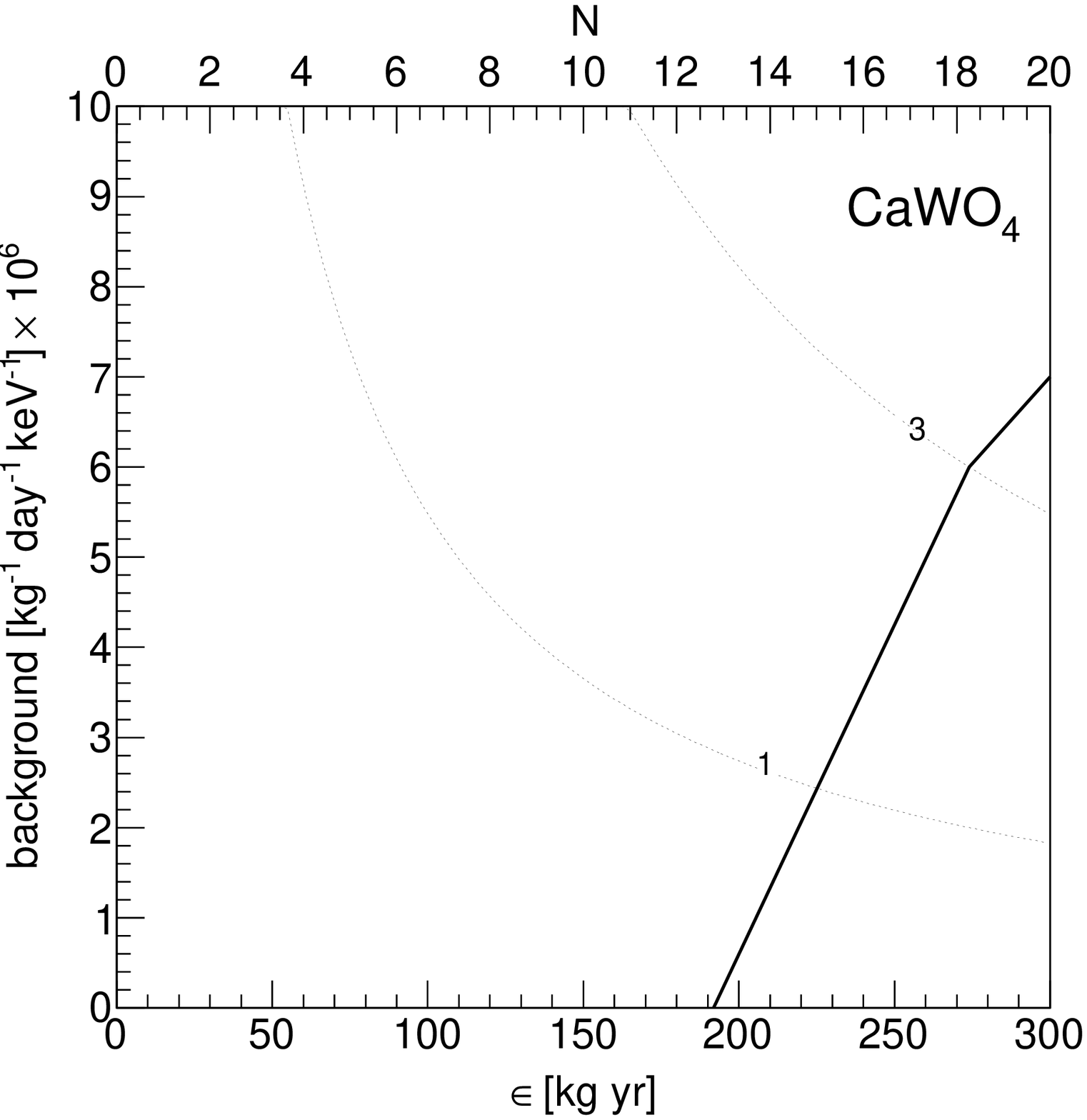,width=5.4cm}\hspace*{-0.3cm}
	\epsfig{file=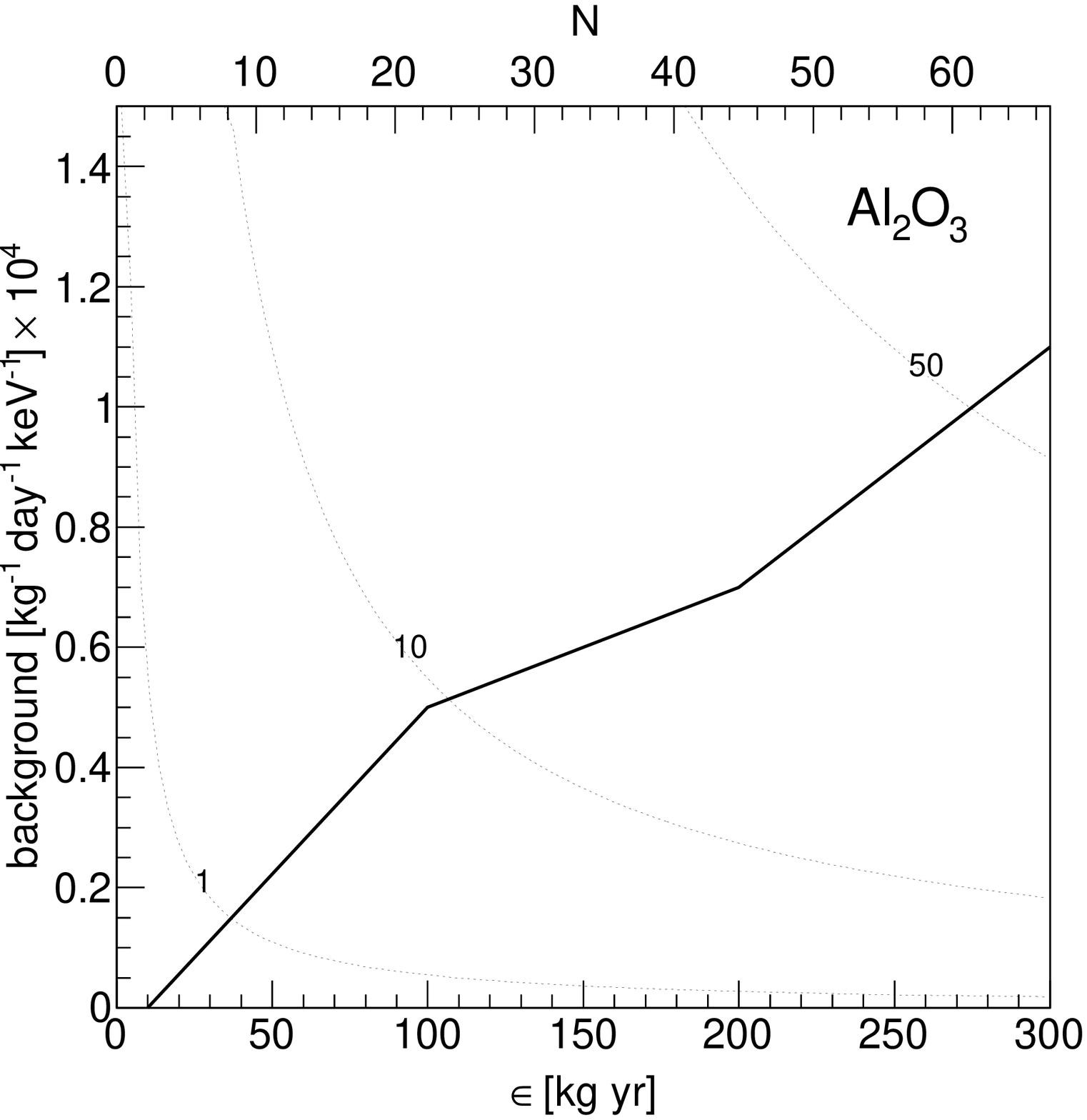,width=5.4cm}\hspace*{-0.3cm}
	\epsfig{file=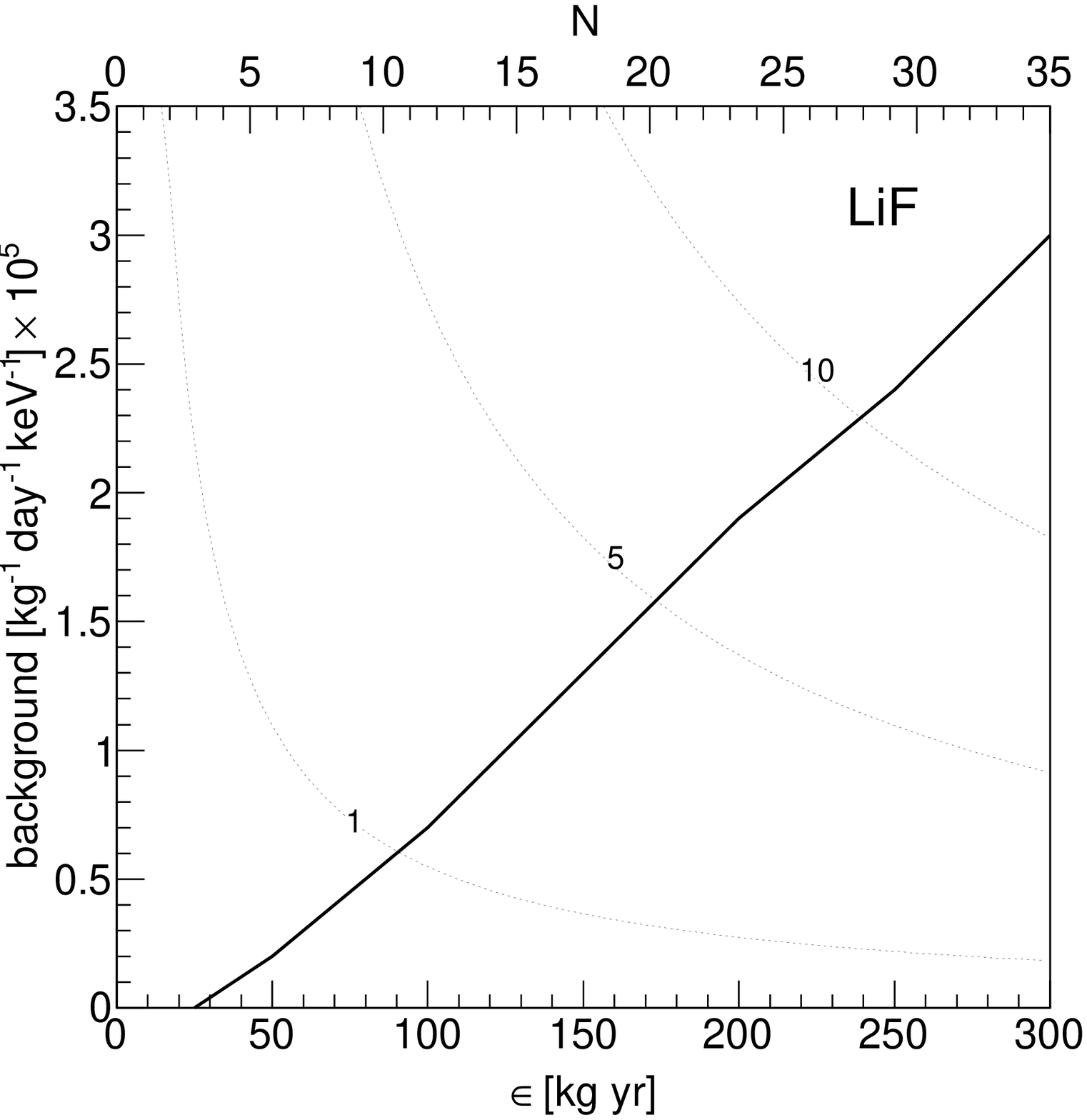,width=5.4cm}
\caption{\small The solid line represents the maximum level of background for which complementarity is attained as a function of the exposure for each of the bolometric targets for representative benchmark points ($(\sigsi,\sigsd)=(6\times10^{-10}, 4\times10^{-4})$ pb for CaWO$_4$, $(10^{-9}, 5\times10^{-5})$ pb for Al$_2$O$_3$, and $(10^{-9},10^{-5})$ pb for LiF, from left to right) with $\mwimp=50$~GeV. The expected total number of WIMP recoil events is indicated in the top horizontal axis. Dotted lines denote the number of background events per bin.
}
\label{fig:background}
\end{figure}

For completeness, we have determined the highest level of background for which complementarity is attained as a function of the exposure for each bolometer in a given benchmark point, and represented the results in Fig.\,\ref{fig:background}. We see how, for zero background, the exposure for the three bolometers can be reduced. For Al$_2$O$_3$ and LiF this reduction can be very significant. For example, in the benchmarks that we have chosen in Fig.\,\ref{fig:background}, it suffices to have a clear signal for DM in Al$_2$O$_3$ or LiF, even with a very reduced number of events (to determine the lower bound we considered that 1 event in a background free experiment is statistically significant). 
From these results we can also conclude that the complementary areas of Fig.\,\ref{fig:compl} would not shrink significantly if a moderate background is included. We have to bear in mind that this computation was carried out assuming a flat background, for which a DM signal can be easily distinguished. 
This assumption can be considered equivalent to the estimate of the sensitivity of an experiment that does not surpass this background level in any bin of the energy window, independently of the background dependence on energy.

\section{Conclusions}
\label{sec:conclusions}

In this work we have investigated the determination of DM parameters from a combined use of different targets in direct detection experiments. More specifically, we study the determination of the WIMP mass and the SI and SD components of its scattering cross section off nucleons. We apply the method to a set of bolometric targets developed and characterised by the ROSEBUD collaboration (Al$_2$O$_3$ and LiF) or in use in CRESST (CaWO$_4$). We investigate the conditions under which the DM parameters can be obtained unambiguously when combining data from them and germanium and xenon experiments. In doing so, we take into account astrophysical uncertainties in the DM halo properties and nuclear uncertainties in the SD structure functions.

We first show how one ton scale germanium and xenon targets, which might excel in providing the first measurement of a WIMP, might not be able to measure all the DM properties. This is due to a degeneracy in the SI-SD plane that might be unresolved since both targets are mostly sensitive to the SI component of the WIMP cross section. Interestingly, SD-sensitive targets might provide extra information with which this degeneracy can be broken. 
We have studied the combination of germanium and xenon based targets with the one used by the COUPP collaboration (CF$_3$I). The presence of fluorine improves the reconstruction of DM parameters, however, COUPP does not provide information about the recoil spectrum, which limits its complementarity capability. 

The situation is much more interesting for experiments which are sensitive to the SD WIMP-nucleon interaction and which can provide an accurate measurement of the energy spectrum, such as the scintillating bolometric targets developed and studied by the ROSEBUD and CRESST collaborations.
We observe how the combined use of these detectors with germanium and xenon experiments can break the degeneracies in the determination of DM parameters and provide a good reconstruction of the WIMP mass and SI and SD scattering cross section.
In particular, the inclusion of CaWO$_4$ can lead to a better determination of the WIMP mass and SI cross section. Nevertheless, since its interaction with DM is dominated by the SI contribution, it only leads to a complementary results in a small window of the parameter space in which the rate in germanium and xenon is mainly due to SD interactions. This area is in fact very close to the region already excluded by XENON100.
On the other hand, Al$_2$O$_3$ and LiF (being more sensitive to the SD contribution) can be complementary targets to germanium and xenon in regions of the parameter space where the rate in the latter is dominated by SI contributions. This can happen for values of the cross section as small as $\sigsi\gsim2\times 10^{-10}$~pb and $\sigsd\gsim 10^{-5}$~pb for a WIMP with a mass $\mwimp=50$~GeV. 

In some regions of the DM parameter space the exposure can be reduced to approximately 50 kg yr for Al$_2$O$_3$ and LiF without loosing complementarity under the assumption of zero background. Finally, we investigate the effect of the background and observe that complementarity can be achieved for large exposures with a background level as large as $10^{-4}$~kg$^{-1}$ day$^{-1}$ keV$^{-1}$ for Al$_2$O$_3$ or $10^{-5}$~kg$^{-1}$ day$^{-1}$ keV$^{-1}$ for LiF.
In the case of CaWO$_4$ a larger exposure and smaller background level is required.

\appendix

\section{Experimental features of the different targets}
\label{sec:targets}
In this Appendix we detail the experimental parameters that has been used for each detector. We have considered natural abundances of the different isotopes for a given target. 

\begin{table}
\begin{center}
\begin{tabular}{|c|cccc|}
\hline
Target 
&\begin{minipage}{2cm}\begin{center}$\{E_T,\, E_{max}\}$\\ (keV )\end{center}\end{minipage}
&\begin{minipage}{1cm}\begin{center}$\sigma(E)$\\ (keV)\end{center}\end{minipage}
&\begin{minipage}{1cm}\begin{center}$\Delta E$\\ (keV)\end{center}\end{minipage}
&\begin{minipage}{4cm}\begin{center}Background (kg$^{-1}$day$^{-1}$keV$^{-1}$)\end{center}\end{minipage}\\
\hline
Ge &$\{10,\,100\}$ & $\sqrt{(0.3)^2+(0.06)^2\, E/{\rm keV}}$ &5&$4 \times 10^{-8}$\\
Xe &$\{8.4,\,44.6\}$ & $0.6\sqrt{E/{\rm keV}}$ &3.64&$4\times10^{-9}$\\
C$_3$FI &$\{10,\,200\}$ & - &- &$4.1\times10^{-8}$\\
\hline
CaWO$_4$ &$\{10,\,100\}$ &  5\% FWHM & 5&\\
Al$_2$O$_3$ &$\{10,\,100\}$ &  5\% FWHM & 5&\\
LiF &$\{10,\,100\}$ &  5\% FWHM & 5&\\
\hline
\end{tabular}
\end{center}
\caption{Energy range, parameterization of the resolution and background for the different detectors considered in this work. The background level for the first three targets is inspired on estimates for SuperCDMS \cite{talk}, XENON1T \cite{Selvi:2011zz,Levy:2011zz},  and COUPP \cite{coupp:taup11}, that we consider energy-independent for simplicity.  }
\label{tab:targets}
\end{table}

We divide the energy window sensitive to recoils (from $E_{T}$ to $E_{ max}$) into $N_{bin}$ evenly spaced bins with a size of $\Delta E=5$ keV, with $N_{bin}=(E_{max}-E_{T})/{\Delta E}$. The estimated number of events in each bin is kept as a decimal number, without rounding it to an integer. This is, in principle, not physical and might overestimate the ability to discriminate different spectra. However it allows us to neglect the dependence of the reconstructed parameters on the particular realization chosen for the nominal number of events in the different bins (see Ref.\,\cite{Strege:2012kv}). This is an important source of uncertainty that has to be taken into account when dealing with real data, but we decide to neglect it here in order to study and estimate complementarity in a scenario uncompromised by statistical fluctuations, and also because, as found in Ref. \cite{Strege:2012kv} the relevance of those statistical fluctuations decreases as the number of experiments increases, thus we assume that coverage is good enough when dealing with signals from three detectors. Notice that, in the case of LiF, the thresholds obtained to this date are far from the 10 keV value used in this work. Additional R\&D is needed on this target before using it for DM detectors.

In our analysis the energy resolution of the detector $\sigma$ is included as a convolution of the differential rate with a Gaussian function with a variance $\sigma^2$ which depends on the recoil energy and on the particular experiment (see Table\,\ref{tab:targets}).

\section{Uncertainties in the spin-dependent structure function}
\label{app:sdsf}

We have incorporated nuclear uncertainties in the SD structure functions for each nucleus. Following the procedure described in Ref.\,\cite{Cerdeno:2012ix} we use a three-parameter function to describe the various components of the SD structure functions for the various target nuclei,
\begin{equation}
	S_{ij}(u)=N\left((1-\beta)e^{-\alpha u} + \beta\right)\,.
	\label{eqn:family}
\end{equation}
The parameters $N$, $\alpha$ and $\beta$ are varied in such a way that the family of curves provides an envelope for the existing theoretical calculations. These quantities are computed using nuclear physics models, and the results may differ, depending on the methodology and the potential used to describe the nuclear interaction. For the targets under consideration we extracted the predictions from Refs.\,\cite{Bednyakov:2006ux,Bednyakov:2004xq} and references therein. Since we have concentrated on the case $a_p/a_n=-1$, only the $S_{11}$ component is important in our analysis.\footnote{A prescription which is independent of the DM model is possible at zero momentum transfer \cite{Tovey:2000mm}. It has also been argued that the similar momentum dependence of the spin-dependent structure functions $S_{ij}(q)$ can be used to extract a common form factor \cite{Cannoni:2011iu}.} 
We summarise in Table\,\ref{tab:sdsf} the resulting ranges for $N$, $\alpha$ and $\beta$ used in our calculations. For some isotopes shell model computations of the form factors are not available. In these cases we use the approximation 
\begin{equation}
S_{ij}(u) = S_{ij}(0) e^{-q^2 R^2/4}\,,
\end{equation}
which works well in the low momentum transfer regime, and where $R$ is an effective radius which is a function of the atomic number $A$. In these cases a fixed value of $\alpha$ is used and $\beta=0$ and the only quantity that is varied is $N$, which parameterizes the zero momentum value of the SD structure function.

\begin{table}
\begin{center}
\begin{tabular}{|c|ccc|}
\hline
Isotope &$N$	 &$\alpha$	 &$\beta$\\
\hline
$^{7}$Li &$0.129-0.077$	&$2.25$	&$0$\\
$^{17}$O &$0.133-0.010$	&$3.16$	&$0$\\
$^{19}$F &$0.144-0.110$	&$3.03$	&$0.009-0$\\
$^{27}$Al &$0.202-0.146$	&$1.54$	&$0$\\
$^{73}$Ge &$0.206-0.117$	&$6.00-5.04$	&0.04-0.02\\
$^{129}$Xe &$0.052-0.029$	&$4.66-4.20$	&0.007-0.001\\
$^{131}$Xe &$0.025-0.017$	&$5.00-4.28$	&$0.061-0.042$\\
$^{183}$W &$0.015-0.0005$	&$3.61$	&$0$\\
\hline\end{tabular}
\end{center}
\caption{\label{tab:sdsf} Ranges of the inputs $N$, $\alpha$ and $\beta$ used in the parameterization of the $S_{11}$ term of the SD structure function of the different target nuclei used in the text. A flat probability distribution is assumed for each parameter.}
\end{table}

\bigskip

\noindent{\bf \large Acknowledgments}

The authors would like to thank A.M.~Green, A.~Ibarra, B.~Kavanagh and M.~Pato for useful comments. D.G.C. is supported by the Ram\'on y Cajal program of the Spanish MICINN and also thanks the Technische Universit\"at M\"unchen and the Excellence Cluster Universe for their hospitality and partial support during a stage of this project. 
M.F is supported by a Leverhulme Trust grant.
C.C. and M.P. are supported by a MultiDark Scholarship.
Y.O. is supported by a MultiDark Fellowship. 
We also thank the support of the Consolider-Ingenio 2010 programme under grant MULTIDARK CSD2009-00064, the Spanish MICINN under Grants No. FPA2009-08958 and FPA2012-34694, the Spanish MINECO ``Centro de excelencia Severo Ochoa Program" under grant No. SEV-2012-0249, the Community of Madrid under Grant No. HEPHACOS S2009/ESP-1473, the Spanish and the European Regional Development Fund MINECO-FEDER under grant FPA2011-23749, the Government of Arag\'on, and the European Union under the Marie Curie-ITN Program No. PITN-GA-2009-237920.

\end{document}